\documentclass[twocolumn,trackchanges]{aastex63}

\usepackage{graphicx}
\usepackage{amsmath}
\usepackage{breqn}
\usepackage{amssymb}
\usepackage{hyperref}
\usepackage{xcolor}
\usepackage{multirow}
\usepackage{booktabs}
\usepackage{float}

\usepackage{graphics}

\usepackage{amsmath}
\usepackage{amssymb}

\usepackage{needspace}

\usepackage{mathrsfs}

\newcommand{\descendantseedmass}{M_{\mathrm{ESD}}}

\newcommand{\seedmass}{M_{\mathrm{seed}}}

\newcommand{\massembly}{M^{\mathrm{galaxy}}_{\mathrm{total}}}

\newcommand{\environmentseedprobability}
{P_{\mathrm{seed}}^{\mathrm{env}}}

\begin{document}

\title[BH assembly under heavy seeds]
{Supermassive Black Hole Assembly from Heavy Seeds with Dynamical Friction in the \texttt{BRAHMA} Simulations: Implications for JWST, LISA, and the Local Universe}

\correspondingauthor{Aklant K. Bhowmick}
\email{aklant.app@gmail.com}

\author[0000-0002-7080-2864]{Aklant K. Bhowmick}
\affiliation{Department of Astronomy, University of Virginia, 530 McCormick Road, Charlottesville, VA 22904}
\affiliation{Virginia Institute for Theoretical Astronomy, University of Virginia, Charlottesville, VA 22904, USA}
\affiliation{The NSF-Simons AI Institute for Cosmic Origins, USA}

\author[0000-0002-2183-1087]{Laura Blecha}
\affiliation{Department of Astronomy, University of California Berkeley, Gainesville, FL 32611, USA}

\author[0000-0002-5653-0786]{Paul Torrey}
\affiliation{Department of Astronomy, University of Virginia, 530 McCormick Road, Charlottesville, VA 22904}
\affiliation{Virginia Institute for Theoretical Astronomy, University of Virginia, Charlottesville, VA 22904, USA}
\affiliation{The NSF-Simons AI Institute for Cosmic Origins, USA}

\author{Luke Zoltan Kelley}
\affiliation{Department of Astronomy, University of California Berkeley, CA 22904, USA}

\author{Rachel S. Somerville}
\affiliation{Center for Computational Astrophysics, Flatiron institute, New York, NY 10010, USA} 

\author[0000-0001-6260-9709]{Rainer Weinberger}
\affiliation{Leibniz Institute for Astrophysics Potsdam (AIP), An der Sternwarte 16, 14482 Potsdam, Germany}

\author[0000-0001-6950-1629]{Priyamvada Natarajan}
\affiliation{Dept. of Astronomy, Yale University, 219 Prospect Street, New Haven, CT 06511, USA}
\affiliation{Dept. of Physics, 217 Prospect Street, Yale University, New Haven, CT 06511,USA}
\affiliation{Black Hole Initiative, Harvard University, 20 Garden Street, Cambridge, MA 02138, USA}

\author{Tiziana Di Matteo}
\affiliation{McWilliams Center for Cosmology, 
Carnegie Mellon  University, Pittsburgh, PA 15213, USA} 

\author[0000-0001-6950-1629]{Lars Hernquist}
\affiliation{Harvard-Smithsonian Center for Astrophysics, Harvard University, Cambridge, MA 02138, USA}

\author[0000-0001-8593-7692]{Mark Vogelsberger}
\affiliation{Department of Physics and Kavli Institute for Astrophysics and Space Research, Massachusetts Institute of Technology,Cambridge, MA 02139, USA}

\author[0000-0002-8111-9884]{Alex M. Garcia}
\affiliation{Department of Astronomy, University of Virginia, 530 McCormick Road, Charlottesville, VA 22904}
\affiliation{Virginia Institute for Theoretical Astronomy, University of Virginia, Charlottesville, VA 22904, USA}
\affiliation{The NSF-Simons AI Institute for Cosmic Origins, USA}

\begin{abstract}
The \textit{JWST} discoveries of supermassive black holes~(BHs) at $z \gtrsim 5$ may provide key insights into their seeding origins. Using new $[18{-}72~\rm Mpc]^3$ \texttt{BRAHMA} cosmological simulations, we investigate how variations in heavy-seed prescriptions, coupled with a subgrid dynamical friction model, shape BH populations at $z \sim 5$ and $z \sim 0$. We consider two ``lenient" seed models, in which all halos containing sufficient dense \& metal-poor gas form $\sim10^4$ and $\sim10^5~M_{\odot}$ seeds, and a ``strict" seed model, in which $\sim10^5~M_{\odot}$ seeds form only under additional constraints motivated by direct collapse black hole formation. By $z \sim 5$, all models produce  $M_*-M_{\rm BH}$ relations broadly consistent with the observed local Universe for $M_*\gtrsim10^9~M_{\odot}$ galaxies, but only the lenient scenarios generate systems near the upper envelope of the observed local scatter. In galaxies hosting $M_{\rm BH} \sim 10^8$--$10^9~M_{\odot}$ BHs, lenient production of $\sim10^5~M_{\odot}$ seeds also produces multiple overmassive systems with $M_{\rm BH}/M_* \gtrsim 0.01$. Although their growth is dominated by seeding and mergers, these systems reach luminosities of $\sim10^{43}$--$10^{45}~\mathrm{erg~s^{-1}}$, comparable to those inferred for \textit{JWST}-detected BHs. As a key observational signature, the lenient seed models yield merger rates of $\gtrsim100~\mathrm{yr^{-1}}$ and near-unity local BH occupation fractions even in galaxies with $M_* \lesssim 10^7~M_{\odot}$. In contrast, the strict seed model produces merger rates of only $\sim1~\mathrm{yr^{-1}}$ and local occupation fractions of $\lesssim10\%$ for galaxies with $M_* \lesssim 10^8~M_{\odot}$. Future gravitational-wave event rates and measurements of local BH occupation fractions will therefore provide strong constraints on the dominant pathways responsible for high-redshift BH assembly.
\end{abstract}


\section{Introduction}

The origin of the ``seeds” of supermassive black holes~(SMBHs) remains elusive. The traditional candidates are Population III stellar remnants (Pop III seeds), with characteristic masses of $\sim 10^2~M_{\odot}$~\citep{2001ApJ...550..372F,2001ApJ...551L..27M,2013ApJ...773...83X,2018MNRAS.480.3762S}. However, the discovery of billion-solar-mass quasars at $z \gtrsim 6$~\citep{2001AJ....122.2833F,2010AJ....139..906W,2011Natur.474..616M,2015MNRAS.453.2259V,2016ApJ...833..222J,2016Banados,2017MNRAS.468.4702R,2018ApJS..237....5M,2018ApJ...869L...9W,2018Natur.553..473B,2019ApJ...872L...2M,2019AJ....157..236Y,2021ApJ...907L...1W} presented significant challenges for “light” seeds, as reaching the observed masses would require sustained super-Eddington accretion over several hundred million years. To alleviate this stringent growth requirement, scenarios involving more massive initial seeds have been proposed. A popular scenario is Direct Collapse Black Hole formation~(DCBH seeds), which suggests that under specific conditions primordial gas clouds could collapse directly into seeds with masses of $\sim 10^4$–$10^5~M_{\odot}$~\citep{2003ApJ...596...34B,2006MNRAS.371.1813L,2007MNRAS.377L..64L,2006MNRAS.370..289B,2018MNRAS.476.3523L,2019Natur.566...85W,2020MNRAS.492.4917L,2023MNRAS.526L..94B}. These “heavy” seeds are expected to be rare, as avoiding fragmentation may for instance, require the gas to be metal-free and exposed to sufficiently strong Lyman–Werner (LW) radiation to suppress molecular hydrogen formation as well as other proposed mechanisms for preventing fragmentation. Nevertheless, since $z \gtrsim 6$ quasars are themselves rare ($\sim 1~\mathrm{Gpc}^{-3}$), DCBHs remain compelling candidates for explaining their existence.

\subsection{JWST and the Rapidly Evolving Observational Landscape}

The \textit{James Webb Space Telescope} (\textit{JWST}) is radically transforming the observational landscape of the earliest SMBHs, revealing a much larger population of lower-luminosity systems at $z \sim 4$–$7$ that were previously inaccessible~\citep{2023ApJ...942L..17O,2023ApJ...959...39H,2023ApJ...954L...4K,2023arXiv230801230M,2023ApJ...953L..29L,2023arXiv230905714G,2024arXiv240403576K,2024A&A...685A..25A,2025ApJ...991...37A,2025ApJ...985..169D,2026ApJ..1002..129B}. In addition, JWST has extended the redshift frontier by detecting a handful of BHs out to $z \sim 9$–$11$~\citep{2023ApJ...953L..29L,2024Natur.627...59M,2024NatAs...8..126B,2023ApJ...955L..24G,2024ApJ...965L..21K}. One of the detected sources, UHZ1 at $z=10.1$, accompanied by a \textit{Chandra} detection,  also appears to be consistent with predictions for DCBHs \citep{2024NatAs...8..126B,2024ApJ...960L...1N}.  

Several unexpected features have emerged from these recent observations. First, the inferred luminosity functions lie significantly above extrapolations from pre-JWST quasars at similar redshifts. Second, a substantial fraction of these sources exhibit highly compact morphologies and red rest-frame UV colors, often referred to as “Little Red Dots (LRDs)”~\citep{2023ApJ...959...39H,2023ApJ...954L...4K,2023arXiv230801230M,2023arXiv230905714G,2024ApJ...963..129M,2024A&A...691A..52K,2024ApJ...968...38K,2025ApJ...978...92L,2025NatAs...9.1732S,2025ApJ...979..138H}. Third, many of these BHs appear “overmassive”~($M_{\mathrm{BH}}/M_* \gtrsim 0.01$) relative to their inferred host galaxy stellar masses compared to expectations based on local BH–galaxy scaling relations~\citep{2023ApJ...957L...7K,2024NatAs...8..126B,2024ApJ...960L...1N,2024ApJ...968...38K,2024arXiv240403576K,2024arXiv240610329D}.

Several works have also suggested that current JWST BH mass measurements may be overestimated by factors of $\sim10$--$100$, particularly for LRDs if they are enshrouded by dense gas and undergoing accretion rates close to or exceeding the Eddington limit~\citep{2025arXiv250316596N,2026Natur.649..574R}. In the case of super-Eddington accretion, it is also possible that the standard single-epoch virial relations used to estimate BH masses may no longer hold~\citep{2024A&A...689A.128L}. In such scenarios, these high-$z$ JWST BHs could readily fall onto the local relations. There is also the possibility that the stellar masses are underestimated due to the difficulty of separating the stellar and AGN components in the SED~\citep{2020MNRAS.499.4325R}.

In this work, we will be drawing implications under the assumption that the current measurements of overmassive BHs can be taken at face value. Such overmassive systems have been predicted as a potential signature of heavy-seed formation scenarios~\citep{2013MNRAS.432.3438A,2017ApJ...838..117N,2018ApJ...865L...9V,2023MNRAS.519.2155S,2024ApJ...960L...1N,2024MNRAS.531.4584S}. Taken together with their inferred (over) abundances, these potentially overmassive BHs raise the possibility that heavy-seed formation may be more common than previously assumed, operating through additional pathways beyond the restrictive high redshift direct-collapse conditions. If a sufficiently large population of heavy seeds formed at early times, their subsequent mergers could generate gravitational waves detectable out to $z \sim 15$–$20$ by the upcoming \textit{Laser Interferometer Space Antenna} (\textit{LISA})~\citep{2017arXiv170200786A}.

\subsection{Summary of previous theoretical work}

Over the years, several promising pathways have been identified to enhance the efficiency of heavy-seed formation. Runaway stellar collisions in dense nuclear star clusters (NSCs) were originally proposed as a pathway to produce intermediate-mass seeds of $\sim 10^3$–$10^4~M_{\odot}$~\citep{2011ApJ...740L..42D,2014MNRAS.442.3616L,2020MNRAS.498.5652K,2021MNRAS.503.1051D,2021MNRAS.tmp.1381D}. However, ultra-dense NSCs, with stellar densities $\gtrsim 10^{8}~\mathrm{pc}^{-3}$, could further boost initial seed masses to $\sim 10^{4}$–$10^{6}~M_{\odot}$~\citep{2023PhRvD.108h3012K}. \cite{2025ApJ...994...40P} also showed that seeds with masses $\gtrsim 10^{4}~M_{\odot}$ may be produced in LRDs through rapid stellar collisions and mergers. High redshift gas rich NSCs could also act as incubators for amplifying seed masses by wind-fed accretion that circumvents the Eddington limit \citep{TalPN2014}. Meanwhile, \citet{2021MNRAS.501.1413N} showed that NSCs can induce supra-exponential BH growth, again yielding seed masses of $\sim 10^{4}$–$10^{5}~M_{\odot}$ across cosmic time, suggesting that BH seed formation could be an extended process. More recently, \citet{2025A&A...695A..97D} demonstrated that feedback-free starbursts, often invoked to explain the highest-redshift JWST galaxies, could form star clusters in which global dynamical instabilities~(gravo-gyro instability) assemble $\sim 10^{4}~M_{\odot}$ seeds that may subsequently merge into more massive BHs. In terms of enhancing the frequency of DCBH formation channels, \citet{2019Natur.566...85W} and \citet{2020OJAp....3E..15R} showed that dynamical heating during rapid halo assembly~(e.g. during major mergers) can sustain large gas inflow rates and substantially relax the requirement for extremely strong Lyman–Werner (LW) fluxes ($\gtrsim 10^3~J_{21}$) traditionally assumed in such scenarios. Finally, recent high-resolution zoom simulations suggest that a small fraction of Pop III remnants may undergo a hyper-Eddington accretion episode right after formation, allowing them to grow rapidly to $\gtrsim 10^{4}~M_{\odot}$~\citep{2026NatAs.tmp...21M}.

Large-volume cosmological simulations provide a powerful framework for predicting the statistical imprint of different seeding scenarios on BH populations and their observational signatures. Yet, the physical processes relevant for seed formation operate on scales that cannot be resolved in such simulations. Moreover, in many of the largest-volume runs, the baryonic resolution elements are themselves more massive ($\gtrsim 10^6~M_{\odot}$) than the postulated seed masses of $\sim 10^2$–$10^5~M_{\odot}$, preventing these simulations from capturing the earliest stages of seed BH growth. Therefore, while many simulations implement physically motivated prescriptions inspired by Pop III, NSC, or DCBH formation scenarios~\citep{2014MNRAS.442.2751T,2016MNRAS.463..529H,2017MNRAS.467.4739K,2017MNRAS.470.1121T,2024arXiv240218773J,2025MNRAS.542.2597C}, the relevant small-scale physics (e.g., cooling, fragmentation, star formation, angular momentum transport, and disk instabilities) must be accounted for via subgrid prescriptions that may fail to capture their effects and hence carry substantial uncertainties

To address the impact of these small-scale uncertainties, we developed a set of novel gas-based seeding prescriptions for cosmological simulations that depend on a wide range of gas properties, including gas density, metallicity, Lyman–Werner (LW) radiation intensity, angular momentum, and environmental richness. By conditioning seed formation on combinations of these properties, we constructed a flexible and unified seeding framework that enables exploration of multiple seed formation channels within a single simulation suite through systematic variations of the adopted criteria. For example, some prescriptions rely solely on the presence of actively star-forming, low-metallicity gas, thereby approximating Pop III or NSC seeding~\citep{2021MNRAS.507.2012B}. Other models additionally require enhanced LW radiation and/or low gas angular momentum, more closely reflecting canonical DCBH scenarios~\citep{2022MNRAS.510..177B}. To emulate dynamical-heating-induced DCBH formation, we further introduced a criterion to seed BHs in rich environments comprising merging galaxies~\citep{2024MNRAS.533.1907B}. Within each channel, small-scale uncertainties in seed formation translate into seeding parameters whose values are not known {\it a priori}; we therefore explore a range of plausible choices and their observable consequences. Finally, we developed a stochastic seeding model that statistically maps these physically motivated gas-based prescriptions into larger cosmological volumes where the target seed masses cannot be resolved directly~\citep{2024MNRAS.529.3768B}. Together, these models form the basis of the \texttt{BRAHMA} suite of cosmological simulations~\citep{2024MNRAS.531.4311B,2024MNRAS.533.1907B,2025MNRAS.538..518B}, enabling a systematic study of how uncertainties in BH seeding propagate into the observable properties of BH populations.

In \cite{2024MNRAS.533.1907B}, we showed that an overmassive $M_*–M_{\mathrm{BH}}$ relation at $z \sim 5$ can be produced with heavy ($\sim 10^5~M_{\odot}$) seeds only if two conditions are simultaneously satisfied. First, the seeds must form in abundant numbers only requiring moderate amounts of LW radiation~($\lesssim 10~J_{21}$).  
Second, BH mergers must occur efficiently within $\lesssim 750~\mathrm{Myr}$ following the merger of their host halos. However, because those original \texttt{BRAHMA} simulations employed BH repositioning, we were unable to directly model the merger delay times. We have since implemented a subgrid dynamical-friction prescription based on \cite{2023MNRAS.519.5543M} within the \texttt{BRAHMA} framework. Using this model, we predicted the merger time-scales and rates of lower-mass ($\sim 10^3~M_{\odot}$) seeds~\citep{2025ApJ...991...81B}. In our most recent work~\citep{2026ApJ...997..187B}, we applied this framework to study the assembly of the earliest BHs at $z \sim 9$–$11$, finding that mergers dominate the initial growth phases and are critical for assembling systems such as GN-z11 and CEERS-1019 under standard assumptions for AGN and stellar feedback. However, incorporating explicit dynamical friction significantly reduces the efficiency of BH–BH mergers relative to repositioning-based models. 
As a result, assembling the most massive candidate BHs observed at $z \sim 9$–$11$, such as UHZ1, GHZ9, and CAPERS-LRD-z9, becomes substantially more challenging.     

\subsection{Focus of this work}

In this work, we expand our \texttt{BRAHMA} simulation suite with new simulation boxes designed to probe the assembly of $z\sim5$ BHs and their evolution to $z\sim0$. Compared to our previous study with a similar focus~\citep{2024MNRAS.533.1907B}, this work introduces three key advances. First, we investigate the role of mergers in the assembly of $z\sim5$ BHs using a subgrid dynamical friction model rather than BH repositioning, allowing us to assess the impact of dynamical delays and explore the implications of current \textit{JWST} measurements for future \textit{LISA} detections. Second, we employ our stochastic BH seeding framework~\citep{2024MNRAS.529.3768B} to represent heavy-seed models in larger-volume, lower-resolution simulations extending to $[72~\rm Mpc]^3$. Third, we simultaneously examine the $z\sim5$ and $z\sim0$ BH populations, enabling us to assess the implications of current \textit{JWST} measurements all the way to the local Universe. 

\section{Methods}
\label{Methods}

Our simulations used the \texttt{AREPO} gravity + magneto-hydrodynamics~(MHD) solver~\citep{2010MNRAS.401..791S,2011MNRAS.418.1392P,2016MNRAS.462.2603P,2020ApJS..248...32W} that uses the PM Tree~\citep{1986Natur.324..446B} based gravity solver coupled with ideal MHD equations for the gas. The MHD solver uses a dynamic unstructured grid generated via a Voronoi tessellation of the domain. Our simulations are initialized at $z=127$ using the \texttt{MUSIC}~\citep{2011MNRAS.415.2101H} initial condition generator with the underlying cosmology adopted from the \cite{2016A&A...594A..13P} ~i.e.~$\Omega_{\Lambda}=0.6911, \Omega_m=0.3089, \Omega_b=0.0486, H_0=67.74~\mathrm{km}~\mathrm{s}^{-1}\mathrm{Mpc}^{-1},\sigma_8=0.8159, n_s=0.9667$. Halos and Subhalos are identified using the friends of friends~(FOF, ~\citealt{1985ApJ...292..371D}) and the SUBFIND algorithm~\citep{2001MNRAS.328..726S} respectively. The FOF finder uses a linking length of 0.2 times the mean particle separation.  

With the exception of BH physics, we adopt the \texttt{IllustrisTNG} galaxy formation model~\citep{2018MNRAS.473.4077P, 2017MNRAS.465.3291W} which itself is built upon the Illustris model~\citep{2013MNRAS.436.3031V,2014MNRAS.438.1985T}.
The cooling modules include contributions from both primordial species~($\mathrm{H},\mathrm{H}^{+},\mathrm{He},\mathrm{He}^{+},\mathrm{He}^{++}$) and metals. Primordial cooling rates are calculated based on \citealt{1996ApJS..105...19K}) whereas the metal cooling rates are interpolated from pre-calculated tables as in \cite{2008MNRAS.385.1443S} in the presence of a spatially uniform, time dependent UV background. Star formation occurs at densities exceeding $0.13~\mathrm{cm}^{-3}$ with a time scale of $2.2~\mathrm{Gyr}$. The star forming gas represents an unresolved multiphase interstellar medium described by an effective equation of state~\citep{2003MNRAS.339..289S,2014MNRAS.444.1518V}. Star particles represent unresolved single stellar populations that are characterized by their ages and metallicities. The underlying initial mass function is adopted from \cite{2003PASP..115..763C}. The subsequent stellar evolution is modeled based on \cite{2013MNRAS.436.3031V} with modifications for \texttt{IllustrisTNG} as described in \cite{2018MNRAS.473.4077P}. The stellar evolution leads to chemical enrichment of stars, which is modeled by following the evolution of seven species of metals~(C, N, O, Ne, Mg, Si, Fe) in addition to H and He. Stellar and Type Ia/II Supernova feedback are modeled as galactic scale winds~\citep{2018MNRAS.475..648P} that deposit mass, momentum and metals on to the gas surrounding the star particles. This leads to the chemical enrichment of gas, which is otherwise assigned an initial metallicity of $7\times10^{-8}~Z_{\odot}$. For readers interested in further details, please refer to \cite{2018MNRAS.473.4077P}.

\subsection{Black hole seeding}
\label{Black hole seed models}

The most distinct feature of the \texttt{BRAHMA} simulations is their novel implementation of BH seeding. \texttt{BRAHMA} employs two distinct sets of seeding prescriptions depending on both the simulation’s mass resolution and the target seed mass, corresponding to the physical seeding channel being modeled.

\subsubsection{Direct-Gas-Based~(DGB) Seed Models}
When the target seed mass is well resolved by the simulation’s gas mass resolution~($\rm \seedmass \gtrsim M^{gas}_{\rm target}$), we adopt a class of prescriptions that seed BHs directly based on the local gas properties within halos. These ``direct-gas-based" or ``DGB" seed models are motivated by the physical conditions expected at the formation sites of Pop~III, NSC, and DCBH seeds, and have been developed and extensively validated in our previous high-resolution zoom-in studies~\citep{2021MNRAS.507.2012B,2022MNRAS.510..177B,2024MNRAS.529.3768B}. For example, all three seeding channels are expected to operate predominantly within dense and pristine gas environments. NSC seeding  may be expected to additionally require sufficiently deep gravitational potentials. Finally, DCBH seeding may require conditions of strong LW radiation, low gas spins or dynamical heating during major mergers to combat fragmentation from $H_2$ cooling. To capture these environments, we have designed the following seeding criteria: 

\textbf{Dense and metal-poor gas mass criterion.} Halos are required to contain a minimum gas mass ($5~M_{\rm seed}$) that is simultaneously dense ($>0.13~\mathrm{cm}^{-3}$; the star formation threshold) and extremely metal poor ($Z < 10^{-4} Z_\odot$). 

\textbf{Halo mass criterion.} Seeds are placed only in halos that exceed a critical threshold in total halo mass. In practice, this threshold is naturally imposed by the simulation owing to the mass resolution and halo identification limits~($>32$ DM particles). 
For the DGB-based simulations, this corresponds to $6.1\times10^7~M_{\odot}$ and $7.8\times10^6~M_{\odot}$ for the two adapted resolutions.   

\textbf{Lyman--Werner flux criterion.} This criterion requires that dense, metal-poor gas satisfying the seeding conditions also be exposed to a Lyman--Werner (LW) radiation field exceeding a critical threshold, $J_{\rm crit}$. Star formation is suppressed in such regions. In this work we adopt $J_{\rm crit} = 10~J_{21}$. This value is substantially lower than the canonical thresholds of $\gtrsim 1000~J_{21}$ inferred from small-scale radiation--hydrodynamics and one-zone models of DCBH formation~\citep{2010MNRAS.402.1249S,2014MNRAS.445..544S}, which are too restrictive to produce any seeds within our cosmological volumes~\citep{2022MNRAS.510..177B}. Instead, our adopted value is motivated by dynamical-heating-induced DCBH formation scenarios, which make lower critical fluxes $\sim 10~J_{21}$ feasible~\citep{2019Natur.566...85W,2020OJAp....3E..15R}. Because we do not explicitly solve radiative transfer, the LW radiation field is computed using the semi-empirical prescription described in Section~2.1.2 of \cite{2022MNRAS.510..177B}.  

\textbf{Gas spin criterion.} This criterion restricts seeding based on the specific angular momentum of gas within the halo. A seed forms if
\begin{equation}
\lambda = \frac{|\vec{\mathbf{J}}{\rm spin}|}{\sqrt{2}, M_{\rm gas} R_{\rm vir} V_{\rm vir}} < \lambda_{\rm max},
\end{equation}
where $\lambda$ is the dimensionless spin parameter, and $\lambda_{\rm max}$ corresponds to the maximum spin allowing Toomre-unstable collapse. $M_{\rm gas}, R_{\rm vir}~\&~V_{\rm vir}$ are the gas mass, virial radius and the virial velocity of the halos. This approach follows stability analyses of primordial disks~\citep{2006MNRAS.371.1813L} and subsequent implementations in cosmological simulations~\citep{2012MNRAS.422.2051N,2020MNRAS.491.4973D}. Details are provided in Section 2.1.1 of \cite{2022MNRAS.510..177B}.

\textbf{Halo environment criterion.} To identify candidate halos for dynamical-heating-induced DCBH formation, we use this criterion to restrict seed formation in halos with at least one neighboring halo of comparable or greater mass within $5~R_{\rm vir}$. Although this distance is somewhat arbitrary, it is small enough to effectively select halos in rich environments. A full description is given in \cite{2024MNRAS.529.3768B}. Since we use very low critical LW fluxes of $10~J_{21}$, our environmental criterion is intended to identify halos where DCBH formation could be induced under such low fluxes. Therefore, we always apply this criterion together with the LW flux criterion.

\subsubsection{Stochastic Seed Model for Extrapolated Seed Descendants (ESDs)}

When the target seed mass falls below the gas-mass resolution of the simulation 
($\seedmass \lesssim M_{\rm target}^{\rm gas}$), we employ the Stochastic Seed Model 
introduced in \citet{2024MNRAS.529.3768B}. In this framework, we do not attempt to seed 
BHs directly at the unresolved target mass. Instead, we stochastically initialize the
smallest \emph{resolvable} seed descendants based on how they naturally emerge in 
higher-resolution simulations that use the DGB seed models and resolve our target 
seed masses. We refer to these descendant BHs as \textit{Extrapolated Seed Descendants}
(ESDs). To this end, for every high-resolution simulation we run using the DGB model 
(hereafter DGB simulations), we calibrate and execute corresponding simulations at 
lower resolution and larger volume that employ the stochastic seed model (ESD 
simulations), such that they represent the same underlying seeding scenario.

The ESDs are placed within subhalos identified using the same FOF 
algorithm employed for halo identification, but with the linking length reduced by a 
factor of three. We refer to these short-linking-length FOF groups as 
``best friends-of-friends'' or ``bFOFs''. See section 4.1 of \cite{2024MNRAS.529.3768B} for a detailed study of the bFOF properties and their relationship to their host halos. For simplicity, we will hereafter refer to these 
bFOFs as ``galaxies.'' ESDs are initialized with a mass 
$\rm M_{ESD}$~(close to the gas-mass resolution) inside 
galaxies according to two key criteria:

\textbf{Stochastic Galaxy Mass Criterion.}  
ESDs are seeded in galaxies whose total mass exceeds a threshold that is 
stochastically drawn from a redshift-dependent distribution. This distribution is 
calibrated using higher-resolution DGB simulations that capture the earliest growth 
phase from $\seedmass$ to $\mathrm{M}_{\mathrm{ESD}}$. The calibration ensures that, in 
lower-resolution cosmological volumes where $\seedmass$ is not directly resolved, 
ESDs are seeded in galaxies that would host descendants (with mass 
$M_{\mathrm{ESD}}$) of the target seed mass in the higher-resolution DGB simulations.

\textbf{Stochastic Galaxy Environment Criterion.}
ESDs are preferentially seeded in galaxies residing in rich environments with multiple neighboring 
galaxies.\footnote{The \textit{Stochastic Galaxy Environment Criterion} used in the ESD simulations is similar to the \textit{Halo Environment Criterion} used in the DGB simulations. However, while the former simply reduces the likelihood of seeding in isolated galaxies, the latter is deterministic and does not allow seeds to form in isolated haloes.} Galaxies with zero or one neighbor of comparable or higher mass are assigned reduced seeding probabilities, while galaxies with more than one such neighbor are assigned a probability of unity. This prescription reflects the behavior observed in the DGB simulations, where seed descendants preferentially assemble in rich environments with enhanced merger-driven growth. The seeding probabilities are calibrated such that the BH clustering in the ESD simulations broadly matches that of the higher-resolution DGB simulations.

Finally, we add an additional contribution to BH growth from unresolved minor mergers, i.e., cases in which the primary BH exceeds the ESD mass while the secondary (unresolved) BH has a mass below the ESD threshold. Further details of the stochastic seed model and its calibration for different seed formation scenarios are provided in Appendix~\ref{appendix_stochastic}.

\subsubsection{Seeding scenarios explored in this work}
\label{Seeding scenarios explored in this work}

In this paper, we explore three DGB seeding scenarios:

\begin{itemize}

\item \texttt{Lenient-Heavy:}
Assumes a seed mass of $\seedmass = 1.5\times10^{5}~M_{\odot}$ with seeds  formed in any halo that satisfies both the \textit{Dense and metal-poor gas mass criterion} and the \textit{Halo Mass criterion}.

\item \texttt{Lenient-LowMass:}
Adopts the same seeding conditions as \texttt{Lenient-Heavy}, but with a lower seed mass of $\seedmass = 1.8\times10^{4}~M_{\odot}$.

\item \texttt{Strict-Heavy:}
Assumes a seed mass of $\seedmass = 1.5\times10^{5}~M_{\odot}$, but requires a more restrictive environment for seed formation. In addition to the \textit{Dense and metal-poor gas mass criterion} and the \textit{Halo Mass criterion}, we also impose the \textit{LW flux criterion} and the \textit{Halo-Environment criterion}.
\end{itemize}

The \texttt{Lenient-Heavy} and \texttt{Lenient-LowMass} seed models most naturally correspond to physical scenarios capable of producing $\sim10^{4}$–$10^{5}~M_{\odot}$ BH seeds either through NSC-based runaway collapse or through rapid, intermittent hyper-Eddington growth experienced by a small fraction of Pop~III remnants~\citep{2024OJAp....7E.107M,2026NatAs.tmp...21M}. In contrast, the \texttt{Strict-Heavy} model is more representative of the dynamical-heating-induced DCBH formation pathway~\citep{2019Natur.566...85W}. 

Finally, for each of these DGB scenarios, we calibrate the parameters of our stochastic seed model using an effective ESD mass of $\mathrm{M_{ESD}} = 8\times\seedmass$. As with the target seed mass $\seedmass$ adopted in the DGB simulations, this ESD mass is chosen to match the mass resolution of the corresponding parent simulations, which are by construction $\sim 8\times$ lower than the DGB runs. The resulting calibrated parameters~(defined in Appendix \ref{appendix_stochastic}) are listed in Table~\ref{tab:brahma_suite}. Notably, the parameters associated with the \textit{stochastic galaxy mass criterion} vary substantially across the different DGB models, reflecting the distinct environments in which seeds form and grow in each case. In contrast, for the \textit{stochastic galaxy environment criterion}, we adopt a single, common set of parameters for all three seed models. This choice is motivated by simplicity and by the fact that these shared parameters are sufficient to broadly reproduce the BH clustering for all the seed models~(see again Appendix \ref{appendix_stochastic}). In Appendix~\ref{Comparing ESD vs DGB model results}, we demonstrate that the calibrated stochastic seeding model can reasonably reproduce the number densities, BH mass functions, and merger rates predicted by the high-resolution DGB simulations in lower-resolution boxes.

\subsection{Modeling Black hole dynamics and mergers}

\label{Black hole dynamics and mergers}

Most of our simulations follow BH dynamics using a subgrid dynamical–friction (DF) prescription. This approach compensates for the unresolved gravitational drag as the limited numerical resolution cannot capture the formation of small-scale density wakes around moving BHs. We adopt the formulation of \citet{2023MNRAS.519.5543M}, which introduces an additional acceleration term,
\begin{equation}
    \mathbf{a}_{\rm df} = 
    \sum_i 
    \frac{\alpha_i b_i}{(1+\alpha_i^2)(r_i+r_{\rm soft})}
    \left(
        \frac{G\,\Delta m_i}{(r_i+r_{\rm soft})^2}
    \right)
    \hat{\mathbf{V}}_i ,
    \label{DF_eqn}
\end{equation}
where $\alpha_i \approx b_i V_i^2/(G M_{\rm bh})$ and $b_i \equiv r_i \left| \hat{\mathbf{r}}_i - (\hat{\mathbf{r}}_i \cdot \hat{\mathbf{V}}_i)\hat{\mathbf{V}}_i \right|$. The summation extends over all resolution elements encountered on the gravitational tree while computing the usual gravitational accelerations. For every element, $\Delta m_i$, $\mathbf{r}_i$, and $\mathbf{V}_i$ denote its mass, displacement, and velocity relative to the BH, and $r_{\rm soft}$ represents the corresponding gravitational softening length.

To avoid artificial kicks from massive DM particles that are $\sim 10$ times more massive than our seed BHs, we assign each newly formed BH a \emph{dynamical seed mass} of $24\,\seedmass$ (i.e.\ $\sim 2.3\,M_{\rm DM}$). This enhanced mass influences only the BH's gravitational response, while the \emph{true} mass used for seeding, accretion, and feedback routines gets initialized at $\seedmass$.

BH pairs are merged when they are gravitationally bound and separated by less than twice the DM softening length. Gravitational boundedness is determined based on the criterion, 
\[
    \frac{|\Delta \mathbf{v}|^2}{2} + \Delta\mathbf{a}\cdot\Delta\mathbf{r} < 0 ,
\]
where $\Delta\mathbf{v}$, $\Delta\mathbf{a}$, and $\Delta\mathbf{r}$ are the pairwise velocity, acceleration, and separation vectors. This subgrid-DF based framework for BH dynamics was investigated and extensively tested under several seed model variations in \citet{2025ApJ...991...81B}.

Finally, to understand the impact of our dynamics model on BH growth, we also run a box in which the BHs are repositioned to the local potential minima at every time-step. While this model leads to unphysically prompt mergers between BHs, it provides a useful benchmark to understand the impact of our subgrid DF model on the merger-driven BH growth.

\subsection{Modeling Black hole accretion and feedback}

To model gas accretion onto BHs, we employ a modified Bondi--Hoyle prescription given by
\begin{eqnarray}
\dot{M}_{\mathrm{bh}} &=& \mathrm{min}\bigl(\dot{M}_{\mathrm{Bondi}},\, f_{\rm Edd}\dot{M}_{\mathrm{Edd}}\bigr), \\
\dot{M}_{\mathrm{Bondi}} &=& \alpha\,\frac{4\pi G^2 M_{\mathrm{bh}}^2 \rho}{c_s^3}, \\
\dot{M}_{\mathrm{Edd}} &=& \frac{4\pi G M_{\mathrm{bh}} m_p}{\epsilon_r \sigma_T\, c},
\label{bondi_eqn}
\end{eqnarray}
where $G$ is the gravitational constant, $\rho$ denotes the local gas density, $M_{\mathrm{bh}}$ is the BH mass, $c_s$ is the sound speed, $m_p$ is the proton mass, and $\sigma_T$ is the Thomson scattering cross section. The model contains three free parameters that vary across our simulation suite: the radiative efficiency $\epsilon_r$, the boost factor $\alpha$, and the Eddington factor $f_{\rm Edd}$, which caps the accretion rate at a fixed multiple of the Eddington limit. The radiative efficiency sets the bolometric luminosity,
\[
L_{\mathrm{bol}} = \epsilon_r\,\dot{M}_{\mathrm{bh}}\, c^2 ,
\]
a fraction of which is coupled to the surrounding gas as AGN feedback. For the \texttt{Lenient-Heavy} and \texttt{Strict-Heavy} models, we adopt the default TNG settings, i.e. $\epsilon_r = 0.2$, $f_{\rm edd} = 1$, and $\alpha = 1$. For the \texttt{Lenient-LowMass} model, we instead adopt a more lenient accretion prescription, with $\epsilon_r = 0.1$, $f_{\rm edd} = 10$, and $\alpha = 100$.\footnote{We adopted a more lenient accretion model for the \texttt{Lenient-LowMass} simulations just to test whether the lower $\sim10^4~M_{\odot}$ seed masses could accrete rapidly enough to catch up with the more massive $\sim10^5~M_{\odot}$ seeds in the other two models. But it does not strongly impact the BH mass assembly at least at the earliest times when the growth is dominated by mergers.}

The feedback implementation follows two distinct modes that depend on the Eddington ratio $\eta$. For high accretion states ($\eta > \eta_{\mathrm{crit}} \equiv \mathrm{min}[0.002(M_{\mathrm{bh}}/10^8\,\mathrm{M_\odot})^2,\,0.1]$), thermal energy is injected with an efficiency of $\epsilon_{\rm f,high}=0.1$. At low accretion states ($\eta < \eta_{\mathrm{crit}}$), feedback operates kinetically by imparting momentum kicks to neighboring gas elements, with directions chosen randomly at irregular intervals.

\subsection{Simulation suite}

\begin{figure*}
\centering

\includegraphics[width=18cm]{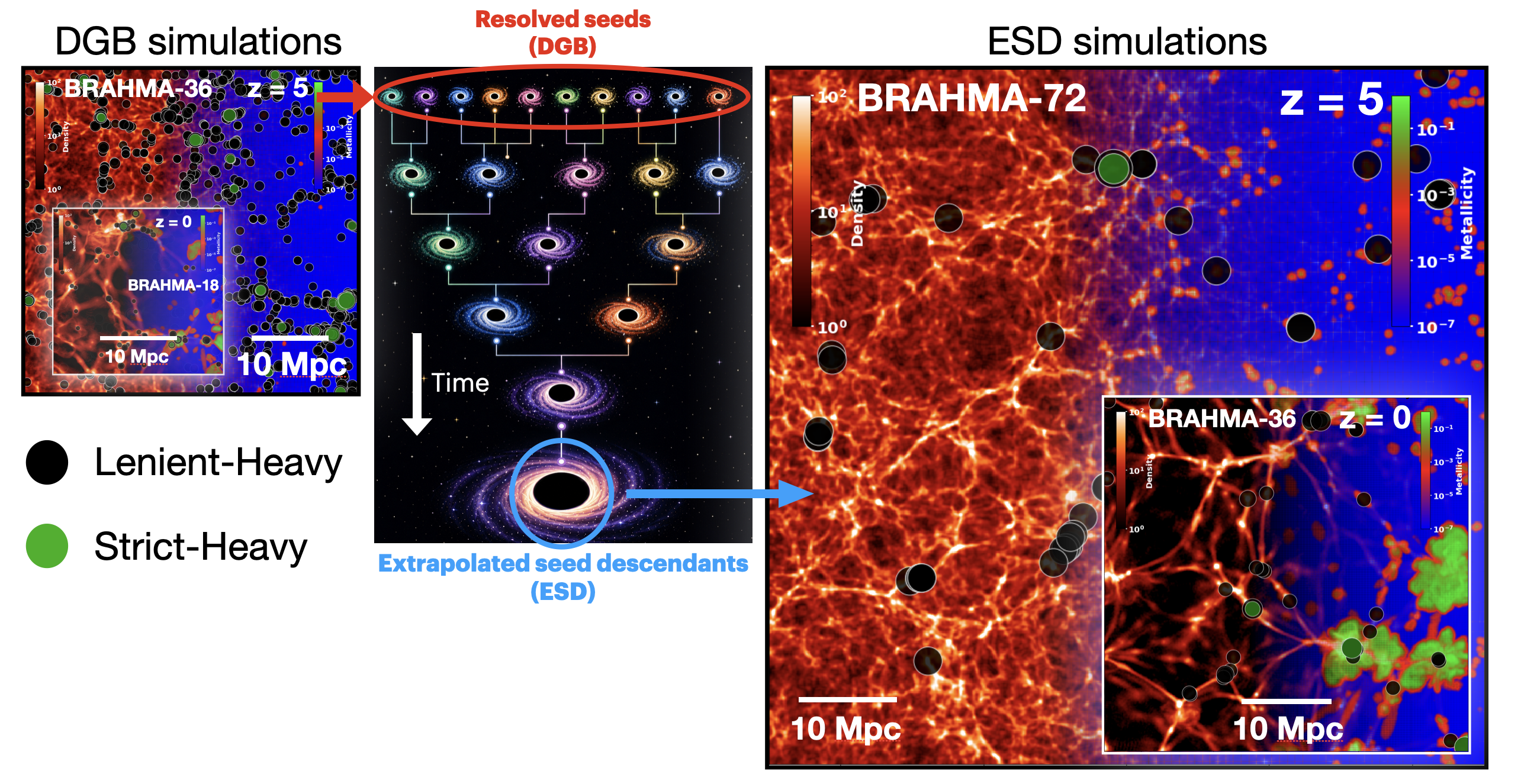}

\caption{Visualizations of the boxes at the final snapshots, illustrating the implementation of the \texttt{Lenient-Heavy} and \texttt{Strict-Heavy} seed models. The larger panels correspond to \texttt{SUITE-z5} at $z=5$, and the inset panels correspond to \texttt{SUITE-z0} at $z=0$. The left halves of the 2D maps show the gas density field, which smoothly transitions to the gas metallicity field on the right halves. Since all simulations with the same volume share identical initial conditions, we overplot the BHs produced by the different seed models on the same panels. BHs are overplotted as circles, colored black for \texttt{Lenient-Heavy} and green for \texttt{Strict-Heavy} seed models. The middle schematic~(generated using OpenAI-GPT5) illustrates how we track BH growth from the DGB simulations along a galaxy~(bFOF) merger tree. Overall, this figure summarizes our two-step seeding strategy: for each seed model, we first run high-resolution DGB boxes~(left) that resolve the target seed mass and apply the DGB seeding prescription. We then follow the subsequent growth of these seeds and identify the galaxies in which they assemble higher-mass descendant BHs. These descendants are initialized as ESDs in lower-resolution boxes~(right) with eight times larger volume.}
\label{visualizations_fig}
\end{figure*}

In this work, we have expanded our existing \texttt{BRAHMA} simulation suite to include several new uniform cosmological boxes with two key advancements:

\begin{itemize}
\item Heavy-seed DGB models from \cite{2024MNRAS.533.1907B,2025MNRAS.538..518B}, now evolved using our subgrid-DF BH dynamics model in place of BH repositioning.
\item Stochastic seed models calibrated against simulations with these resolved heavy seeds, and subsequently applied to cosmological volumes up to $[72~\mathrm{Mpc}]^{3}$. These are the largest boxes so far within the \texttt{BRAHMA} suite.
\end{itemize}

In terms of resolution and target redshift, we ran two classes of simulations. The first consists of boxes with $N = 512^{3}$ dark-matter resolution elements evolved to $z = 0$ (hereafter \texttt{SUITE-z0}), while the second consists of boxes with $N = 1024^{3}$ resolution elements evolved to $z = 5$ (hereafter \texttt{SUITE-z5}). 

\textit{DGB simulations:} For both the \texttt{SUITE-z0} and \texttt{SUITE-z5} boxes, we first ran a set of simulations that resolve our target seed masses and adopt our Direct-Gas-Based seeding prescriptions. These include the \texttt{Lenient-Heavy}, \texttt{Strict-Heavy}, and \texttt{Lenient-LowMass} models described in Section~\ref{Seeding scenarios explored in this work}. For the two heavy-seed prescriptions with $\sim 10^{5}~M_{\odot}$ seeds, the simulation volumes are $[\texttt{36~Mpc}]^{3}$ for \texttt{SUITE-z5} and $[\texttt{18~Mpc}]^{3}$ for \texttt{SUITE-z0}. For the \texttt{Lenient-Heavy} case, we also ran three additional variants to explore differences in BH dynamics, accretion, and feedback modeling with the aim of quantifying the individual impact of each subgrid choice on the BH population at $z \sim 5$ and $z\sim0$. For the \texttt{Lenient-LowMass} seed model with $\sim 10^{4}~M_{\odot}$ seeds, we ran a $[\texttt{18~Mpc}]^{3}$ box within \texttt{SUITE-z5}.

\textit{ESD simulations:} Finally, for all three of our DGB seed models, we calibrated our stochastic seed models for ESDs that are $8\times$ larger than the original seed mass. Following this, for each of our DGB boxes, we ran stochastic seed models at $8\times$ lower mass resolution inside $8\times$ larger volumes. For the models calibrated from the \texttt{Lenient-Heavy} and \texttt{Strict-Heavy} DGB simulations, we ran two $[\texttt{36~Mpc}]^{3}$ boxes for \texttt{SUITE-z0} and two $[\texttt{72~Mpc}]^{3}$ boxes for \texttt{SUITE-z5}. Likewise, for the \texttt{Lenient-LowMass} model, we ran a $[\texttt{36~Mpc}]^{3}$ box within \texttt{SUITE-z5}. Since the \texttt{Lenient-LowMass} boxes are at 8 times higher resolutions~(to resolve the lower seed masses) compared to the other two seed models, we did not run any versions of them all the way to $z=0$ due to high computational expense. 

Our overall approach for building the simulation suite 
is schematically illustrated in Figure~\ref{visualizations_fig}. The figure illustrates this procedure 
for the \texttt{Lenient-Heavy} and \texttt{Strict-Heavy} models, but the same approach 
is also applied to the \texttt{Lenient-LowMass} models. In summary, for each seed model, we first 
run the DGB simulations and track the growth of the seeds along galaxy merger trees 
to identify the galaxy masses at the epochs when the BHs exceed the ESD mass. We then 
use this information to calibrate the stochastic seed model and run the corresponding 
ESD simulations. Table \ref{tab:brahma_suite} summarizes the full simulation suite.

\begin{table*}[t]
\centering 
\hspace{-0.2 cm}

\begin{tabular}{|c|c|c|c||c|c|c|c|c|c|c|}
\hline
\multicolumn{11}{|c|}{
    \textbf{$N = 2 \times 512^{3}$ \quad ; \quad $z_{\rm final} = 0$~(\texttt{SUITE-z0})}
} \\ \hline

\multicolumn{4}{|c||}{\textbf{DGB simulations}} &
\multicolumn{7}{c|}{\textbf{ESD simulations~($M_{\rm ESD}= 8\times \seedmass$)}} 
\\ \hline

\textbf{Model} & \textbf{Seed Label} & $\seedmass$ & $N_{\rm box}$ 
& \textbf{Model} & $\rm \log M_{trans}$ & $\rm z_{trans}$ & $\sigma$ & $\alpha$ & $\beta$ & $N_{\rm box}$ 
\\ \hline

\texttt{BRAHMA-18-D5} & Lenient-Heavy  & $\sim10^5$ & 1+3
& \texttt{BRAHMA-36-E6} & 9.5 & 3 & 0.18 & -0.13 & -0.02 &  1 \\ \hline

\texttt{BRAHMA-18-D5} & Strict-Heavy  & $\sim10^5$ & 1
& \texttt{BRAHMA-36-E6} & 10.4 & 5 & 0.25 & -0.12 & 0 & 1 
\\ \hline

\multicolumn{11}{c}{} \\[-6pt]   
\multicolumn{11}{c}{\rule{0pt}{20pt}} \\[-6pt]  

\hline
\multicolumn{11}{|c|}{
    \textbf{$N = 2 \times 1024^{3}$ \quad ; \quad $z_{\rm final} = 5$}~(\texttt{SUITE-z5})
} \\ \hline

\multicolumn{4}{|c||}{\textbf{DGB simulations}} &
\multicolumn{7}{c|}{\textbf{ESD simulations~($M_{\rm ESD}= 8\times \seedmass$)}} 
\\ \hline

\textbf{Model} & \textbf{Seed Label} & $\seedmass$ & $N_{\rm box}$ 
& \textbf{Model} & $\rm \log M_{trans}$ & $\rm z_{trans}$ & $\sigma$ & $\alpha$ & $\beta$ & $N_{\rm box}$ 
\\ \hline

\texttt{BRAHMA-36-D5} & Lenient-Heavy & $\sim10^5$ & 1+3
& \texttt{BRAHMA-72-E6} & 9.5 & 3 & 0.18 & -0.14 & -0.02 &  1 \\ \hline

\texttt{BRAHMA-36-D5} & Strict-Heavy  & $\sim10^5$ & 1
& \texttt{BRAHMA-72-E6} & 10.4 & 5 & 0.25 & -0.12 & 0 & 1 
\\ \hline

\texttt{BRAHMA-18-D4} & Lenient-LowMass & $\sim10^4$ & 1
& \texttt{BRAHMA-36-E5} & -- & -- & -- & -- & -- &  1 \\ \hline

\end{tabular}

\caption{Overview of the new \texttt{BRAHMA} boxes created in this work. \texttt{SUITE-z0} simulations~(upper half) with $N_{\rm DM} = 512^3$ are run to $z=0$ for the \texttt{Lenient-Heavy} and \texttt{Strict-Heavy} seed models. \texttt{SUITE-z5} simulations~(lower half) with $N_{\rm DM} = 1024^3$ are run to $z=5$ for all three seed models. Due to the lower seed mass, the  \texttt{Lenient-LowMass} simulations are 8 times higher in resolution and smaller in volume compared to the other models. DGB simulations $[18-36~\rm Mpc]^3$~(left side) are denoted as \texttt{BRAHMA-*-D*}, where the first `*' represents the box-size, and the second `*' represents the approximate gas mass resolution in $\log10[M_{\odot}]$ units. For all the DGB simulations, there are corresponding ESD simulations~(right side) that are 8 times larger~($[36-72~\rm Mpc]^3$), that are similarly denoted by \texttt{BRAHMA-*-E*}. For these ESD boxes, we show values of the 5 parameters that determine the \textit{stochastic galaxy mass criterion}, calibrated from the higher resolution DGB simulations. Finally, for the \texttt{Lenient-Heavy} DGB simulations, `$N_{\rm box} = 1+3$' means that we run three additional simulations with dynamics, accretion and feedback model variations.}
\label{tab:brahma_suite}
\end{table*}

\section{Results}
\label{Results}

\subsection{BH number density evolution and seeding models}

\begin{figure*}
\centering
\includegraphics[width= 18cm]{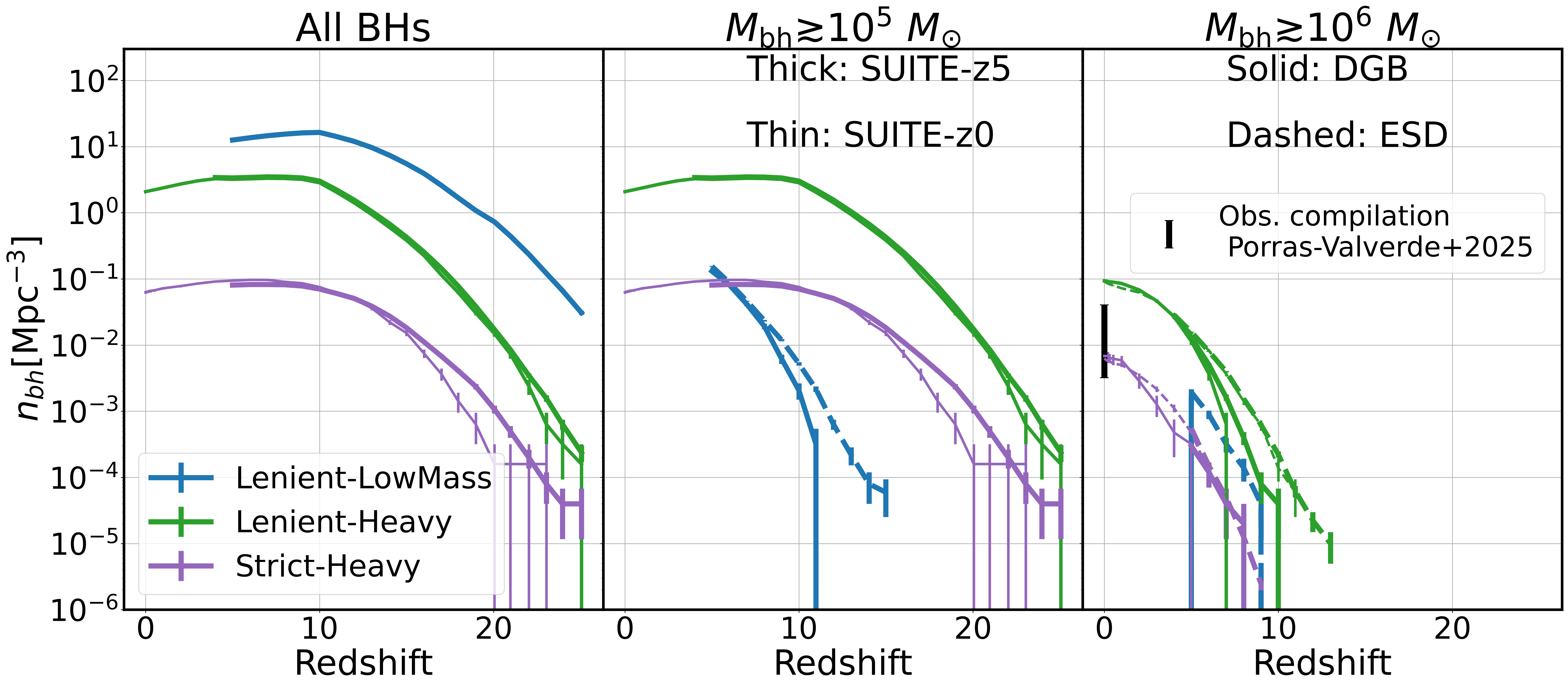}
\caption{Evolution of the BH number density for our three different seed models. The leftmost panel shows the overall number density, whereas middle and right panels show the number density of $\gtrsim10^5~M_{\odot}$ and $\gtrsim10^6~M_{\odot}$ BHs respectively. In the rightmost panel, the black marker shows the range of observational constraints obtained by integrating the range of $z=0$ BHMF measurements compiled by \cite{2026ApJ...998...48P} and shown by the grey region in Figure~\ref{BHMF_AGNLF_a}. While the \texttt{Lenient-LowMass} model produces the highest numbers of seeds at $\sim10^4~M_{\odot}$, it starts to assemble $\gtrsim10^5~M_{\odot}$ BHs much later than the \texttt{Lenient-Heavy} and \texttt{Strict-Heavy} models that seed at $\sim10^5~M_{\odot}$.}
\label{Number_density}
\end{figure*}

In Figure~\ref{Number_density}, we show the evolution of the BH number density for our three seed models. For a given seed model, the results from the \texttt{SUITE-z5} (solid lines) and \texttt{SUITE-z0} (thinner lines) boxes are naturally very similar, with small differences arising from their different volumes due to cosmic variance. This cosmic variance is strongest at the highest redshifts. As intended, we also find that our ESD-based lower-resolution boxes (dashed lines) reproduce number densities that are reasonably close to those of the DGB-based higher-resolution boxes. The differences between the ESD and DGB boxes (for the same seed model) are largest at the highest redshifts. This may be partly driven by cosmic variance, since the ESD boxes are larger than the DGB boxes. But in Appendix~\ref{Comparing ESD vs DGB model results}, we perform supplementary ESD runs at the DGB volumes to demonstrate that this discrepancy is largely due to imperfect calibration of the stochastic seeding model at the highest redshifts, where BH statistics are limited. At lower redshifts, where robust calibration is possible, the differences between the ESD and DGB boxes become negligible. Even at the highest redshifts, these differences remain much smaller than the differences between the seed models themselves. Therefore, our stochastic seed model performs sufficiently well for the purposes of this work, although we plan a more precise calibration in future work using a substantially larger suite of simulations. Notably, our initial studies of the stochastic seed model \citep{2024MNRAS.529.3768B,2024MNRAS.531.4311B} tested this framework only in simulations with BH repositioning, whereas here we demonstrate that it also performs reasonably well when using subgrid dynamical friction.  

The leftmost panel of Figure \ref{Number_density} presents the number densities of all BHs present in the simulations across different snapshots. Initially, the number densities increase for all seed models up to $z \sim 10$, driven primarily by the formation of new seeds. As demonstrated in our previous works~\citep{2025MNRAS.538..518B,2025ApJ...991...81B}, seed formation becomes suppressed at $z \lesssim 10$ due to a combination of metal enrichment and a reduction in the gas content of low-mass haloes caused by stellar feedback. In this regime, the number densities exhibit a mild decline as BHs merge with one another.

As expected, the \texttt{Lenient-LowMass} seed model produces the highest number of seeds at $\sim10^{4}~M_{\odot}$, reaching a peak number density of $\sim20~\mathrm{Mpc}^{-3}$ by $z \sim 10$. This is because the dense, metal-poor gas mass threshold in our model scales with the seed mass. The \texttt{Lenient-Heavy} seed model yields lower peak abundances of $\sim4~\mathrm{Mpc}^{-3}$ at $z \sim 10$, as the formation of more massive $\sim10^{5}~M_{\odot}$ seeds requires substantially larger reservoirs of dense~\&~metal-poor gas. Finally, the \texttt{Strict-Heavy} seed model produces the lowest peak BH number densities, reaching only $\sim0.08~\mathrm{Mpc}^{-3}$ at $z \sim 10$, since the additional requirements imposed by the \textit{Lyman–Werner flux}, \textit{gas spin}, and \textit{galaxy environment} criteria collectively act to strongly suppress seed formation~\citep{2024MNRAS.533.1907B}.

Despite producing the highest number of seeds overall, the \texttt{Lenient-LowMass} seed model begins producing $\gtrsim10^{5}~M_{\odot}$ BHs only at $z\lesssim10$, significantly later than the \texttt{Lenient-Heavy} and \texttt{Strict-Heavy} models (second panel of Figure~\ref{Number_density}). This delay indicates that assembling $\sim10^{5}~M_{\odot}$ BHs from lower mass $\sim10^4~M_{\odot}$ seeds takes longer than forming them directly as $\sim10^5~M_{\odot}$ even under the \texttt{Strict-Heavy} model. As we shall see, the delays in BH mergers under our subgrid-DF model play a dominant role in slowing down early BH growth. Consequently, at $z\gtrsim7$, the \texttt{Lenient-Heavy} and \texttt{Strict-Heavy} models yield substantially higher abundances of $\gtrsim10^{5}~M_{\odot}$ BHs than the \texttt{Lenient-LowMass} model. At lower redshifts, however, the heavy-seed models stop forming new seeds and their number densities begin to decline, whereas the \texttt{Lenient-LowMass} model continues to rise and eventually surpasses the \texttt{Strict-Heavy} model at $z\lesssim6$. This is because, in the \texttt{Lenient-LowMass} model, $\gtrsim10^{5}~M_{\odot}$ BHs can continue to emerge via growth of the pre-existing $\sim10^4~M_{\odot}$ seeds long after the formation of new seeds is suppressed.

Finally, for $\gtrsim10^{6} M_{\odot}$ BHs, even the \texttt{Lenient-Heavy} seed model—which produces the highest abundance of $\gtrsim10^{5} M_{\odot}$ BHs—begins assembling this more massive population only after $z\sim13$. This again, is largely due to the delayed BH mergers under our subgrid-DF model. The corresponding number density then increases monotonically, reaching $\rm \sim10^{-1} Mpc^{-3}$ by $z\sim0$. The \texttt{Strict-Heavy} seed model starts producing $\gtrsim10^{6} M_{\odot}$ BHs later, at $z\sim10$, and yields number densities of $\rm \sim10^{-2} Mpc^{-3}$ at $z\sim0$. For the \texttt{Lenient-LowMass} model, the statistics are poorer owing to the smaller simulated volumes; nevertheless, the resulting number densities appear to lie between those of the \texttt{Lenient-Heavy} and \texttt{Strict-Heavy} models.

\begin{figure}
\centering
\includegraphics[width= 8cm]{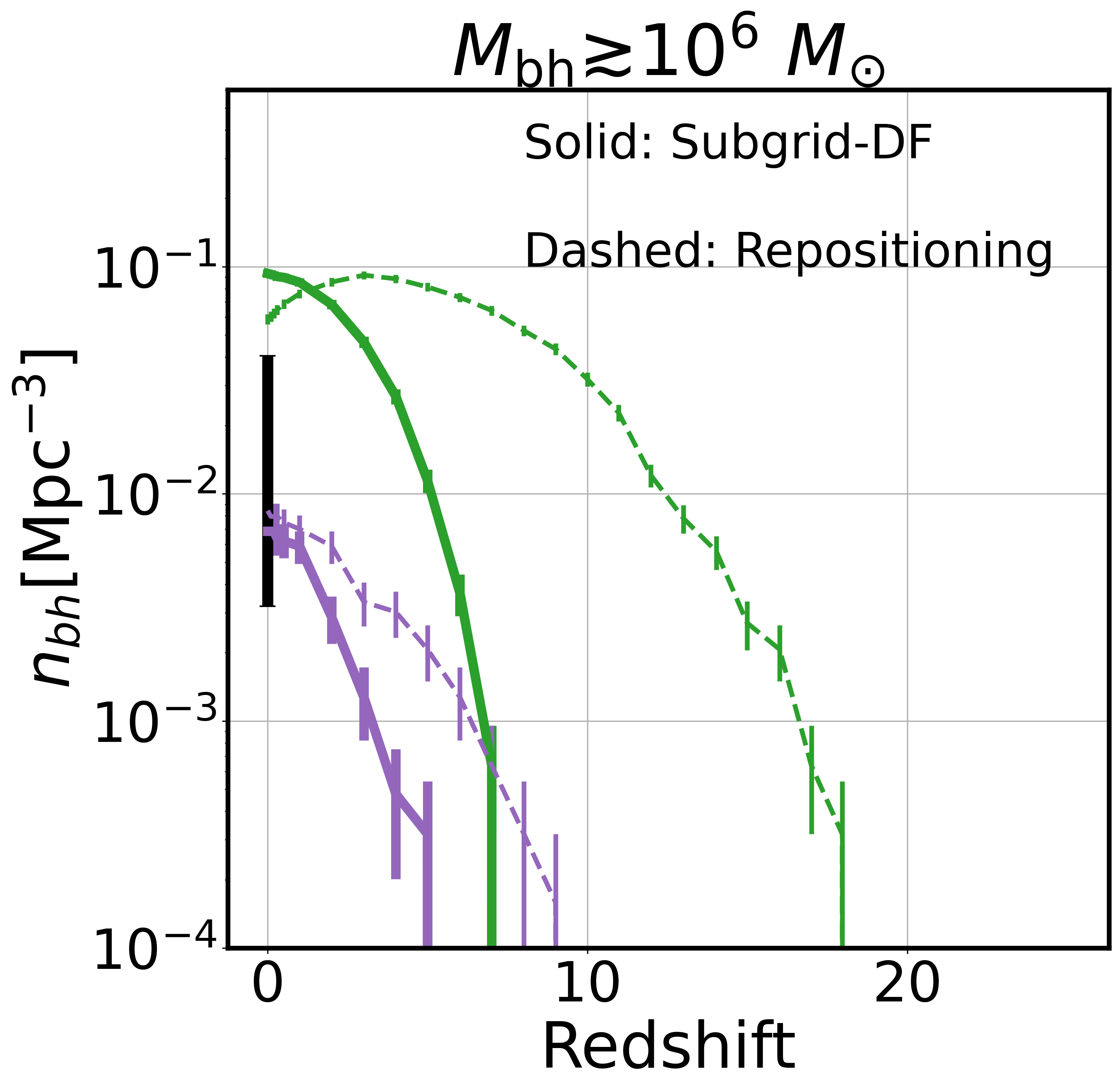}
\caption{Similar to the previous figure, but here we compare the number density evolution produced by our subgrid-DF based \texttt{BRAHMA-18-D5} boxes vs. prior results using repositioning~\citep{2025MNRAS.538..518B} for our \texttt{Lenient-Heavy} and \texttt{Strict-Heavy} seed models. Under subgrid-DF, there is a substantial delay  in the assembly of $\gtrsim10^{6}~M_{\odot}$ BHs compared to respositioning.}
\label{Number_density_zero}
\end{figure}

To further assess the impact of subgrid dynamical friction (DF), Figure~\ref{Number_density_zero} compares the number density evolution of $\gtrsim10^{6} M_{\odot}$ BHs with our previous results that employed BH repositioning \citep{2025MNRAS.538..518B}. We find that such massive BHs emerge significantly earlier when BH repositioning is used. In principle, BH growth under subgrid DF may be suppressed through two channels—reduced accretion rates and delayed BH mergers—since BHs are no longer artificially forced to reside in the densest gas regions. As we explicitly demonstrate in the next section, the dominant effect arises from delayed mergers. Consequently, the discrepancy between the number densities obtained with subgrid DF and BH repositioning is most pronounced at high redshifts. By $z\sim0$, however, both schemes yield similar number densities, as the majority of the delayed mergers have occurred by the present day. 

Observational constraints on the local number density of $\gtrsim10^{6} M_{\odot}$ BHs span the range $\rm \sim5\times10^{-3} Mpc^{-3}$ to $\rm 6\times10^{-2} Mpc^{-3}$. While the \texttt{Lenient-Heavy} models tend to lie mildly above the upper end of these constraints, the \texttt{Strict-Heavy} models generally fall toward the lower end. We emphasize again that these two prescriptions were deliberately chosen to bracket a plausible range of heavy seeding scenarios in order to assess the impact of seeding on BH population statistics. In \cite{2025MNRAS.538..518B}, we also presented intermediate seed models that yield local BH number densities well within the range of observational constraints. We note, however, that these observational constraints are obtained by integrating the BH mass function constraints discussed in Section~\ref{BH mass functions sec}, which remain highly uncertain, particularly at the low-mass end.

\subsection{BH mass assembly from mergers and seeding vs gas accretion}
\begin{figure*}
\centering
\includegraphics[width= 18cm]{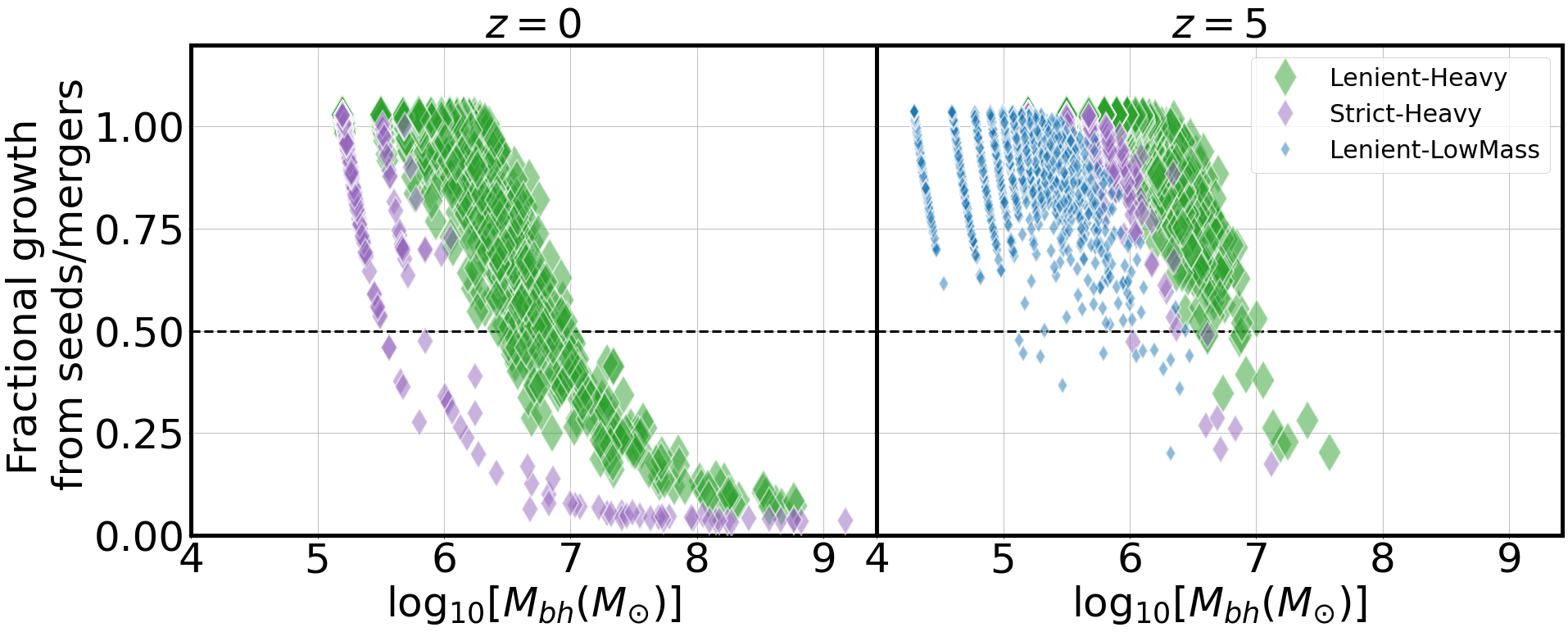}
\caption{For the DGB simulations, we show the fractional contribution to BH growth from seeding and mergers as opposed to gas accretion. For the different seed models, this is shown as a function of BH mass for populations at $z=0$~(left) and $z=5$~(right). In the early phase of BH growth, the majority ($\gtrsim50\%$) of the BH mass is contributed by seeding and mergers, while gas accretion becomes dominant in the later phase. }
\label{mergers_vs_accretion_z0}
\end{figure*}

Figure \ref{mergers_vs_accretion_z0} uses the DGB models to quantify the fractional contribution to BH mass assembly from BH seeding and mergers, as opposed to gas accretion. We show both local ($z=0$) and high-redshift ($z=5$) BH populations for the \texttt{Lenient-Heavy} and \texttt{Strict-Heavy} seed models using the \texttt{SUITE-z0} and \texttt{SUITE-z5} boxes respectively. For the \texttt{Lenient-LowMass} seed model, we show only the $z=5$ results since these simulations were not evolved to $z=0$.

At $z=5$, the $[36~\rm Mpc]^3$ \texttt{SUITE-z5} boxes produce maximum BH masses spanning $3\times10^6 M_{\odot}$ to $3\times10^7 M_{\odot}$ depending on the seed model. For all seed models, mergers and seeding contribute $\gtrsim50 \%$ of the mass assembly for the bulk of the $z=5$ BH population. Only for the most massive BHs does gas accretion become the dominant growth channel. The efficiency of this seeding \& merger driven growth is naturally set by both the abundance of seeds and the merger efficiency. Consequently, in the \texttt{Lenient-Heavy} model, accretion becomes dominant only above $\gtrsim10^7 M_{\odot}$. In contrast, in the \texttt{Strict-Heavy} model, accretion dominates already above $\sim4\times10^6 M_{\odot}$ reflecting the much smaller seed abundance and hence fewer mergers. Finally, for the \texttt{Lenient-LowMass} model, accretion is already dominant above $\sim10^6 M_{\odot}$ since the smaller initial seed masses require a larger number of successive mergers to build up comparable BH masses. Additionally, the smaller seed masses enhance the merger delay time-scales due to weaker BH dynamical friction. 

We also reported the relative dominance of mergers during the early phases of BH growth in our previous work~\citep{2024MNRAS.533.1907B}, but those simulations employed BH repositioning, which represents the most optimistic scenario for merger efficiency. Here, we show that even with subgrid DF, the earliest phases of BH growth remain dominated by mergers in our simulations. In a recent paper \citep{kho2026learninguniversehighredshifts}, we further show that this is primarily because gas accretion is strongly suppressed at these early times due to feedback from both AGN and stars. However, owing to the longer merger delay times introduced by subgrid DF (see Section~\ref{mergers_time_scales_sec}), the relative contribution of mergers is lower than in simulations that employ BH repositioning.

At $z=0$, the dominance of mergers during the early phases of BH growth persists for the \texttt{Lenient-Heavy} seed model up to $\sim6\times10^6 M_{\odot}$, similar to $z\sim5$. For the \texttt{Strict-Heavy} seed model, mergers dominate over a much narrower mass range, up to a few $\sim10^5 M_{\odot}$. Compared to $z=5$, we can probe significantly higher BH masses at $z=0$, reaching up to $\sim10^9 M_{\odot}$. For the most massive BHs with masses $\sim10^8-10^9 M_{\odot}$, the contribution from mergers becomes negligible relative to gas accretion.

Finally, several recent ``resolved-ISM" simulations~\citep{2024OJAp....7E.107M,2025arXiv251109640P,2025ApJ...993L..48W,2026A&A...708A...7Z,2026NatAs.tmp...21M}, have found brief episodes of hyper-Eddington accretion that can boost the seed mass by several orders of magnitude up to $\sim10^4-10^6~M_{\odot}$ during the earliest stages of seed BH growth. Due to the effective equation-of-state treatment of the ISM, resolving such episodes is not possible in the \texttt{BRAHMA} simulations. However, these high resolution simulations also find that BH growth is subsequently suppressed by feedback or gas depletion shortly after these initial hyper-Eddington phases, similar to the earliest resolvable BH growth regime in \texttt{BRAHMA} simulations~\citep{2026ApJ...997..187B}. It is precisely during this subsequent feedback-regulated phase that mergers~(when they occur) can become an important channel for BH assembly, as it does in \texttt{BRAHMA}. We therefore clarify that the \texttt{BRAHMA} simulations do not demonstrate that mergers are the only mechanism contributing to the earliest stages of BH growth. Rather, we show that the earliest \textit{resolvable} phase of BH growth in \texttt{BRAHMA} simulations is often merger dominated. In this interpretation, the BH seeds inserted in \texttt{BRAHMA} could represent the end products of unresolved hyper-Eddington growth episodes of smaller initial seeds prior to their regulation by feedback.

\subsection{BH mass functions and AGN luminosity functions}

\begin{figure*}
\centering

\begin{minipage}[t]{0.49\textwidth}
\centering
\includegraphics[width=\linewidth]{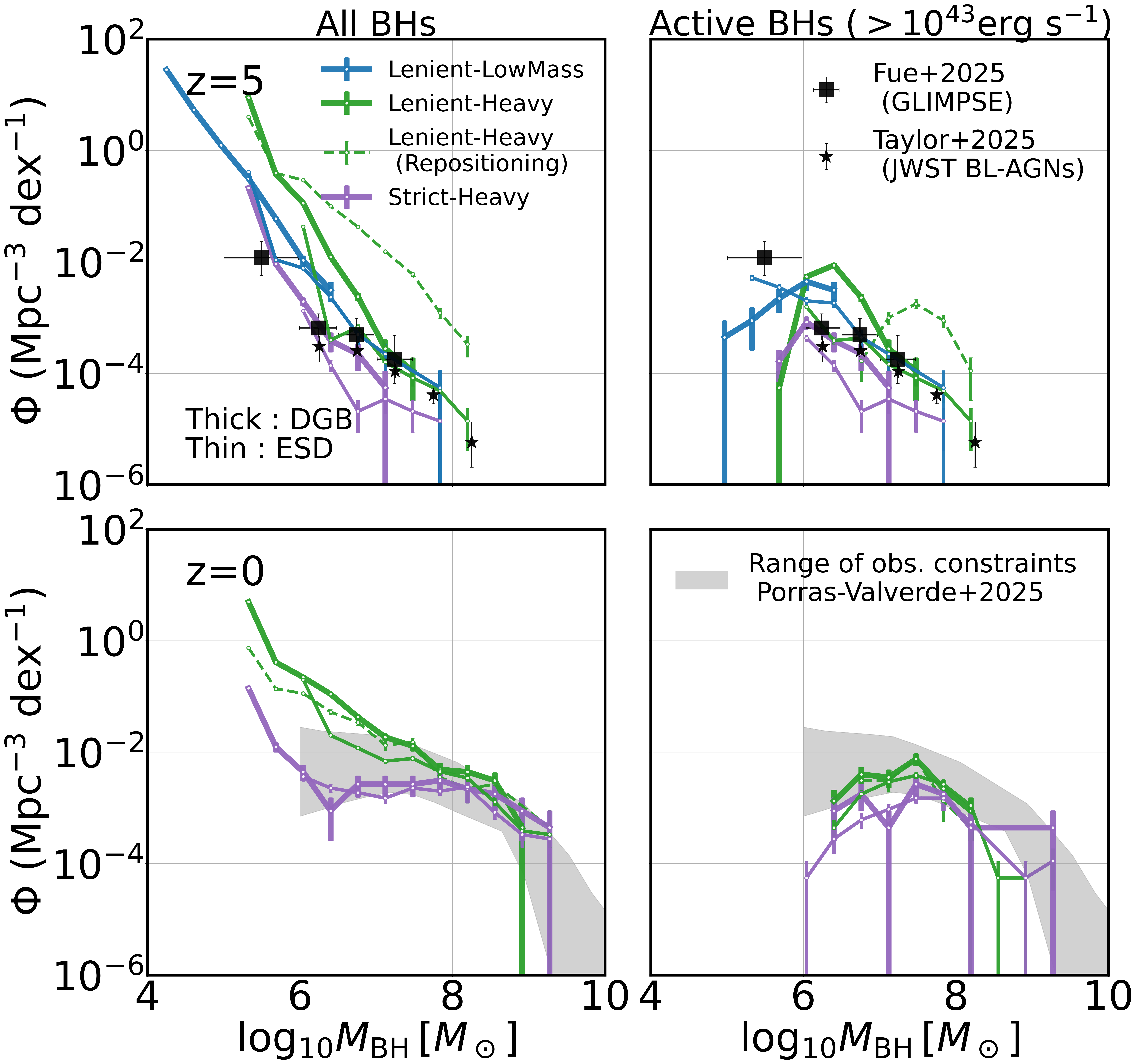}
\caption{BHMFs predicted by different seeding models at $z=5$ (top row) and $z=0$ (bottom row). Thick lines show the DGB predictions, while thin lines show the calibrated ESD predictions. The dashed lines indicate the repositioning-based predictions for the \texttt{Lenient-Heavy} seed model. The left column shows the full BH population, while the right column includes only active BHs with $L_{\rm bol} \gtrsim 10^{43}\ \mathrm{erg\ s^{-1}}$. Black data points show current observational constraints from \textit{JWST} for $z \sim 4$--$7$ BHs in the top row~\citep{2025arXiv250920452F,2024arXiv240906772T}, and the grey region spans the current spread of local constraints as compiled by \cite{2026ApJ...998...48P}. Our subgrid-DF based predictions are in broad agreement with current $z\sim5$ JWST constraints at the massive end.}
\label{BHMF_AGNLF_a}
\end{minipage}
\hfill
\begin{minipage}[t]{0.49\textwidth}
\centering
\includegraphics[width=\linewidth]{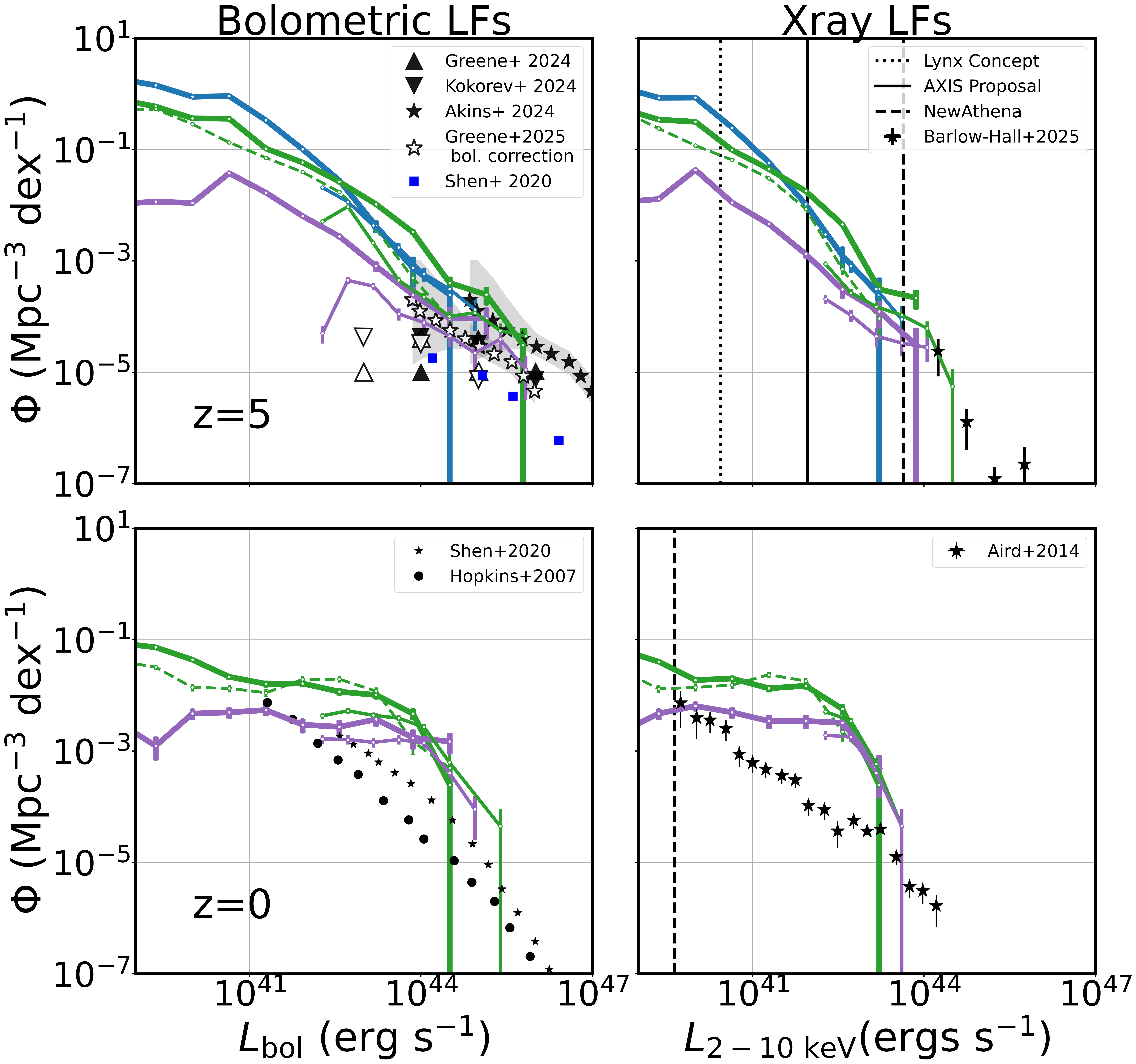}
\caption{Bolometric (left) and X-ray (right) AGNLFs at $z=5$ (top) and $z=0$ (bottom) for the DGB~(thick lines) and ESD~(thin lines) predictions. We note here that the ESD boxes produce smaller luminosities compared to the DGB boxes at $z=5$ because the BH accretion rates are not well converged at these resolutions. For the X-ray AGNLFs, we apply bolometric corrections derived by \cite{2020MNRAS.495.3252S}. The plot legends for the simulations are the same as those in the BHMF panel. Black data points correspond to observational constraints. We compare our $z\sim5$ predictions with high-$z$ bolometric AGNLFs derived from \textit{JWST}~\citep{2024ApJ...964...39G,2025ApJ...991...37A,2024ApJ...968...38K,2025arXiv250905434G}, as well as recent X-ray constraints from \cite{2025arXiv250616145B}. For the \textit{JWST} constraints, we show the original measurements (filled black points) and those adjusted using updated bolometric corrections~(open black points) from \cite{2025arXiv250905434G}. At $z\sim0$, we compare to local constraints~\citep{2007ApJ...654..731H,2015MNRAS.451.1892A,2020MNRAS.495.3252S}. Our high-$z$ AGN populations broadly align with current X-ray constraints, and are somewhat higher than the JWST constraints from LRDs.}
\label{BHMF_AGNLF_b}
\end{minipage}

\end{figure*}

Having examined the relative roles of mergers and gas accretion in BH growth, we now turn to the BH mass functions (BHMFs) and AGN luminosity functions (AGNLFs) produced by the different seeding models. For each model, we present both the higher-resolution DGB predictions and the lower-resolution ESD predictions (thick and thin lines of the same color, respectively). While the ESD simulations are designed to reproduce the DGB results at lower resolution, we nevertheless find some small differences between their predictions, which we discuss in more detail in Appendix~\ref{Comparing ESD vs DGB model results}. Despite this, the impact of seed model variations on the BHMFs and AGNLFs remains clearly visible.

\subsubsection{BH mass functions}
\label{BH mass functions sec} 

Let us first focus on the BHMFs~(Figure \ref{BHMF_AGNLF_a}). The \texttt{Lenient-Heavy} model naturally produces more BHs than the \texttt{Strict-Heavy} model across the entire BH mass range at both redshifts. However, these differences decrease with increasing BH mass. This trend reflects the changing balance between seeding and merger driven vs accretion driven growth. At low BH masses, growth is dominated by seeding and mergers and therefore highly sensitive to the total number of seeds formed. At high masses, growth is increasingly driven by gas accretion, which depends primarily on the availability of sufficiently dense, gas-rich environments. If such environments are rare, only a limited subset of seeds are able to accrete efficiently, regardless of how many seeds form overall. As a result, the massive end of the BH mass function is less sensitive to the global seeding abundance. 

The \texttt{Lenient-LowMass} model BHMF lies below that of the \texttt{Lenient-Heavy} model, but above that of the \texttt{Strict-Heavy} model in the mass range $\sim10^{5}$–$10^{7}~M_{\odot}$ at $z=5$. Additionally, the low-mass-end slopes of both the \texttt{Strict-Heavy} and \texttt{Lenient-Heavy} models are noticeably steeper than that of the \texttt{Lenient-LowMass} model. This reflects the initially faster merger-driven growth of $\sim10^{4}~M_{\odot}$ seeds compared to the $\sim10^{5}~M_{\odot}$ seeds in the heavy-seed models. This is due to two reasons: First, the lower-mass $\sim10^{4}~M_{\odot}$ seeds form in greater numbers than in either of the heavy-seed models. Second, they form in systematically lower-mass halos than the $\sim10^{5}~M_{\odot}$ seeds, and these lower-mass halos experience significantly shorter merger timescales (as we show later in Section~\ref{mergers_time_scales_sec}). Consequently, by $z=5$, the $\sim10^{4}~M_{\odot}$ seeds in the \texttt{Lenient-LowMass} model “catch up to” and surpass the $\sim10^{5}~M_{\odot}$ population produced by the \texttt{Strict-Heavy} model. However, their growth is still insufficient to overtake the more numerous $\sim10^{5}~M_{\odot}$ seeds in the \texttt{Lenient-Heavy} model.

There have been several attempts to constrain the (active) high-z BHMFs with \textit{JWST} observations~\citep{2024arXiv240906772T,2025arXiv250920452F}. While there is substantial uncertainty in these constraints at this early stage due to observational incompleteness as well as selection biases, it is still insightful to compare our predicted $z=5$ BHMFs with them. At the high-mass end ($\sim10^{7}$–$10^{8}~M_{\odot}$), we find generally good agreement with the lenient seeding models. The strict seeding model tends to lie somewhat below the observed number densities. However, observational constraints at the massive end may themselves be overestimated due to Eddington bias. At the low-mass end ($\sim10^{5}$–$10^{7}~M_{\odot}$), the lenient seeding models predict significantly higher number densities than inferred from current \textit{JWST} measurements, while the strict model is closer to the observations. However, \textit{JWST} can only detect BHs that are sufficiently luminous. Based on the latest revised bolometric corrections \citep{2025arXiv250905434G}, the observed BHs typically have bolometric luminosities in the range $\sim10^{43}$–$10^{45}~\mathrm{erg~s^{-1}}$. Motivated by this, we impose a luminosity threshold of $L_{\rm bol} > 10^{43}~\mathrm{erg~s^{-1}}$ and compute the corresponding active BHMFs (right panels of Figure~\ref{BHMF_AGNLF_a}). As expected, the active BHMFs are strongly suppressed relative to the total BHMFs at the low-mass end, bringing the predictions of the lenient seeding models closer to the observations. A fully self-consistent comparison, however, would require applying synthetic photometry and selecting simulated AGNs using the same observational criteria employed in the \textit{JWST} analyses. We therefore defer a more quantitative comparison with \textit{JWST} to future work. 

Finally, for the \texttt{Lenient-Heavy} model, we compare our subgrid-DF results with those obtained using BH repositioning (green solid and dashed lines in Figure~\ref{BHMF_AGNLF_a}). At $z\sim5$, BH repositioning produces a substantially higher normalization of the BHMF at the massive end, by factors of $\sim30$. This primarily occurs because merger-driven growth is significantly enhanced under BH repositioning. Indeed, in \cite{2024MNRAS.533.1907B}, we showed that, for the lenient seed models, nearly the entire contribution to BH mass assembly at $z\sim5$ arises from mergers when BH repositioning is adopted. As a result, the repositioning-based BHMFs substantially exceed the \textit{JWST}-inferred BHMFs for both active and inactive BHs. But again, the massive end of the observed high-$z$ BHMFs may also be affected by observational incompleteness and selection effects such as the Eddington bias. It therefore remains to be seen how these measurements will be revised with future observations from deeper and wider surveys. Nevertheless, our results underscore the importance of accurately modeling BH dynamics when interpreting observational constraints on BHMFs, particularly at high redshift.

By $z\sim0$, the BHMFs predicted by subgrid-DF and BH repositioning become much closer at the high-mass end, while subgrid-DF produces a higher normalization at the low-mass end owing to a larger population of wandering BHs. For $M_{\rm BH}\gtrsim10^{7}~M_{\odot}$, variations among the different seed models are relatively small, but become significant at lower masses ($M_{\rm BH}\lesssim10^{7}~M_{\odot}$), as also shown by \cite{2025MNRAS.538..518B} using BH repositioning. Notably, even in the local Universe, the observed BHMF remains significantly uncertain, as indicated by the gray region in Figure~\ref{BHMF_AGNLF_a} compiled by \cite{2026ApJ...998...48P} from various measurements~\citep{2008MNRAS.388.1011M,2013MNRAS.436..697B,2023MNRAS.518.2123Z,2024MNRAS.531.4503H,2024ApJ...971L..29L,2025ApJ...978...77B} . This is because local BHMF estimates are not derived from complete direct BH counts; instead, they are typically inferred by convolving galaxy stellar-mass or velocity functions with empirical scaling relations, such as $M_{\rm BH}$--$M_{*}$ or $M_{\rm BH}$--$\sigma$. Uncertainties in each of these ingredients therefore propagate into the inferred BHMF. Additionally, estimates of the low-mass end of the BHMF rely on assumptions about the BH occupation fraction, which remains poorly constrained, as discussed in Section~\ref{BH occupation fractions}. Since seed models have their strongest impact in this low-mass regime, improved observational constraints from ongoing and future surveys~\citep[e.g.][for type-1 local AGNs]{2024ApJ...969...93C} will provide valuable insights into the physics of BH seeding.

\subsubsection{AGN lumonsity functions}

Figure~\ref{BHMF_AGNLF_b} shows the AGNLFs. As expected, the strongest variations between seed models appear at the faint end, while differences diminish toward higher luminosities. This is because only a small subset of regions can sustain the accretion rates required to power the brightest AGN, largely independent of the total number of BH seeds formed. 

At $z=5$, we compare our AGNLF predictions with current \textit{JWST} measurements derived from various LRD samples (see the top-left panel). Since not all high-z BHs are LRDs, these measurements should only be interpreted as lower limits. Also, many of the original estimates \citep{2024ApJ...964...39G,2024ApJ...975..178K,2025ApJ...991...37A} were obtained using standard low-redshift bolometric corrections (filled black points). We also show revised estimates (open black points) in which these luminosities are reduced by a factor of $\sim10$, following the updated bolometric corrections of \cite{2025arXiv250905434G}. Relative to these corrected observational LFs, our predicted bolometric LFs are generally higher than the estimates of \cite{2024ApJ...964...39G,2024ApJ...975..178K}, and \cite{2025arXiv250905434G}. The LF estimates of \cite{2025ApJ...991...37A} are closer to our predictions, although they rely on the optimistic assumption that every LRD is powered by an AGN whose accretion disk emission dominates the observed continuum. Overall, these comparisons indicate that our predicted bolometric LFs are so far not in conflict with LRD-based lower limits. 

We also compare our predicted $z=5$ X-ray luminosity functions with the most recent high-$z$ measurements of \citealt{2025arXiv250616145B} (top-right panel). The overlap is limited due to the relatively small simulation volume, but it is nevertheless encouraging that at the brightest end probed by our simulations ($L_{2-10\rm keV}\sim10^{44}~\rm erg~s^{-1}$), the X-ray AGNLFs are broadly consistent with their observations. Here again, it is important to note that the low-z bolometric corrections~\citep{2020MNRAS.495.3252S} we assumed may not hold at high-z. This seems to be at least true for the LRDs, the vast majority of which are not detected in X-rays. Also, the bright end of the AGNLFs may be impacted by Eddington bias. Keeping these caveats in mind, it is nevertheless encouraging that our AGNLF predictions remain broadly consistent with current \textit{JWST} and X-ray observations.

Finally, for the \texttt{Lenient-Heavy} seed model, we also compare the AGNLFs produced by subgrid-DF and BH repositioning (green solid vs.\ dashed lines in Figure~\ref{BHMF_AGNLF_b}). Notably, while BH repositioning leads to a substantial boost in the BHMF at the massive end at $z=5$, the resulting AGNLFs are much closer to the subgrid-DF based predictions. This likely reflects the fact that the systems are in the feedback-regulated regime, wherein despite the increase in BH mass under repositioning, the accretion rates are limited by an equilibrium between the feedback energy and the gravitational binding energy of the halo. 

In fact, we find that the subgrid-DF model produces slightly higher AGNLFs (by factors of $\sim3$--$4$) across the full range of luminosities probed by our simulations. This is expected at the faint end due to the larger population of wandering BHs produced by the longer merger delay times under subgrid-DF. However, it is somewhat counter-intuitive at the bright end, since one might expect repositioning to immediately place BHs in the densest regions and therefore lead to higher accretion rates. In practice, this does occur during the earlier stages of gas accretion. However, the enhanced early accretion also leads to stronger feedback, which subsequently reduces the central gas densities at later times and ultimately results in lower BH luminosities under repositioning at $z=5$. In contrast, under subgrid-DF, accretion rates are lower at earlier stages due to the longer sinking times of BHs, but this allows the gas densities to continue increasing without encountering strong AGN feedback. As a result, when the BHs finally reach the halo centers, the resulting accretion rates can exceed those produced under BH repositioning.

By $z\sim0$, however, the bright end of the AGNLFs becomes consistent between subgrid-DF and repositioning. This may be because sufficient time has passed for the brightest AGN to settle into halo centers and remain there under both dynamical treatments, such that differences in their earlier dynamical histories are largely washed out. Consequently, our conclusions regarding the $z\sim0$ AGNLFs do not significantly differ from the repositioning-based predictions of \cite{2025MNRAS.538..518B}, to which we refer the reader for additional discussion and analysis.

\begin{figure*}[t]
\centering

\includegraphics[width= 18cm]{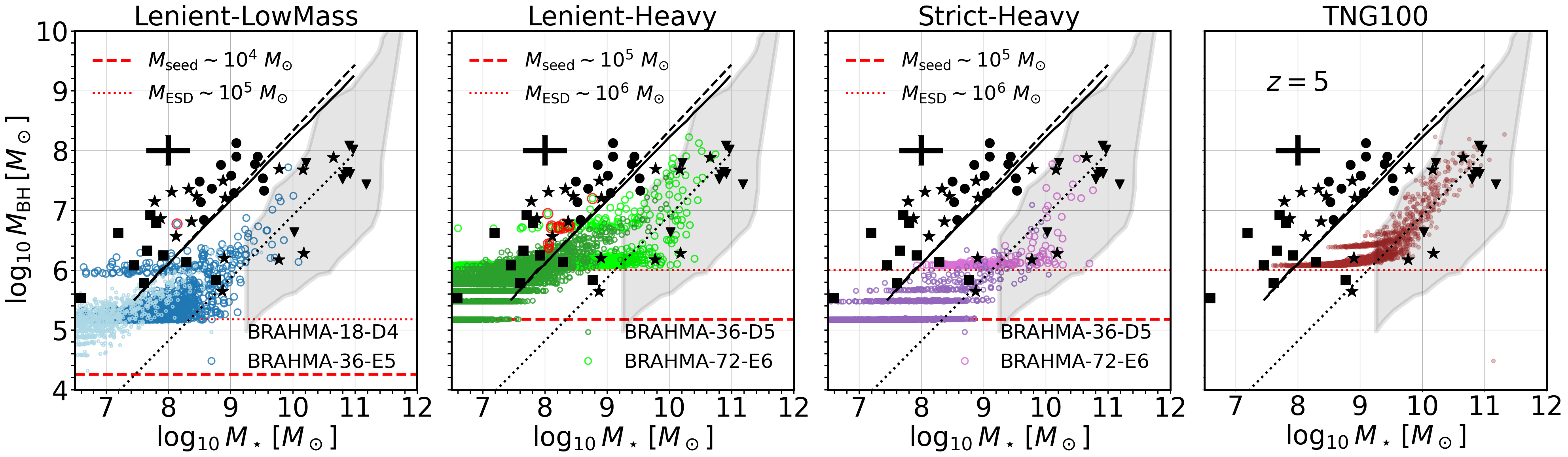}
\includegraphics[width= 18cm]{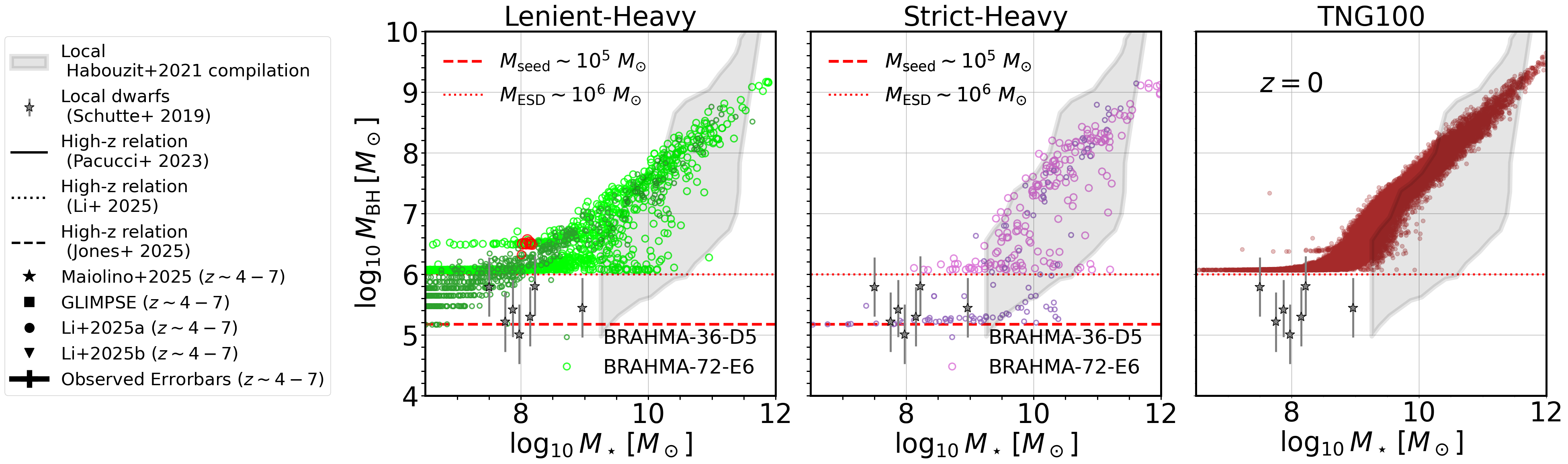}
\caption{Predictions for the $M_{*}$–$M_{\rm BH}$ relation for different seeding models. We focus exclusively on central BHs, plotting the mass of the most massive BH in each halo against the total stellar mass. Each panel corresponds to a given seeding model; smaller circles show DGB boxes and larger circles show calibrated ESD boxes. The top and bottom rows show results at $z=5$ and $z=0$, respectively. On the top row, black points show JWST measurements at $z\sim4$–$7$, while the black lines show the intrinsic high-$z$ relations inferred by \citealt{2023ApJ...957L...3P}~(solid), \citealt{2025ApJ...981...19L}~(dotted) and \citealt{2025arXiv251007376J}~(dashed) after accounting for selection biases. On the bottom row, grey stars show measurements in local dwarf galaxies~\citep{2019ApJ...887..245S}. The shaded grey region broadly represents the scatter in the local BH population compiled by \cite{2021MNRAS.503.1940H} by combining BH samples from \cite{2013ARA&A..51..511K} and  \citet{2015ApJ...813...82R}. At $z\sim5$, central BHs are broadly consistent with the local relation at $M_*\gtrsim10^{9}~M_{\odot}$. Only the lenient models produce overmassive BHs ($M_{\rm BH}/M_*\gtrsim0.01$) in $M_*\sim10^{8}$–$10^{9}~M_{\odot}$ galaxies, highlighted in red.} 
\label{mstar-mbh_lenient}
\end{figure*}

\subsection{$M_*-M_{\rm BH}$ scaling relations: Impact of BH seeding}
\label{relations: Impact of BH seeding}

We now examine the implications for the stellar mass--BH mass ($M_{*}$--$M_{\rm BH}$) relation in Figure~\ref{mstar-mbh_lenient}. For each model, we show both DGB and ESD predictions, which broadly agree as intended. However, the ESD models produce a larger number of points scattered away from the median trends, owing to their larger simulated volumes. Most importantly, the large volumes of the ESD simulations enable us to probe higher mass galaxies compared to the DGB simulations. Merger-dominated growth is evident in both cases, particularly near the seed masses where BHs populate horizontal sequences corresponding to successive merger generations. In the DGB models, the spacing between these sequences equals the seed mass. These horizontal features also arise from our simplifying assumption of a fixed seed mass; in reality, a broader seed-mass distribution is expected. In the ESD models, the spacing between sequences exceeds the nominal ESD seed mass due to additional contributions from unresolved light minor mergers accompanying each resolved heavy merger. 

We compare our predictions to BHs in the local Universe and to JWST-inferred BHs at $z\sim4$–$7$. For local benchmarks, Figure \ref{mstar-mbh_all_models} shows the gray region originally drawn by \cite{2021MNRAS.503.1940H} to broadly represent the observed scatter of local BHs. This region includes BHs with dynamical mass measurements in classical bulges and elliptical galaxies~(\citealt{2013ARA&A..51..511K}), as well as BHs with single-epoch virial mass estimates in local active galaxies~(\citealt{2015ApJ...813...82R}). For the $z\sim4$–$7$ JWST BHs, we compare against populations~(black points) detected in recent surveys which include systems reported to be overmassive \citep{2023ApJ...959...39H,2024A&A...691A.145M,2025arXiv250512867L} as well as those consistent with local BH scaling relations \citep{2025arXiv250205048L,2025arXiv250920452F}. Given that the observed distribution of JWST BHs on the $M_*-M_{\rm BH}$ plane may be affected by selection biases arising from detection limits~\citep{2007ApJ...670..249L}, we also compare our results against inferences of the intrinsic high-$z$ $M_*-M_{\rm BH}$ relation after accounting for these biases. While some studies infer that the high-$z$ $M_*-M_{\rm BH}$ relation is intrinsically overmassive relative to the local relation~\citep[][black solid and dashed lines]{2023ApJ...957L...3P,2025arXiv251007376J}, others find that it is broadly consistent with the local relation~\citep[][black dotted line]{2025ApJ...981...19L}. 

\subsubsection{Predictions of BH scaling relations at $z\sim5$}

We start with focusing on the simulation predictions at $z=5$ shown in the top panels of Figure \ref{mstar-mbh_lenient}. Our \texttt{Lenient-Heavy} and \texttt{Strict-Heavy} simulations can probe $M_* - M_{\rm BH}$ relations up to stellar masses of a few $\times 10^{10}~M_{\odot}$. Since the \texttt{Lenient-LowMass} boxes have a smaller volume, they can only probe up to a few $\times 10^{9}~M_{\odot}$. The \texttt{TNG100} simulation has a larger volume than all of the \texttt{BRAHMA} boxes and can probe galaxies up to $10^{11}~M_{\odot}$.  At the lower-mass end, if we assume that galaxies are ``well-resolved'' when they contain at least $\gtrsim 100$ star particles, then the \texttt{Lenient-Heavy} and \texttt{Strict-Heavy} DGB boxes can probe the relation down to $M_*\sim 10^{7} M_{\odot}$ galaxies, while the \texttt{Lenient-LowMass} DGB box can probe down to $M_*\sim 10^{6} M_{\odot}$.

The strongest differences between the seeding models naturally appear in the least massive galaxies ($M_* \sim 10^{7}$–$10^{8} M_{\odot}$). The \texttt{Strict-Heavy} model produces BH masses that lie closest to (though still somewhat above) a simple extrapolation of the local relation, whereas both the \texttt{Lenient-Heavy} and \texttt{Lenient-LowMass} models yield a large population of BHs that lie substantially above this extrapolation. These differences are primarily driven by the stronger merger-driven BH growth fueled by more abundant seed formation in the lenient models. The recent BHs probed by the GLIMPSE survey \citep[][black squares in Figure~\ref{mstar-mbh_lenient}]{2025arXiv250920452F} show substantially better overlap with the \texttt{Lenient-Heavy} model compared to the \texttt{Strict-Heavy} model.

Toward higher-mass galaxies, we find that $M_{\rm BH}/M_*$ generally decreases and exhibits a slope shallower than the 1:1 relation. This behavior is a natural consequence of merger-dominated BH growth and relatively weak BH accretion, which is outpaced by galaxy growth driven by both mergers and in-situ star formation. In galaxies with $M_* \sim 10^{8}$--$10^{9}\,M_{\odot}$, the BHs produced by the \texttt{Lenient-Heavy} seed model remain significantly above the extrapolation of the local region, whereas the \texttt{Strict-Heavy} model produces BHs that are largely consistent with this extrapolation. The \texttt{Lenient-LowMass} model yields $M_{\rm BH}/M_*$ ratios that lie between those of the \texttt{Lenient-Heavy} and \texttt{Strict-Heavy} models. Notably, this galaxy mass range broadly coincides with that in which \textit{JWST} has uncovered the largest populations of overmassive BHs \citep{2023ApJ...959...39H,2024A&A...691A.145M,2025arXiv250512867L}. Even in our \texttt{Lenient-Heavy} seed model, the median BH masses in these galaxies are a factor of $\sim10$ smaller than those inferred for the most extreme overmassive systems.

However, due to the significant intrinsic scatter in the $M_* - M_{\rm BH}$ relation, we still find several upscattered overmassive BHs~(with $M_{\rm BH}/M_* \gtrsim 0.01$, marked in red) in $M_*\sim 10^{8}$–$10^{9} M_{\odot}$ galaxies in the \texttt{Lenient-Heavy} model. There are 14 overmassive BHs in the larger-volume ESD box, and 5 overmassive BHs in the smaller-volume DGB box at $M_*\gtrsim10^{8}~M_{\odot}$\footnote{Note that there are even more overmassive BHs in galaxies with $M_*\lesssim10^{8}~M_{\odot}$, but we do not focus on them because our results may be affected by limited resolution in this regime, particularly in the ESD models. For example, stellar mass assembly can be delayed at lower resolution, which could artificially enhance the $M_{\rm BH}/M_*$ ratios. We plan to investigate this regime in future higher-resolution simulations.}. For the \texttt{Lenient-LowMass} model, we find only one overmassive BH in the ESD box, compared to five within the \texttt{Lenient-Heavy} DGB box with identical volume. Therefore, even though the \texttt{Lenient-LowMass} model produces larger number of seeds than the  \texttt{Lenient-Heavy} model, their smaller seed masses makes it harder to assemble overmassive BHs. Lastly, neither the \texttt{Strict-Heavy} model nor TNG100 (despite its larger volume) contains any such overmassive BHs. This is again mergers play a substantial role in the assembly of these overmassive BHs in the \texttt{Lenient-Heavy} model, while they are far less common in the \texttt{Strict-Heavy} model and in \texttt{TNG100}. 

In the most massive galaxies probed by our simulations ($M_*\sim10^{9}$--$10^{10}~M_{\odot}$), all seed models produce BH populations that are broadly consistent with locally observed systems at $z=5$. Even in this regime, however, the \texttt{Strict-Heavy} model yields a systematically lower normalization than the lenient models. While the \texttt{Strict-Heavy} model lies well within the local scatter, the lenient models populate the upper envelope of the scatter. None of our seed models produce the most extreme overmassive BHs in this regime. This may simply reflect the limited volume of our simulations, which are unlikely to sample the rarest systems. Alternatively, it may indicate that additional BH assembly channels proposed in the literature (discussed in Section~\ref{implications}) are required to explain these outliers.

\subsubsection{Predictions of BH scaling relations at $z\sim0$}

Next, we focus on $z \sim 0$, where the $M_* - M_{\rm BH}$ relation can be probed up to $M_* \sim 10^{12} M_{\odot}$~(lower panels of Figure \ref{mstar-mbh_lenient}). In the \texttt{Lenient-Heavy} model, we don't find a strong evolution in the relation relative to $z \sim 5$ for galaxies with $M_* \sim 10^{7}$--$10^{10} M_{\odot}$ 
This model also produces a small population of upscattered, overmassive BHs at $z \sim 0$ in $M_* \sim 10^{8} M_{\odot}$ galaxies. In general, identifying such upscattered BHs at $z \sim 0$ is more challenging in the local Universe due to the smaller comoving volumes probed per unit area, requiring systematic searches over large surveys. With that being said, the very recent work of \cite{2026arXiv260317967I} has indeed found a couple of overmassive BH candidates in $M_*\sim10^7~M_{\odot}$ galaxies much closer to the local Universe~($z\sim0.7$), alongside previous detections of overmassive BHs in cosmic noon~($z\sim1-3$, \citealt{2024ApJ...966L..30M}).  

In contrast, the \texttt{Strict-Heavy} model produces significantly lower-mass BHs in $M_* \sim 10^7-10^{10} M_{\odot}$ galaxies compared to the \texttt{Lenient-Heavy} model, again due to the reduced mergers in this scenario. The results of \cite{2019ApJ...887..245S} local dwarfs lie in between the predictions of the \texttt{Lenient-Heavy} and \texttt{Strict-Heavy} models; one of our intermediary models~(but still very lenient) in \citealt{2025MNRAS.538..518B}~(see Figure 8) produces the best overlap with these observations. 

At higher stellar masses ($M_* \gtrsim 10^{10} M_{\odot}$), the seeding models converge, producing BHs with $M_{\rm BH} \gtrsim 10^{8} M_{\odot}$ consistent with local BHs. In this regime, gas accretion dominates BH growth and the memory of the initial seeding is largely erased. As a result, both seeding models also agree with \texttt{TNG100} at the high-mass end. 

\subsubsection{Luminosities of the $z\sim5$ overmassive BHs produced by the lenient seed models}

\begin{figure}
\centering
\includegraphics[width= 8 cm]{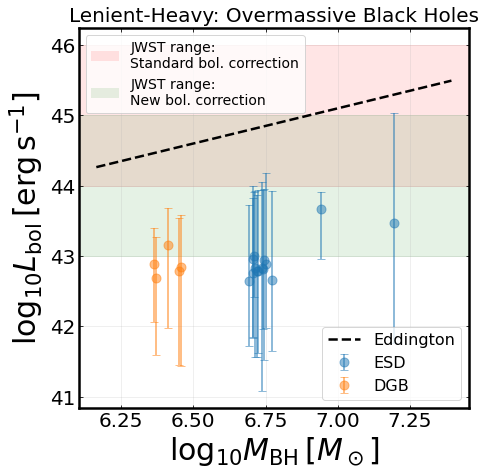}

\caption{Bolometric luminosity versus BH mass for the overmassive $z=5$ BHs in the \texttt{Lenient-Heavy} boxes. Blue and orange symbols correspond to BHs in the ESD and DGB boxes, respectively. The error bars indicate luminosity variability, estimated as the median values of the maximum and minimum luminosities in $50~\rm Myr$ time bins over $z=4.8$--$5.2$. The dashed line marks the Eddington limit. Although mass assembly in these systems is driven primarily by BH mergers, their brightest phases reach luminosities of $\sim 10^{43.5}-10^{45}~\mathrm{erg~s^{-1}}$, overlapping with luminosity estimates for JWST BHs inferred using the revised high-redshift bolometric corrections of \citet{2025arXiv250905434G}.}
\label{overmassive_luminosity_evolution}
\end{figure}

Given that the upscattered, overmassive BHs ($M_{\rm BH}/M_* \gtrsim 0.01$) produced by our lenient seed models are assembled primarily through mergers, it is instructive to examine their AGN bolometric luminosities and compare them with those inferred for $z\sim4$--7 AGNs observed by \textit{JWST}. In Figure~\ref{overmassive_luminosity_evolution}, we show the luminosity evolution of the $z\sim5$ overmassive BHs produced in the \texttt{Lenient-Heavy} boxes, together with their variability on $\sim50~\rm Myr$ timescales. The luminosities exhibit strong variability, spanning nearly two orders of magnitude. In the ESD box (top three rows), the most massive overmassive BH ($\sim2\times10^{7}~M_{\odot}$) reaches the highest luminosities, with typical values of $\sim10^{43.5}~\mathrm{erg~s^{-1}}$ and peaks approaching $\sim10^{45}~\mathrm{erg~s^{-1}}$. The remaining lower-mass overmassive BHs in the ESD box ($\sim10^{6.6}$--$10^{6.8}~M_{\odot}$) exhibit typical luminosities of $\sim10^{42.5}$--$10^{43}~\mathrm{erg~s^{-1}}$, with peak luminosities reaching $\sim10^{44}~\mathrm{erg~s^{-1}}$. Overmassive BHs in the DGB boxes (bottom two rows) are somewhat less luminous owing to their lower BH masses ($\sim10^{6.4}~M_{\odot}$) and less massive host galaxies. Their peak luminosities reach $\sim10^{43.5}~\mathrm{erg~s^{-1}}$.

Overall, the bolometric luminosities of our overmassive BHs span a broad range, fluctuating between $\sim10^{42}-10^{45}~\mathrm{erg~s^{-1}}$, depending on both BH mass and host-galaxy mass. This range is generally lower than the original luminosity estimates inferred for \textit{JWST} AGNs ($\sim10^{44}$--$10^{46}\,\mathrm{erg\,s^{-1}}$), which were based on standard low-redshift bolometric corrections (pink region in Figure~\ref{overmassive_luminosity_evolution}). However, our predicted luminosities substantially overlap with the revised estimates ($\sim10^{43}$--$10^{45}\,\mathrm{erg\,s^{-1}}$; green region in Figure~\ref{overmassive_luminosity_evolution}) that adopt the new high-redshift bolometric corrections for LRDs proposed by \cite{2025arXiv250905434G}.

Lastly, we note that our overmassive BHs accrete at rates substantially below the Eddington limit, with Eddington ratios $\eta_{\rm edd}\sim0.01$, even during their most luminous phases ($\eta_{\rm edd}\sim0.1$). That said, several studies have proposed super-Eddington accretion episodes as a pathway for assembling overmassive BHs~\citep{2024arXiv241214248T,2024Natur.636..594J,2026NatAs.tmp...66L}. In our simulations, however, BH accretion is strongly regulated by feedback (see also \citealt{kho2026learninguniversehighredshifts}), and we therefore do not find such super-Eddington episodes. We note, however, that our finite numerical resolution and effective-equation-of-state treatment of the ISM smooth the gas density field on unresolved scales, thereby damping luminosity variability and potentially suppressing short-lived super-Eddington bursts. Consequently, our simulations cannot definitively rule out such episodes. Nevertheless, our results suggest that if sufficiently large numbers of heavy seeds can be produced---whether via stellar collisions or via a single episode of hyper-Eddington accretion onto light seeds immediately after their formation~\citep{2024OJAp....7E.107M,2025ApJ...993L..48W,2026A&A...708A...7Z,2026NatAs.tmp...21M}---their subsequent growth through mergers can account for at least some of the currently observed overmassive BHs without requiring sustained or repeated episodes of super-Eddington accretion.

\subsection{Impact of BH dynamics, accretion and feedback on the $M_*-M_{\rm BH}$ scaling relations}

\begin{figure*}
\centering
\includegraphics[width= 16cm]{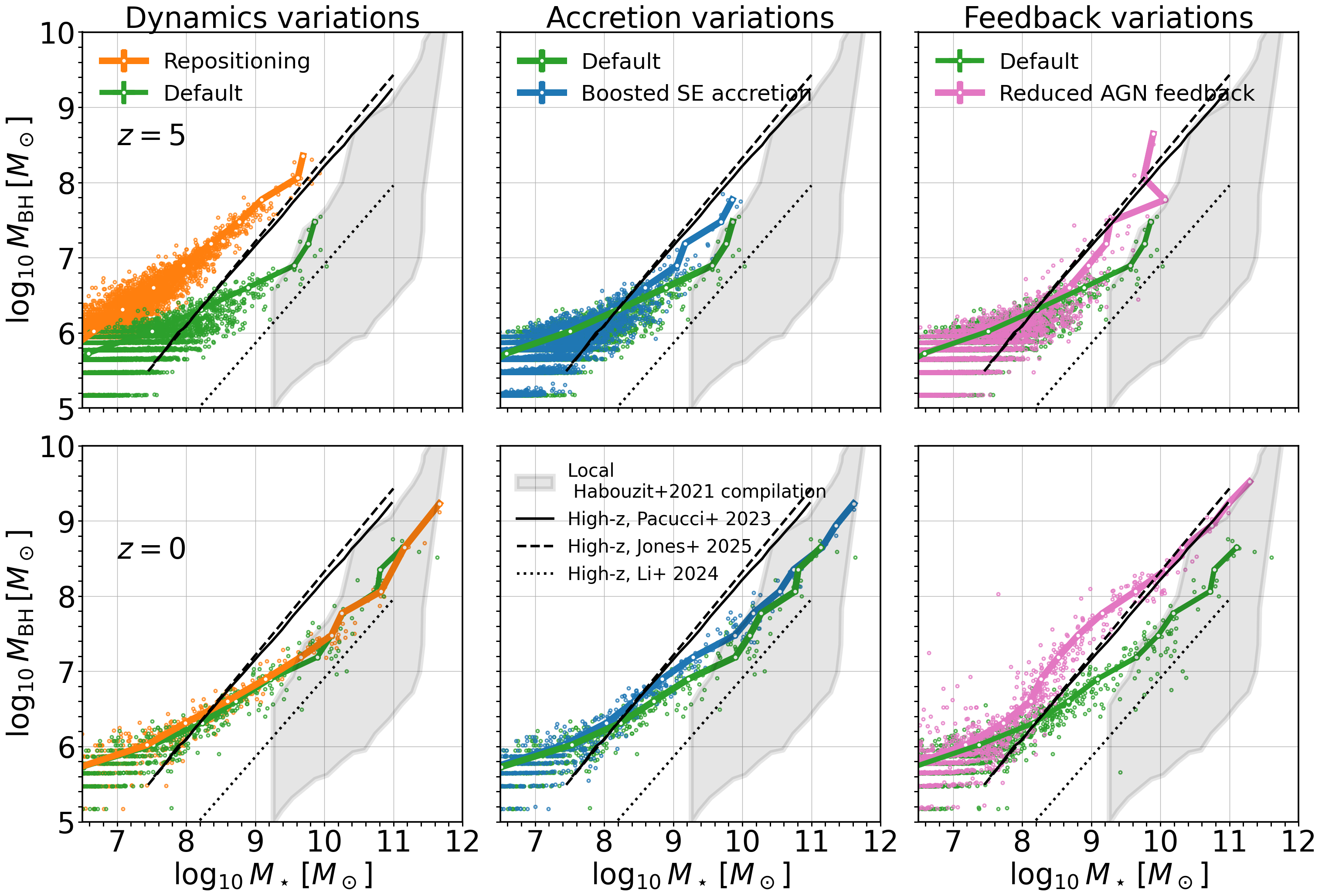}
\caption{Similar to the Figure \ref{mstar-mbh_lenient}, but comparing variations in BH dynamics, accretion, and feedback within the DGB boxes. The left panels compare our fiducial model to simulations that use BH repositioning instead of subgrid dynamical friction. The middle panels compare the default TNG accretion model~($\alpha=1,\epsilon_r=0.2,f_{\rm edd}=1$) to a boosted one~($\alpha=100,\epsilon_r=0.1,f_{\rm edd}=10$; Boosted SE accretion), while the right panels compare the default simulations with one where AGN feedback efficiency reduced by a factor of 10. Boosting mergers with repositioning can increase the normalization of the intrinsic $z\sim5$ $M_{*}$–$M_{\rm BH}$ relation without significantly affecting the $z\sim0$ relation. But boosting accretion or reducing feedback increases the $M_{*}$–$M_{\rm BH}$ normalization at both $z\sim5~\&~0$.}
\label{mstar-mbh_all_models}

\centering
\includegraphics[width= 18cm]{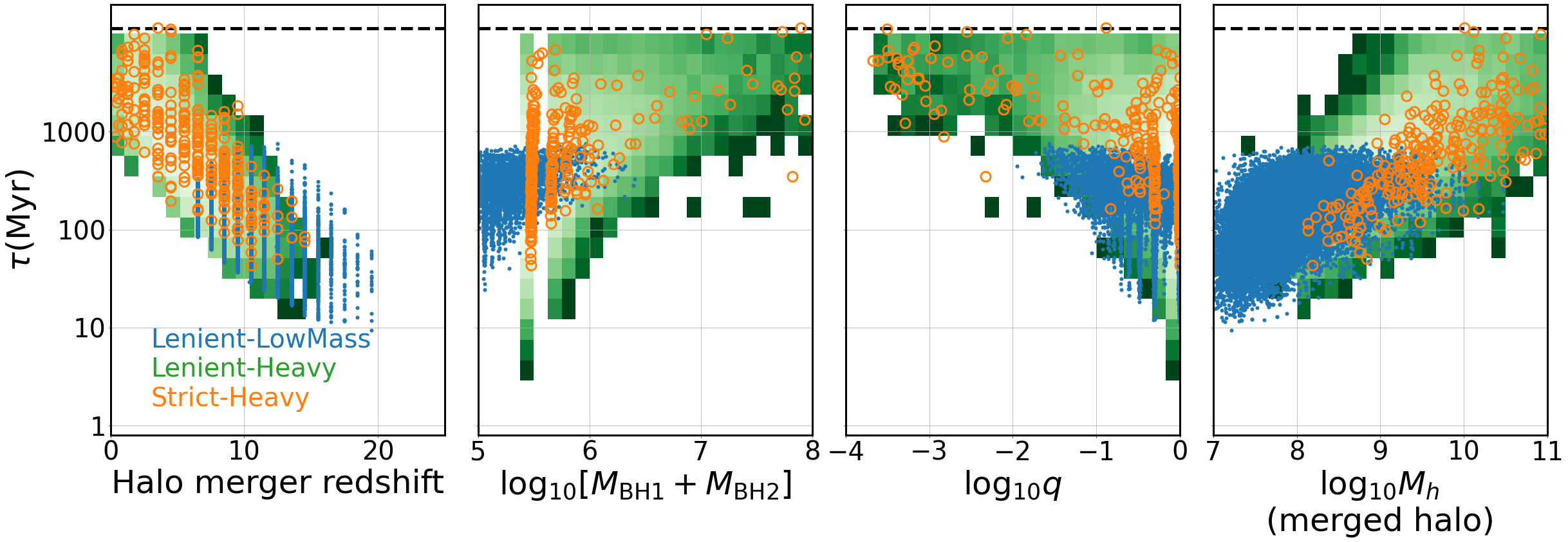}
\caption{
Delay times between BH mergers and their corresponding host galaxy mergers in the \texttt{Lenient-Heavy}, \texttt{Strict-Heavy}, and \texttt{Lenient-LowMass} seed models within the DGB boxes. From left to right, the panels show the delay time as a function of halo merger redshift, total BH mass, BH mass ratio, and merged host halo mass. Merger delays increase toward lower redshifts and in more massive halos. As a result, the lenient seed models produce BHs that lie above extrapolations of local scaling relations in the lowest-mass galaxies, but converge toward the local relations in the most massive galaxies..}
\label{merger_time_scale}
\end{figure*}

We have so far examined how different BH seed models affect the $M_\ast$--$M_{\rm BH}$ relations. We have also shown that the relative importance of mergers versus gas accretion plays a key role in determining how variations in the seed models imprint on these observables. That said, both merger-driven and accretion-driven BH growth are also influenced by the modeling of BH dynamics, accretion, and feedback. We therefore now turn to an exploration of how these additional model variations impact the $M_\ast$--$M_{\rm BH}$ relation (Figure~\ref{mstar-mbh_all_models}). Notably, we perform this exercise using the \texttt{Lenient-Heavy} seed model, primarily because this model produces a higher number of BH mergers.

\subsubsection{Implementation of dynamics: Subgrid-DF vs Repositioning}
\label{Dynamics variations: Subgrid-DF vs Repositioning}

To assess the impact of our BH dynamics modeling, we compare our subgrid-DF based simulation with one that uses BH repositioning~(left panels of Figure \ref{mstar-mbh_all_models}). At $z=5$, the repositioning model produces dramatically enhanced BH growth, driven primarily by the nearly instantaneous BH-BH mergers. Indeed, in \cite{2024MNRAS.533.1907B}, we showed that under BH repositioning, BH growth in this simulation is entirely merger-dominated up to BH masses of $\sim10^8~M_\odot$. In contrast, the subgrid-DF model delays BH-BH merging, thereby increasing the relative importance of BH accretion at earlier times. This effect is especially pronounced in higher-mass galaxies, which experience longer delays between BH-BH mergers (as we shall see in Section~\ref{mergers_time_scales_sec}). As a consequence, merger-driven growth under repositioning leads to substantially enhanced (by factors $\sim10-20$) BH masses at the massive end compared to our subgrid-DF based simulations. This leads to an intrinsically overmassive $M_*-M_{\rm BH}$ relation at $z\sim5$, unlike the case of subgrid-DF wherein the overmassive BHs are formed as rare tip of the iceberg up-scattered BHs~(as shown in Section \ref{relations: Impact of BH seeding}). 

While the relation produced by BH repositioning is even more overmassive than the ones derived by \cite{2023ApJ...957L...3P} and \cite{2025arXiv251007376J}, this should only be interpreted as an extreme upper limit under the assumption of zero delay times for BH-BH mergers. It simply means that if the delay times were short enough, the \texttt{Lenient-Heavy} seed model produces enough seeds and mergers to assemble an overmassive $M_*-M_{\rm BH}$ relation consistent with JWST at $z\sim5$. 

At $z=0$, the repositioning and subgrid-DF models yield very similar $M_\star$--$M_{\rm BH}$ relations for central BHs, because the majority of mergers are delayed beyond $z=5$ in the subgrid-DF model have occurred by $z=0$. Thus, while BH dynamics modeling strongly influences central BH populations at high redshift through merger delays, this effect largely diminishes by the present day. Consequently, even if the repositioning model produces an intrinsically overmassive $M_\star$--$M_{\rm BH}$ relation at $z=5$, by $z=0$ it converges toward the observed local relation that the underlying accretion and feedback model~(inherited from IllustrisTNG) was calibrated to reproduce~\footnote{Notably, TNG was calibrated to reproduce the \cite{2013ARA&A..51..511K} relation, which has a higher normalization than the \cite{2015ApJ...813...82R} as well as \cite{2025ApJ...981...19L} relations}.

\subsubsection{Impact of Accretion and Feedback Prescriptions}

Next, we assess the impact of accretion and feedback modeling. For the accretion model variation (middle column of Figure~\ref{mstar-mbh_all_models}), we run additional DGB simulations with three modifications: (1) a boost factor of 100 applied to the Bondi accretion rate, (2) allowing super-Eddington accretion up to 10 times the Eddington limit, and (3) reducing the radiative efficiency from 0.2 to 0.1. We refer to this as the ``boosted SE accretion" model. For the feedback model variation (right column of Figure~\ref{mstar-mbh_all_models}), we run additional DGB simulations in which the AGN thermal feedback coupling efficiency is reduced by a factor of 10 (hereafter, the ``reduced AGN feedback model").

At $z = 5$, we find that both the boosted SE accretion and reduced AGN feedback models have little impact on the $M_\star$–$M_{\rm BH}$ relation in lower-mass galaxies ($M_\star \lesssim 10^9~M_\odot$), where BH growth is dominated by mergers. In contrast, for more massive galaxies ($M_\star \gtrsim 10^9~M_\odot$), where gas accretion becomes the dominant growth channel, both models produce a significant upward shift in the $M_\star$–$M_{\rm BH}$ relation. For the most massive systems in our simulations at $z = 5$ ($M_\star \sim 10^{10}~M_\odot$), the reduced AGN feedback model yields a stronger enhancement than the boosted SE accretion model.

However, the reduced AGN feedback model also produces a dramatically elevated local $M_\star$–$M_{\rm BH}$ relation—by factors of $\sim 10$—relative to our fiducial setup, rendering it strongly inconsistent with local observations. This outcome is expected, as the fiducial feedback prescription was calibrated to reproduce the observed local scaling relation~\citep{2018MNRAS.479.4056W}. The boosted SE accretion model results in a smaller, though still non-negligible increase in BH masses for galaxies with $M_\star \sim 10^9$–$10^{10}~M_\odot$.

\subsection{BH merger time-scales and their impact on the $z=5$ $M_*-M_{\rm BH}$ scaling relation}
\label{mergers_time_scales_sec}
In Section~\ref{Dynamics variations: Subgrid-DF vs Repositioning}, we demonstrated that BH dynamics modeling plays a critical role in shaping the $M_*-M_{\rm BH}$ relation at $z=5$. In \cite{2024MNRAS.533.1907B}, we further showed that producing an intrinsically overmassive $M_*-M_{\rm BH}$ relation at $z\sim5$ requires not only abundant heavy-seed formation, but also merger delays of $\lesssim750~\rm Myr$ relative to BH repositioning. In this work, we can now directly estimate the merger time-scales predicted by our subgrid-DF model. Figure~\ref{merger_time_scale} shows these merger time-scales, defined as the time elapsed between a halo merger and the subsequent BH--BH merger. These time-scales provide an approximate measure of the delay introduced by subgrid-DF relative to BH repositioning, since BH repositioning typically results in much shorter merger delays following halo mergers. The detailed methodology used to estimate these time-scales is described in \cite{2025ApJ...991...81B}. The merger time-scales are shown as functions of redshift, host halo mass, BH mass ratio, and total BH mass.

Using a random forest regression model, we find that the strongest predictor of merger delay time is the halo merger redshift~(leftmost panel of Figure \ref{merger_time_scale}). This is likely because the higher background matter densities at earlier cosmic times enhance dynamical friction and accelerate orbital decay. The second strongest predictor is the halo mass~(rightmost panel). This is likely because lower-mass halos can bring their central BHs to smaller separations during the halo-merger stage itself, thereby reducing the subsequent hardening time required once the binary evolution becomes dominated by dynamical friction and other sub-resolution processes. Compared to halo merger redshift and halo mass, the BH mass and BH mass ratio~(middle two panels) play relatively subdominant roles in determining the merger delay times.

Focusing on the actual values, at the highest redshifts~($z\sim15-20$) and in the smallest halos~($\sim10^7-10^8~M_{\odot}$), our merger time-scales generally range from $\sim10-100~\mathrm{Myr}$. Close to $z\sim0$ and within more massive halos~($\sim10^{11}~M_{\odot}$), the merger time-scales are $\gtrsim1000~\rm Myr$. These values fully explain the trends in our $z=5$ $M_*-M_{\rm BH}$ results. Mergers with time-scales $\lesssim750~\rm Myr$, which are needed for overmassive BH assembly at $z\sim5$~(again, shown in \citealt{2024MNRAS.533.1907B}), occur primarily for the earliest generation of mergers within lower-mass ($\lesssim10^8~M_{\odot}$) halos. However, for later mergers within more massive halos, a substantial fraction have time-scales $\gtrsim750~\rm Myr$. Therefore, for our \texttt{Lenient-Heavy} model, the mean $M_*-M_{\rm BH}$ relations lie clearly above extrapolations from observed local relations for the smallest galaxies, but are no longer above the local relations for the most massive galaxies~(see again the top row of Figure \ref{mstar-mbh_lenient}).
\subsection{BH Merger rates: Implications of JWST on LISA expectations}
\begin{figure*}
\centering
\includegraphics[width= 18cm]{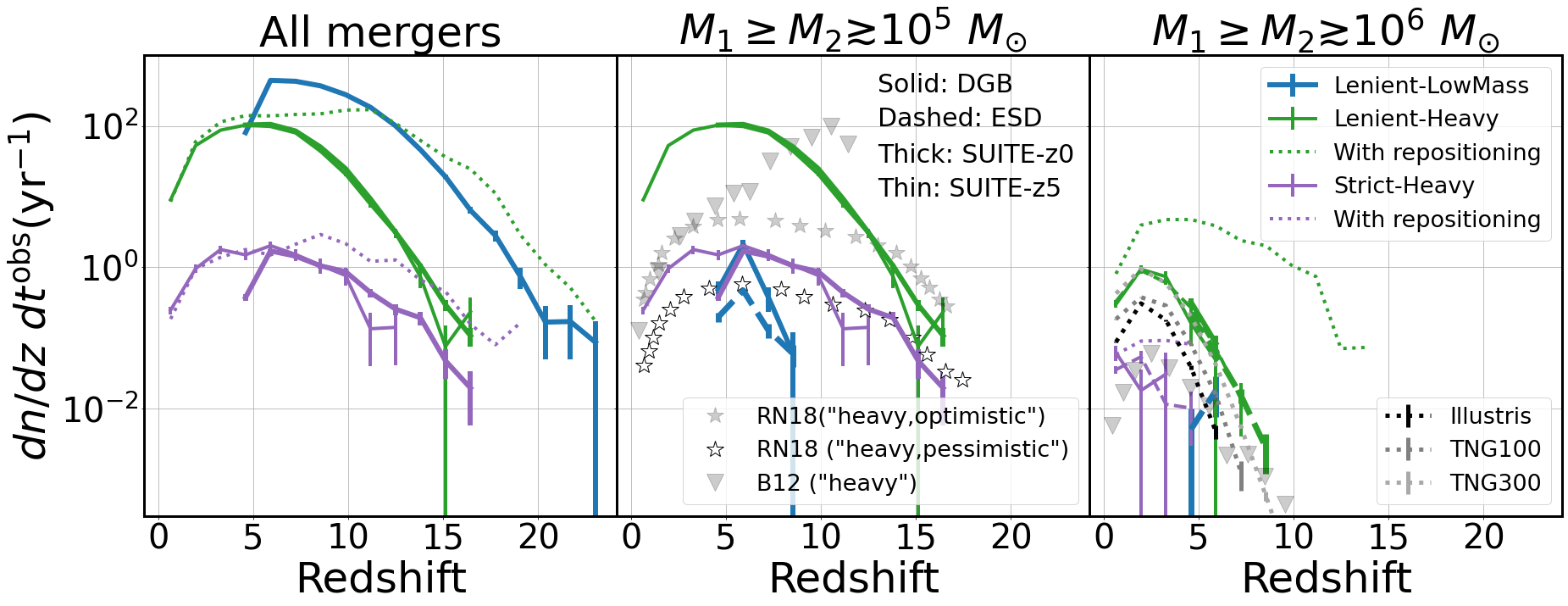}
\caption{BH--BH merger rates predicted by our simulations for the three different seed models. The left panel shows the total merger rates, the middle panel includes only mergers between BHs with masses $\gtrsim 10^{5}~M_{\odot}$, and the right panel shows mergers between BHs with masses $\gtrsim 10^{6}~M_{\odot}$. Colors indicate the different seed models. For each model, thick lines correspond to the \texttt{SUITE-z5} boxes, while thin lines correspond to the \texttt{SUITE-z0} boxes. Solid lines represent the DGB runs and dashed lines represent the ESD runs. Dotted lines show predictions using the repositioning model using boxes from \cite{2024MNRAS.533.1907B}. We also show predictions from other models in literature i.e. SAM predictions from \citealt{2012MNRAS.423.2533B}~(grey triangles) and \citealt{2018MNRAS.481.3278R}~(stars), as well as \texttt{Illustris}~(black dotted line) and \texttt{IllustrisTNG}~(grey dotted line) simulations. Only the lenient seeding models produce peak merger rates exceeding $\gtrsim 100$ events per year.}
\label{merger_rates}
\centering
\includegraphics[width= 17cm]{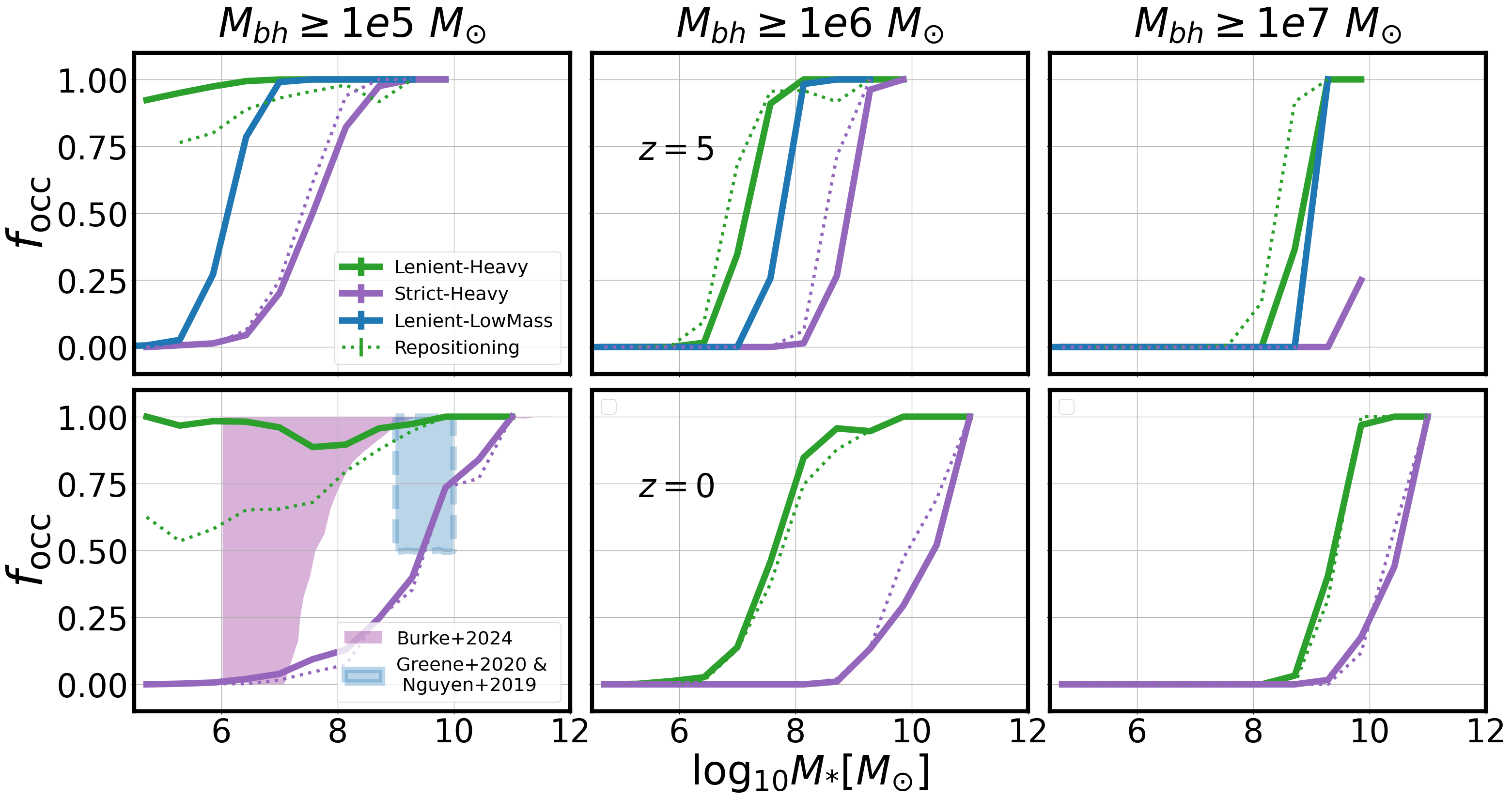}
\caption{BH occupation fractions at $z=5$ (top) and $z=0$ (bottom) predicted by our DGB simulations. Solid lines show the predictions for the different seed models. Dotted lines show predictions from previous \texttt{BRAHMA} boxes that used BH repositioning~\citep{2024MNRAS.533.1907B,2025MNRAS.538..518B}. Grey and pink regions show lower limits on the local occupation fractions inferred from observations by \cite{2015ApJ...799...98M} and \cite{burke2024multiwavelengthconstraintslocalblack}, respectively. The blue region shows local constraints from \cite{2020ARA&A..58..257G} based on dynamical BH mass measurements from \cite{2019ApJ...872..104N}. Our \texttt{Lenient-Heavy} seed models predict near-unity local occupation fractions for $\gtrsim10^5~M_{\odot}$ BHs even for $M_*\lesssim10^7~M_{\odot}$ galaxies, but the \texttt{Strict-Heavy} seed model predicts $\lesssim10~\%$ occupation fractions in $M_*\lesssim10^8~M_{\odot}$ galaxies.}
\label{BH_occupations_fig}

\end{figure*}

Figure~\ref{merger_rates} shows the BH–BH merger rates for our different seed models. The leftmost panels display the full merger population, dominated by first-generation seed BHs. The impact of the merger delays under subgrid-DF can be seen in the comparison against repositioning-based predictions~(solid vs. dotted lines of the same color). Due to these delays, the subgrid-DF based predictions strongly diverge from repositioning at high-$z$, but eventually converge by $z\sim0$. This implies that despite the merger delays under subgrid-DF, most mergers eventually still occur by $z\sim0$. This also explains why the $M_*-M_{\rm BH}$ relation strongly diverges between the two dynamics treatments at $z\sim5$, but becomes similar by $z\sim0$~(see bottom left panel of Figure~\ref{mstar-mbh_all_models}).

The \texttt{Lenient-LowMass} model~(solid blue line in the left panel of Figure~\ref{merger_rates}) naturally produces the highest overall merger rates, peaking at $\sim700$ events per year. The \texttt{Lenient-Heavy} model~(solid green line) reaches peak rates of $\sim100$ per year, while the \texttt{Strict-Heavy} model yields only a few events per year at its peak~(solid pink line). For both the \texttt{Lenient-Heavy} and \texttt{Strict-Heavy} cases, the \texttt{SUITE-z0} simulations show that the merger rate distribution peaks at $z \sim 5$.

The middle panel shows mergers among BHs with masses $\gtrsim 10^5\,M_{\odot}$. For the \texttt{Lenient-Heavy} and \texttt{Strict-Heavy} models, this is still the full population~(same as the left panel) since the seed mass is already $\sim 10^5\,M_{\odot}$. However, the \texttt{Lenient-LowMass} model exhibits substantially lower rates compared to its full merger population, as the $\sim 10^4\,M_{\odot}$ seeds must first grow above $10^5\,M_{\odot}$ before contributing mergers in this mass range; the first such events occur at $z \lesssim 10$. In contrast, even the \texttt{Strict-Heavy} model is able to produce a substantial number of $\gtrsim 10^5\,M_{\odot}$ mergers at $z \gtrsim 10$. Thus, even if only a rare DCBH-like channel (as represented by the \texttt{Strict-Heavy} model) were viable for forming $\sim 10^5\,M_{\odot}$ seeds, it would still dominate the $\gtrsim 10^5\,M_{\odot}$ merger population at $z \gtrsim 10$ despite the presence of more efficient lower-mass seeding channels (e.g., Pop~III or NSC seeds). By $z \sim 5$, both the \texttt{Lenient-LowMass} and \texttt{Strict-Heavy} models predict peak rates of $\sim 1$ event per year above $10^5\,M_{\odot}$, whereas the \texttt{Lenient-Heavy} model yields roughly two orders of magnitude more mergers.

The rightmost panel shows merger rates for BHs with masses $\gtrsim10^6~M_{\odot}$, corresponding to later-generation mergers in all models. Even in the \texttt{Lenient-Heavy} case, the earliest such events appear at $z\lesssim10$, with peak rates of $\sim1$ per year. The \texttt{Strict-Heavy} model predicts $\lesssim0.1$ events per year, although the small number of events limits statistical robustness.  

Overall, the merger rate distributions exhibit two key signatures of efficient heavy-seed formation that can be tested with LISA. First, the \texttt{Lenient-Heavy} seed channel predicts $\gtrsim100$ mergers per year among $\sim10^5~M_{\odot}$ BHs. Second, the detection of even a single merger per year involving $\gtrsim10^6~M_{\odot}$ BHs at high redshift would strongly favor scenarios with abundant formation of $\sim10^5~M_{\odot}$ seeds. Also, these merger rates tend to be higher than those predicted by most simulations~(e.g., \textit{Illustris}, \textit{IllustrisTNG}) and SAMs in the literature~\citep[e.g.][]{2012MNRAS.423.2533B,2018MNRAS.481.3278R}. This is primarily because our lenient seed models produce many more seeds than these models. For example, in \textit{Illustris} and \textit{IllustrisTNG}, BH seeding is restricted to substantially more massive halos than in \texttt{BRAHMA}. Similarly, \cite{2012MNRAS.423.2533B} and \cite{2018MNRAS.481.3278R} allow heavy-seed formation only at $z\lesssim20$, whereas \texttt{BRAHMA} imposes no such redshift restriction.

Finally, regardless of the seed model, the merger population at $z \gtrsim 10$ is almost entirely dominated by first-generation seed BHs. If these events can be localized, they would directly constrain both the maximum seed masses formed in the early Universe and the dominant formation channels of these seeds. This therefore provides a distinguishing observational signature between the \texttt{Lenient-LowMass} and \texttt{Lenient-Heavy} seed models. While the former produces the highest overall merger rates across cosmic time, only the latter can produce mergers involving $\gtrsim10^5~M_{\odot}$ BHs at $z\gtrsim10$.

\subsection{The BH occupation fraction}
\label{BH occupation fractions}

Finally, we look at the occupation fractions of BHs as a function of stellar mass, as shown in Figure \ref{BH_occupations_fig}. For all the seed models, there is a rightward shift in the occupation fractions from $z\sim5$~(top row) to $z\sim0$~(bottom row), i.e. galaxies at fixed stellar masses have higher occupation fractions at higher redshifts. This is most pronounced for smaller BH mass thresholds ($\geq10^5~M_{\odot}$~{\rm and}~$\geq10^6~M_{\odot}$), and is again because galaxy growth is faster than the merger-dominated BH growth in this regime.

As also pointed out in \cite{2025MNRAS.538..518B}, we generally see strong seed model variations in the BH occupation fractions. However, the use of BH repositioning in that work does lead to an artificial suppression of the occupation fractions at the smallest BH mass threshold ($\geq10^5~M_{\odot}$) for the \texttt{Lenient-Heavy} model~(green dotted vs. green solid lines in the leftmost panels). This is likely because repositioning leads to the well-known numerical effect of low-mass galaxies prematurely losing their BHs to higher-mass galaxies during fly-by encounters. For the BHs formed by the \texttt{Strict-Seed} model, and for higher-mass BHs in general, this numerical effect is diminished because these BHs reside in higher-mass galaxies that are less prone to losing them.

Under the new subgrid-DF prescriptions, our \texttt{Lenient-Heavy} seed model produces near-unity occupation fractions for $\geq10^5~M_{\odot}$ BHs all the way down to systems with stellar masses of $M_*\sim10^6~M_{\odot}$. In contrast, the occupation fractions for the \texttt{Strict-Seed} model already decline to $\sim10~\%$ for $M_*\sim10^8~M_{\odot}$ galaxies at $z\sim0$. 

Currently, the most direct constraints on the local BH occupation fractions of $\gtrsim10^5~M_{\odot}$ come from \cite{2020ARA&A..58..257G} based on dynamical BH mass measurements from \cite{2019ApJ...872..104N}. However, this was based on only a small sample of the highest-mass ($\gtrsim10^9~M_{\odot}$) dwarf galaxies, where the gap between our seed model predictions is smaller. The resulting constraints~(blue region) therefore do not currently rule out any of our models. In addition, \cite{burke2024multiwavelengthconstraintslocalblack} recently have inferred a lower limit on the overall occupation fractions based on AGN occupation fractions. While these are indirect measurements that rely on multiple assumptions~(for e.g. the Eddington ratio distribution), it is noteworthy that the most recent \cite{burke2024multiwavelengthconstraintslocalblack} lower-limits favor our \texttt{Lenient-Heavy} seed model.

\section{Discussion}
\label{discussion}

\subsection{Is the high-z $M_*-M_{\rm BH}$ scaling relation intrinsically overmassive, or consistent with the local Universe?}

Several works have shown that even after accounting for potential selection biases, the inferred BH masses and host stellar masses of the \textit{JWST} BHs imply that the ``intrinsic'' mean $M_*-M_{\rm BH}$ relation at high-$z$ is significantly above the local scaling relation~\citep{2023ApJ...957L...3P,2025arXiv251007376J}. In our previous work~\citep{2024MNRAS.533.1907B}, we determined that an intrinsically overmassive $z\sim5$ $M_*-M_{\rm BH}$ relation could only be produced by our simulations if: (1) there is abundant formation of heavy $\sim10^4-10^5~M_{\odot}$ seeds, and (2) these seeds are able to merge within $\lesssim750~\rm Myr$ after their host galaxies merge. But in this work, our subgrid-DF model leads to $\gtrsim750~\rm Myr$ merger time-scales in a substantial fraction of halos, particularly the more massive ones~(revisit the right-most panel of Figure~\ref{merger_time_scale}). Therefore, even for our most lenient models under subgrid-DF, the intrinsic high-$z$ $M_*-M_{\rm BH}$ relation approaches the local relation for  $M_*\gtrsim10^{8}~M_{\odot}$ galaxies. Our \texttt{Lenient-Heavy} seed models do produce overmassive BHs, but as ``tip-of-the-iceberg'' systems up-scattered on the $M_*-M_{\rm BH}$ plane. This picture is more consistent with \cite{2025ApJ...981...19L}, who also derived that JWST BHs are consistent with the local scaling relations after accounting for selection biases. 

But if the high-$z$ $M_*-M_{\rm BH}$ relation in our Universe is indeed intrinsically overmassive, this would be in tension with our subgrid-DF based simulations, particularly when combined with our fiducial BH accretion and feedback model inherited from \texttt{IllustrisTNG}. We showed that adjusting the BH accretion and feedback model does enhance the normalization of the high-$z$ $M_*-M_{\rm BH}$ relation. However, it is well understood that these changes also tend to increase the normalization of the $M_\star-M_{\rm BH}$ relation at $z\sim0$, thereby increasing the tension against observations. On the other hand, even a dramatic shortening of merger time-scales via repositioning can produce a highly overmassive high-$z$ $M_\star$--$M_{\rm BH}$ relation while still leaving the local scaling relations largely unaffected. It is also worth noting that recent evidence for a stochastic gravitational wave background from pulsar timing array experiments~\citep{2023ApJ...951L...8A,2024A&A...685A..94E,Xu_2023} may provide a complementary probe of the $M_\star$--$M_{\rm BH}$ relation at $z>0$, with current measurements of the background amplitude potentially favoring an intrinsically overmassive relation~\citep{2026ApJ...997..188M}. Several works have explored mechanisms for shortening BH merger time-scales through enhanced dynamical friction, for example via the formation of compact stellar galactic disks~\citep{2025A&A...695A..97D} or self-interacting dark matter~\citep{2024PhRvL.133b1401A,2024A&A...690A.299F}. We will explore these possibilities in future work.

Producing an intrinsically overmassive $z\sim5$ $M_*-M_{\rm BH}$ relation~(with shorter merger time-scales) also enhances the predicted BHMF, placing it well above current \textit{JWST}-inferred constraints~(revisit the dotted green line in Figure~\ref{BHMF_AGNLF_a}). This is, of course, inevitable unless an increase in the $M_*-M_{\rm BH}$ normalization is accompanied by a commensurate decrease in either the stellar mass functions or the BH occupation fractions. However, our stellar mass function predictions at $z\sim5$ are already broadly consistent with \textit{JWST} observations~(see Figure~2 of \citealt{2024MNRAS.533.1907B}), while the \texttt{Lenient-Heavy} seed models produce near-unity BH occupation fractions. Overall, this suggests that if the intrinsic $M_*-M_{\rm BH}$ relation at $z\sim5$ is substantially overmassive, then either: (1) the currently inferred BHMFs remain significantly incomplete, even at the massive end, or (2) the true high-$z$ BH occupation fractions are substantially below unity, contrary to the predictions of our \texttt{Lenient-Heavy} models.

\subsection{Implications of JWST discoveries on complementary observable regimes: LISA binaries and low-z BHs}

\label{implications}

Our \texttt{BRAHMA} simulations suggest that abundant formation and mergers of heavy $\sim10^5~M_{\odot}$ seeds could produce at least some of the currently observed overmassive BHs. In our previous work, we also found that similar conditions are required to explain the JWST BH discoveries at much higher redshifts of $z\sim9\text{--}11$~\citep{2026ApJ...997..187B}.

If $\sim10^5~M_{\odot}$ seeds indeed form as abundantly in the Universe as predicted by our \texttt{Lenient-Heavy} seed models, they would leave behind two unique imprints in observational regimes distinct from those currently probed by JWST:
\begin{itemize}
\item Overall merger rates detectable by LISA above $\sim100~\rm yr^{-1}$, and mergers amongst supermassive $\gtrsim10^6~M_{\odot}$ BHs at rates of $\sim1~\rm yr^{-1}$.

\item The local BH occupation fractions of $\gtrsim10^5~M_{\odot}$ BHs are close to unity for all galaxies down to the smallest resolvable systems in our simulations~($M_*\sim10^{7}~M_{\odot}$).  
\end{itemize}
The above two predictions will serve as falsifiable tests in the future, as they would likely differ substantially from the signatures expected from other proposed pathways of early BH assembly explored in the literature. These scenarios include early BH assembly via short episodes of super-Eddington accretion during galaxy mergers~\citep{2023MNRAS.526.3250S}, sustained Eddington accretion onto DCBH seeds~\citep{2025ApJ...979..127J}, ultra-massive $\sim10^6~M_{\odot}$ seeds~\citep{2024ApJ...961...76M,2026arXiv260104955C}, and primordial BHs~\citep{2025JCAP...04..040Z,2026ApJ..1000L..19Z}. Since none of these scenarios require abundant BH mergers to reproduce the JWST BH population, they would likely produce substantially lower merger rates compared to the predictions of our lenient heavy seed models.

Finally, we emphasize that the early BH assembly scenerio proposed by our simulations (i.e., abundant formation and mergers of heavy seeds) do not necessarily rule out the other scenarios discussed above. This is just one possible pathway that was unravelled by our simulations, amongst the limited possibilities that could be probed by our resolution and subgrid physics treatment. Almost all of the above alternative formation and growth pathways require much higher resolutions and detailed physics~(e.g. a resolved ISM or radiative transfer) modeling to effectively probe. Simulations that achieve these resolutions are often idealized or zoom-ins, and often do not have the volume to make a direct comparison against observed number densities, BHMFs, or AGNLFs. Additionally, these high resolution models are also not guaranteed to reproduce a realistic population of local galaxies and BHs. In contrast, with our \texttt{BRAHMA} simulation, we were able to demonstrate that the lenient heavy seed models produce  consistent BHMFs, AGNLFs, as well as galaxy stellar mass functions~(in \citealt{2024MNRAS.533.1907B}) with the JWST measurements at $z\sim5$, while also maintaining good agreement with locally observed BHs~(and galaxies).     

In general, it is possible that multiple scenarios are simultaneously operating to produce the observed JWST BHs. In this context, recall that our subgrid-DF based \texttt{BRAHMA} simulations do not produce any BHs with the highest $M_{\rm BH}/M_*$ ratios measured by JWST, particularly those discovered within more massive~($M_*\gtrsim10^9~M_{\odot}$) galaxies~(revisit upper panels of Figure \ref{mstar-mbh_lenient}). While this may simply be due to our limited simulation volume, it is also possible that one or more of these additional scenarios in the current literature could contribute to producing these BHs and increasing the scatter in the $M_{\rm BH}-M_*$ relation.

\section{Conclusions}
\label{conclusions}
The \textit{JWST} discovery of a large population of supermassive BHs at $z\sim4-7$ provides new insight into the formation and early growth of BH seeds. In this work, we present a new suite of \texttt{BRAHMA} cosmological simulations designed to study the assembly of BH populations at $z\sim5$ and their implications at $z\sim0$, under different heavy-seed formation scenarios coupled with subgrid dynamical friction.

We explore two classes of seeding models: \emph{lenient} and \emph{strict}. In the lenient models, we place $\sim10^{4}~\&~10^{5}\,M_{\odot}$ seeds in every halo that contains sufficient dense~\&~metal-poor gas (at least $5\times$ the seed mass). This scenario is most representative of remnants of stellar collisions or supra-exponential accretion within ultra-dense NSCs~\citep{2021MNRAS.501.1413N,2023PhRvD.108h3012K}, or post-formation hyper-Eddington growth of Pop III seeds~\citep{2026NatAs.tmp...21M}. In the strict model, motivated by DCBH formation scenarios, $\sim10^{5}\,M_{\odot}$ seeds form only in halos exposed to strong Lyman--Werner backgrounds ($\gtrsim10\,J_{21}$), with low gas spin and residing in rich environments. To capture unresolved small-scale BH dynamics, we adopt the subgrid dynamical friction model of \cite{2023MNRAS.519.5543M}.

We perform simulations in $[18\,\mathrm{Mpc}]^{3}$ and $[36\,\mathrm{Mpc}]^{3}$ volumes where these seed masses are explicitly resolved with gas. These runs are used to calibrate a stochastic seeding prescription that allows us to represent these seed models in lower-resolution simulations that cannot resolve the seed masses. Using this calibrated framework, we extend our analysis to larger volumes up to $[72\,\mathrm{Mpc}]^{3}$.

Our main conclusions are as follows:

\begin{itemize}

\item At both $z\sim5$ and $z\sim0$, the earliest resolvable phase in the BH growth in our simulations is primarily driven by seed formation and mergers. Starting from the assumed seed masses in our simulations, BH seeding and mergers dominate the mass assembly of $10^{6}$--$10^{7}\,M_{\odot}$ BHs that typically reside in $\lesssim10^{9}\,M_{\odot}$ galaxies, contributing $\gtrsim50\%$ of their final mass. Gas accretion dominates the growth of more massive ($\gtrsim10^{7}\,M_{\odot}$) BHs hosted by $\gtrsim10^{9}\,M_{\odot}$ galaxies.

\item In $M_{*}\sim10^{9}$--$10^{11}\,M_{\odot}$ galaxies at $z\sim5$, all seed models produce BH populations broadly consistent with observed local BH scaling relations. However, the lenient seed models preferentially populate the upper envelope of the observed local scatter.

\item In lower-mass $M_{*}\sim10^{8}$--$10^{9}\,M_{\odot}$ galaxies at $z\sim5$, only the lenient seed models produce a substantial population of BHs lying above simple extrapolations of the local $M_{*}$--$M_{\rm BH}$ relation. These include rare overmassive BHs with $M_{\rm BH}/M_{*}\gtrsim0.01$ which represent a tip-of-the-iceberg population significantly up-scattered in the $M_{*}$--$M_{\rm BH}$ plane. 
\begin{itemize}
\item Among the two lenient seed models with different seed masses, the model that produces the more massive $\sim10^5~M_{\odot}$ seeds is more effective at producing these up-scattered overmassive BHs, despite the lower-mass $\sim10^4~M_{\odot}$ seeds being nearly an order of magnitude more abundant.

\item These up-scattered overmassive BHs assemble the majority of their mass through mergers. Despite this, their bolometric luminosities fluctuate between $\sim10^{43}$--$10^{45}\,\mathrm{erg\,s^{-1}}$ consistent with observational estimates based on high-redshift AGN bolometric corrections from \cite{2025arXiv250905434G}.
\end{itemize}

\item In the lowest-mass galaxies ($M_{*}\sim10^{7}$--$10^{8}\,M_{\odot}$) at $z\sim5$, most BHs produced by the lenient seed models lie above the extrapolations of the local $M_{*}$--$M_{\rm BH}$ relations. These overlap with recent detections of BHs in these galaxies within the \textit{JWST} GLIMPSE survey.

\item The possible assembly of \textit{JWST} BHs via these lenient seeding channels can be tested with future gravitational-wave observations as well as local electromagnetic surveys of BHs in dwarf galaxies. For $\gtrsim10^5~M_{\odot}$ BHs , the lenient seed models predict peak BH merger rates of $\gtrsim100$ events per year and near-unity local BH occupation fractions even for galaxies with $M_*\lesssim10^7~M_{\odot}$. In contrast, the strict model yields peak rates of only $\sim1~\mathrm{yr^{-1}}$ and occupation fractions of $\lesssim10\%$ for local galaxies with $M_{*}\lesssim10^{8}~M_{\odot}$.

\item For supermassive BH mergers~($\gtrsim10^6~M_{\odot}$ BHs), the lenient model predicts $\sim1~\rm yr^{-1}$, whereas the strict model predicts $\sim0.1~\rm yr^{-1}$. Therefore, the detection of even a handful of supermassive BH mergers during the 4 year LISA mission would strongly favor our lenient seed model.   

\item At $z\gtrsim10$, BH merger events are dominated almost entirely by mergers between first-generation BH seeds. Mergers involving BHs that have grown to $\gtrsim10$ times their initial seed mass occur exclusively at $z\lesssim10$. Consequently, gravitational-wave detections at $z\gtrsim10$, if localized, would provide the strongest constraints on the initial masses and abundances of heavy BH seeds.
\end{itemize}

Our results could be impacted by the modeling choices adopted for SMBH physics, and these caveats have been discussed extensively in our own previous work~\citep{2026ApJ...997..187B}. These include not only the modeling choices for BH accretion, feedback, and dynamical modeling, which directly affect BH growth, but also other aspects of the galaxy formation model such as the treatment of the ISM, star formation, metal enrichment, and stellar feedback, all of which can in turn also influence BH seeding. Future work will therefore explore high-$z$ BH growth under a broader range of galaxy formation models.

Overall, among the formation and growth scenarios of heavy seeds explored by \texttt{BRAHMA}, the current \textit{JWST} discoveries and BH mass measurements at $z\sim5$ tend to favor scenarios in which the most massive ($\sim10^5~M_{\odot}$) seeds form and merge abundantly. While our subgrid dynamical friction model does not produce mergers rapidly enough to assemble overmassive BHs observed by \textit{JWST} in massive ($M_*\gtrsim10^9~M_{\odot}$) galaxies, the lenient production of $\sim10^5~M_{\odot}$ seeds nevertheless produces overmassive BHs in lower-mass ($M_*\lesssim10^9~M_{\odot}$) galaxies. In our previous work, we demonstrated that similarly lenient heavy-seed formation conditions are also required to explain the most distant BHs detected at $z\sim9$--$11$~\citep{2026ApJ...997..187B}. Future gravitational-wave detection rates with LISA, as well as systematic searches for BHs in local dwarf galaxies, will provide key observational tests of this scenario.

\appendix

\section{Calibration of the stochastic seed model}
\label{appendix_stochastic}

\begin{figure*}
\centering
\includegraphics[width= 12cm]{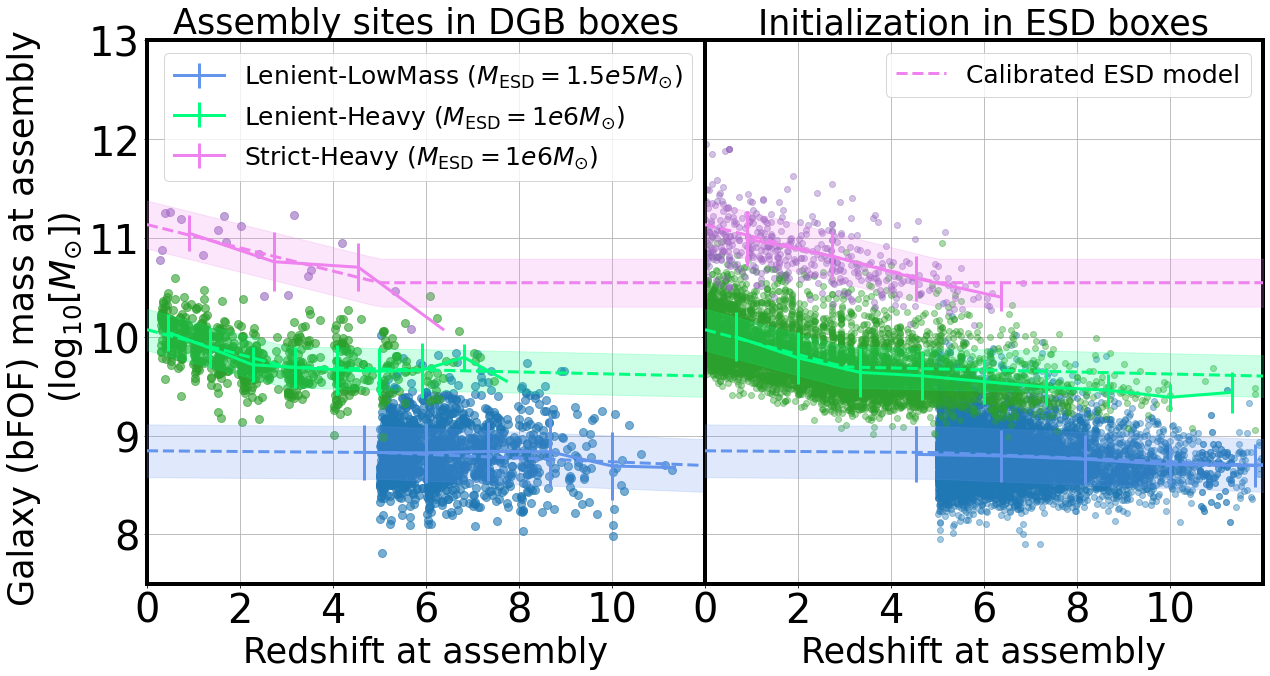}
\caption{\textit{Left Panel:} For the DGB simulations, we plot the galaxy masses and redshifts at which descendant BHs with $M_{\rm bh}=M_{\rm ESD} = 8\,\seedmass$ assemble. Each data point (filled circle) corresponds to an assembly event traced along galaxy merger trees. In the \texttt{Lenient-LowMass} seed model, this represents the assembly of $\sim10^{5}\,M_{\odot}$ descendants from $\sim10^{4}\,M_{\odot}$ seeds (note that this simulation is evolved only to $z=5$). In the \texttt{Lenient-Heavy} and \texttt{Strict-Heavy} seed models, the descendants reach $\sim10^{6}\,M_{\odot}$ starting from $\sim10^{5}\,M_{\odot}$ seeds. Solid lines with error bars show the mean trend and standard deviation. Dashed lines show the best-fit double-power-law relation that defines the \textit{stochastic galaxy mass criterion}, while the shaded regions indicate the average standard deviation across redshift bins. These fits determine the seed parameters $\alpha$, $\beta$, $z_{\rm trans}$, $\log M_{\rm trans}$, and $\sigma$, which are listed in Table~\ref{tab:brahma_suite}. \textit{Right Panel:} We show the same quantities for the ESD simulations, where the seed parameters are used as inputs to the stochastic seed model. Each data point represents an ESD initialization event, designed to statistically reproduce the assembly locations predicted by the DGB simulations. Because the ESD simulations cover larger cosmological volumes, they contain a correspondingly larger number of initialization events compared to the DGB runs.}
\label{bFOFmass_vs_redshift_at_assembly}

\includegraphics[width= 12cm]{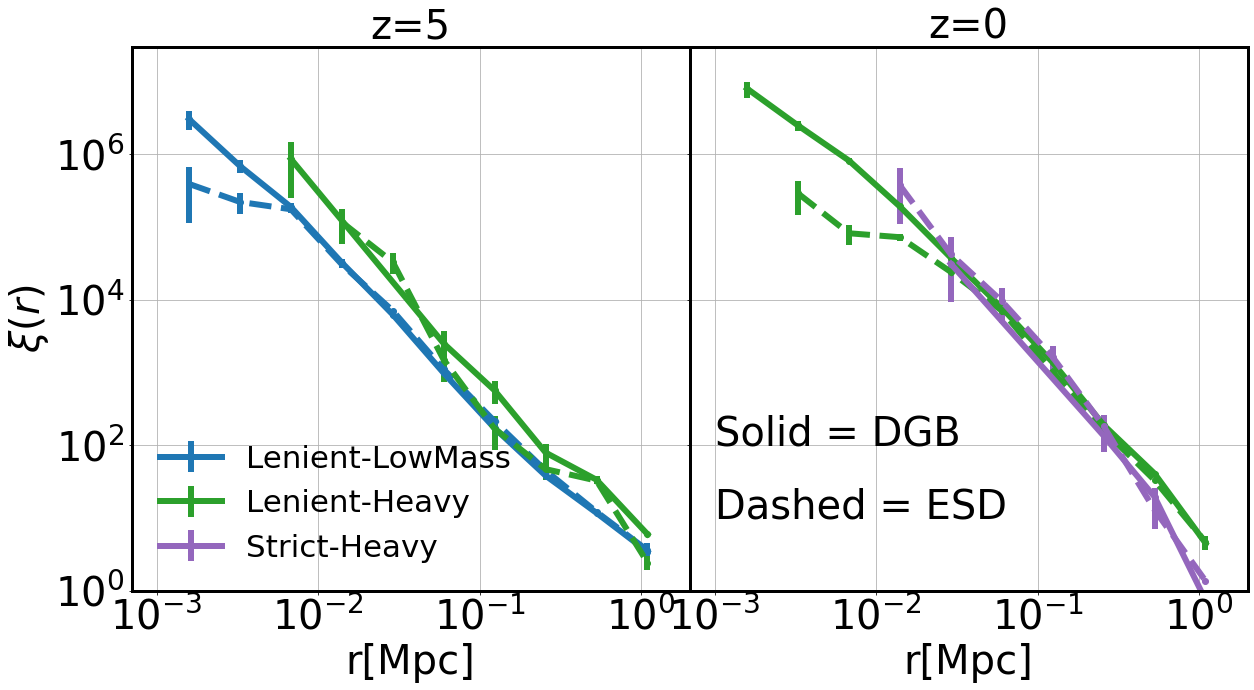}
\caption{Comparison of the two-point correlation function predicted by the DGB (solid) and ESD (dashed) simulations for BHs with $M_{\rm bh} > M_{\rm ESD}$ at $z=0$ (left) and $z=5$ (right). Because the clustering signal can depend on simulation volume—since larger volumes sample more diverse environments across all spatial scales—we performed additional ESD simulations in the same $(18\,{\rm Mpc})$ volume as the DGB runs, albeit at eight times lower mass resolution. For the \texttt{Lenient-LowMass} model (which is evolved only to $z=5$), we apply a threshold of $M_{\rm ESD}=1.5\times10^{5}\,M_{\odot}$. For the \texttt{Lenient-Heavy} and \texttt{Strict-Heavy} models, we adopt $M_{\rm ESD}=1\times10^{6}\,M_{\odot}$. We do not show a prediction for the \texttt{Strict-Heavy} model at $z=5$, as the number of BHs at this epoch is too small to obtain a meaningful clustering signal. Overall, the ESD simulations broadly reproduce the BH clustering predicted by the DGB runs, which is the intended outcome of the \textit{stochastic galaxy environment criterion}.}

\label{small_scale_clustering}
\end{figure*}

\begin{figure*}
\centering
\includegraphics[width= 18cm]{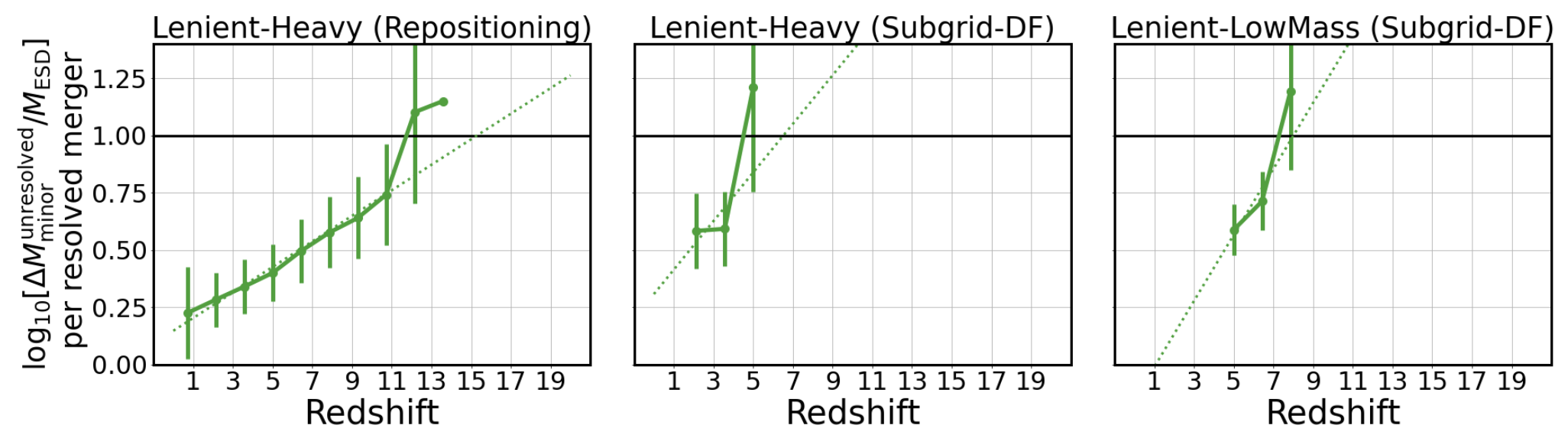}
\caption{\textit{Calibration of the contribution from unresolved minor mergers in the stochastic seed model}: 
The y-axis shows the mass growth due to unresolved minor mergers ($\Delta M^{\rm unresolved}_{\rm minor}$) per resolved merger, in units of the ESD mass ($\descendantseedmass$). This quantity is added to the remnant BH mass for every resolved merger occurring in the ESD simulations. The solid lines show the predicted contribution derived from the DGB simulations, which are able to resolve these minor mergers. The left and middle panels show results for the \texttt{Lenient-Heavy} simulations with BH repositioning and subgrid dynamical-friction (DF) treatments, respectively. The right panel corresponds to the \texttt{Lenient-LowMass} model with subgrid-DF. The contribution decreases with time, which we fit using a power-law model. The black line indicates the maximum allowed contribution of $10\,M_{\rm ESD}$, imposed to prevent unphysical growth in regimes not probed by the ESD simulations. 
}
\label{unresolved_minor_mergers}
\end{figure*}

Here we summarize additional details about the stochastic seed model and how its different components are calibrated using results from higher resolution DGB simulations. The distinct components of the stochastic seed model are calibrated as follows:

\subsection{Stochastic galaxy mass criterion}

In the \textit{stochastic galaxy mass criterion}, the galaxy-mass threshold for seeding is drawn from a log-normal distribution with mean $\left<M^{\rm galaxy}_{\rm seed}\right>$ and scatter $\sigma$. The mean threshold, $\left<M^{\rm galaxy}_{\rm seed}\right>$, is redshift-dependent and is described by a double power-law function,

\begin{eqnarray}
\begin{aligned}
&\log_{10}\left<M^{\rm galaxy}_{\rm seed}\right>= \\
&\left\{
    \begin{array}{lr}
        (z-z_{\mathrm{trans}})\,\alpha + \log_{10}M_{\mathrm{trans}},  & \text{if } z \geq z_{\mathrm{trans}}\\
        (z-z_{\mathrm{trans}})\,\beta + \log_{10}M_{\mathrm{trans}}, & \text{if } z < z_{\mathrm{trans}}
    \end{array}
\right\}.
\label{double_powerlaw_eqn}
\end{aligned}
\end{eqnarray}

Here, $\alpha$ and $\beta$ are the slopes of the relation above and below the transition redshift $z_{\rm trans}$, respectively, while $M_{\rm trans}$ is the mean galaxy-mass threshold for seeding at $z=z_{\rm trans}$. Further details on the motivation for this functional form are given in \cite{2024MNRAS.529.3768B}, where this model was first introduced. The key seed parameters for this criterion are therefore $\alpha$, $\beta$, $z_{\rm trans}$, $M_{\rm trans}$, and $\sigma$.

To calibrate these parameters for the \texttt{Lenient-LowMass}~($M_{\rm ESD} = 1.5\times10^5~M_{\odot}$), \texttt{Lenient-Heavy}, and \texttt{Strict-Heavy} seed models~($M_{\rm ESD} = 1\times10^6~M_{\odot}$), we plot in the left panel of Figure~\ref{bFOFmass_vs_redshift_at_assembly} the host-galaxy masses at the redshifts where descendant BHs first reach a mass of $M_{\rm ESD}$ in the DGB simulations. These galaxy masses are determined by tracking BH growth along galaxy merger trees, as described in Section~2.5 of \cite{2024MNRAS.529.3768B}. The resulting distributions differ substantially between the seed models, enabling them to be distinguished from one another when implemented in the lower-resolution ESD simulations.

We then choose values of $\alpha$, $\beta$, $z_{\rm trans}$, and $M_{\rm trans}$ such that the double power-law model predictions (dashed lines) broadly reproduce the mean trends of these distributions (solid lines). These values are listed in Table~\ref{tab:brahma_suite}. Due to the sparsity of the data, it is not feasible to perform a full MCMC fit to determine all parameters simultaneously. We reserve such an analysis for future work that will involve a larger set of simulations with multiple IC realizations. Instead, we choose suitable values of $z_{\rm trans}$ and $M_{\rm trans}$ by eye and fit $\alpha$ and $\beta$ using \texttt{scipy.optimize.curve_fit}. Lastly, we estimate $\sigma$ (faded regions around the dashed lines) as the average of the standard deviations computed across the redshift bins used to derive the mean trends. The resulting calibrations are expected to be most reliable within the redshift ranges probed by the corresponding DGB simulations. For example, the fit derived for the \texttt{Lenient-LowMass} model cannot be considered reliable below its final redshift of $z=5$. In addition, the data become increasingly sparse at higher redshifts, which reduces the reliability of the fits in that regime as well. 

Having calibrated the \textit{stochastic galaxy mass criterion}, the right panel of Figure~\ref{bFOFmass_vs_redshift_at_assembly} shows the galaxy masses and redshifts at which BHs are initialized in the lower-resolution, larger-volume ESD simulations. As intended, these broadly reproduce the predictions from the DGB simulations.

\subsection{Stochastic galaxy environment criterion}

The \textit{stochastic galaxy environment criterion} is implemented through an environment-dependent seed probability~($\environmentseedprobability$) assigned to each galaxy, given by

\begin{equation}
\environmentseedprobability=
\left\{
    \begin{array}{lr}
        \left(M^{\mathrm{galaxy}}_{\mathrm{total}}-\left<\massembly\right>\right)\gamma~p_0 + p_0 , & \text{if } N_{\mathrm{ngb}}=0\\
        \left(M^{\mathrm{galaxy}}_{\mathrm{total}}-\left<\massembly\right>\right)\gamma~p_1 + p_1 , & \text{if } N_{\mathrm{ngb}}=1 \\
        1, & \text{if } N_{\mathrm{ngb}}>1
    \end{array}
\right\},
\label{environment_based_seed_probability}
\end{equation}
where $N_{\rm ngb}$ is the number of neighboring galaxies with similar or higher mass within a distance of $5~R_{\rm vir}$. To preferentially seed BHs in richer environments, galaxies with fewer than two such neighbors are assigned probabilities smaller than unity, while galaxies with a higher number of neighbors are seeded with unit probability. This prescription introduces three seed parameters: $p_0$, $p_1$, and $\gamma$.

In \cite{2024MNRAS.529.3768B}, we explored a broad range of these parameters and found that values of $p_0 \sim 0.1\!-\!0.2$, $p_1 \sim 0.2\!-\!0.6$, and $\gamma \sim 1.6$ reproduced the BH two-point clustering over a wide range of seed models. In the present work, we found that similar choices also broadly reproduced the two-point BH clustering predicted by the higher-resolution DGB simulations in ESD boxes of the same volume (see Figure~\ref{small_scale_clustering}). Specifically, we adopt $p_0,p_1,\gamma = 0.1,0.3,1.6$ for the \texttt{Strict-Heavy} and \texttt{Lenient-Heavy} seed models, and $p_0,p_1,\gamma = 0.2,0.6,1.6$ for the \texttt{Lenient-LowMass} seed model.  

\subsection{Unresolved minor mergers}

Our ESD simulations, by construction, do not resolve the contribution from minor mergers (where the secondary BH mass is smaller than the ESD mass) to the growth of the ESDs. We therefore include this additional contribution explicitly in the BH growth. This is done by adding an extra ``unresolved merger contribution'' for every resolved merger that occurs in the simulation. In principle, this contribution can be directly calibrated using predictions from the DGB simulations, which resolve minor mergers. These predictions are shown in Figure~\ref{unresolved_minor_mergers} for the \texttt{Lenient-Heavy} and \texttt{Lenient-LowMass} seed models. For the \texttt{Lenient-Heavy} case, we also show results obtained using BH repositioning, as this treatment naturally produces the largest number of mergers and therefore provides stronger statistical power. We find that the mass growth from unresolved minor mergers, when normalized per resolved merger, decreases with time. This trend arises because the number of resolved mergers increases as BHs grow in mass beyond the ESD value, while the number of unresolved mergers is expected to decline as the formation of new seeds (which would serve as secondary BHs in minor mergers) becomes increasingly suppressed. A similar trend is seen when BH repositioning is replaced by the subgrid-DF model, although the statistical power is significantly lower due to the smaller number of mergers. We fit the DGB predictions with a simple power-law model and apply this calibration to the ESD simulations. Specifically, whenever a resolved merger occurs in the ESD simulations, we add an additional mass contribution to the remnant BH using these power-law fits. In Figure~15 of \cite{2024MNRAS.529.3768B}, we show that this correction accounts for the unresolved minor mergers not captured in the ESD simulations and is necessary to match the BHMFs obtained in the DGB simulations. Finally, we note that the \texttt{Strict-Heavy} model (with subgrid-DF) could not be calibrated due to the lack of mergers. We therefore adopt the same calibration as the \texttt{Lenient-Heavy} model as a rough approximation. In practice, this choice has little impact because the ESD model produces an extremely small number of resolved mergers in this case.

\section{Successes and limitations of the stochastic seed model}
\label{Comparing ESD vs DGB model results}

\begin{figure*}
\centering
\includegraphics[width = 5.2cm]{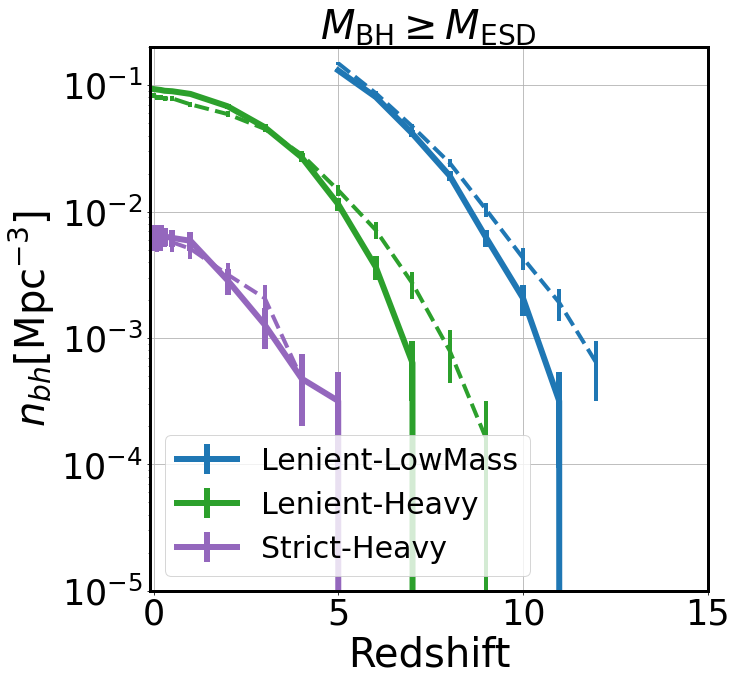}
\includegraphics[width = 12.7cm]{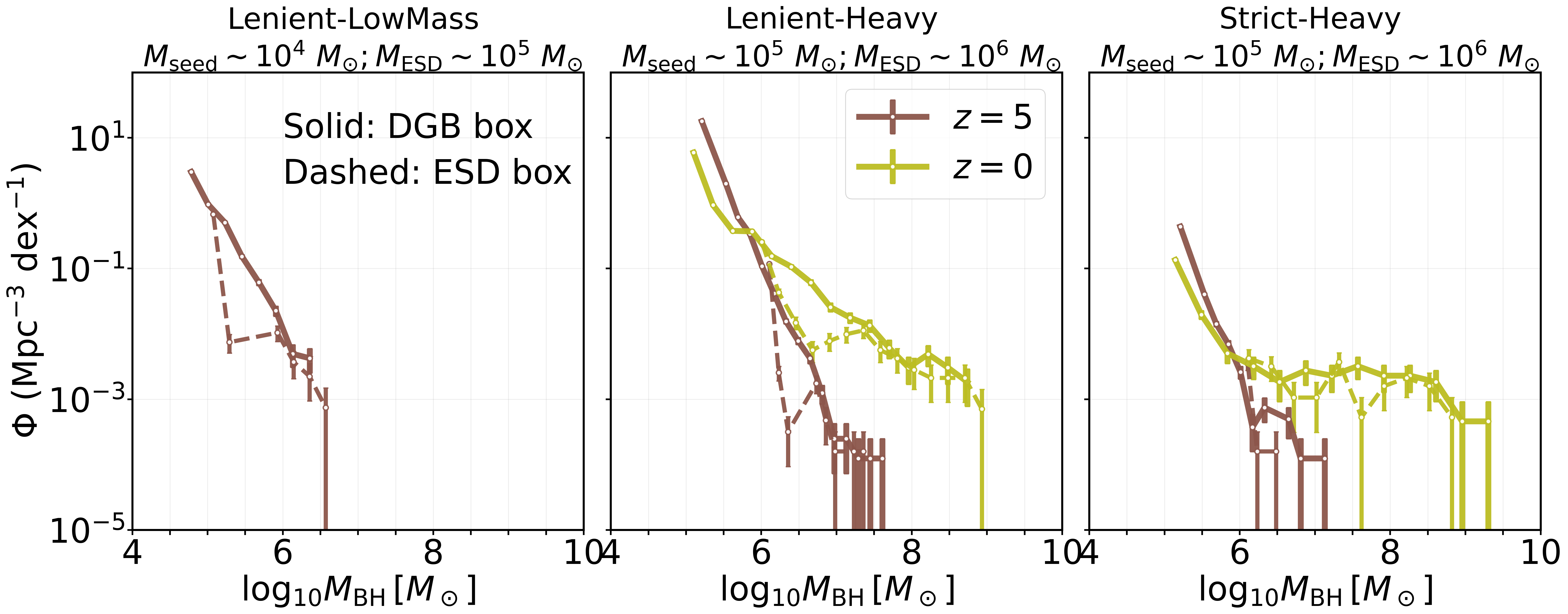}
\caption{Validation of the stochastic seed model: Comparison of the BH number density evolution (leftmost panel) and the BHMFs at $z=5~\&~0$~(remaining three panels) between the DGB (solid) and ESD (dashed) simulations. Importantly, both DGB and ESD simulations have the same volume and initial condition (IC) realization, although the ESD simulations have 8× lower mass resolution. The three seed models are shown with different colors in the number density evolution and in separate panels for the BHMFs. For the number densities, the ESD simulations best reproduce the DGB predictions at $z \lesssim 5$, where a sufficient number of BHs (above the ESD seed mass threshold) exist in the DGB simulations to robustly calibrate the stochastic seed model. The DGB BHMFs at $z\sim5~\&~0$ are also generally well reproduced by the ESD simulations, except at the low-mass end where the model~(by construction) cannot fully account for the missing contribution from unresolved minor mergers.   
}
\label{validating_ESD}
\end{figure*}

Here we assess the ability of our stochastic seed model to capture the higher resolution DGB simulation results within the lower resolution ESD simulations. Note that the stochastic seed model was originally built using simulations that used BH repositioning~\citep{2024MNRAS.529.3768B,2024MNRAS.531.4311B}, but here we have tested it for the first time in simulations that use a subgrid-DF treatment for BH dynamics.

Since our primary ESD simulations are $\sim8$ times larger in volume than the DGB simulations, deviations between their predictions (for the same underlying seed model) could arise from both cosmic variance and imperfect calibration or intrinsic model limitations. Here, we isolate the latter contribution by comparing supplementary ESD simulations that were run with the same volume and IC realizations as the DGB simulations. In Figure~\ref{validating_ESD}, we compare the BH number density evolution and BHMFs between the DGB and these supplementary ESD simulations.

For all three seed models, the ESD number densities reproduce the DGB predictions well at later redshifts, where sufficient statistics (i.e., BHs above the ESD seed mass threshold) exist in the DGB simulations to enable robust calibration. At the earliest redshifts, however, the calibrated ESD simulations tend to overpredict the number densities by factors of $\sim2$–$3$ for the \texttt{Lenient-Heavy} and \texttt{Lenient-LowMass} models. This discrepancy is likely due to the limited number of BHs in the DGB simulations available for calibration at these epochs~(see left panel of Figure \ref{bFOFmass_vs_redshift_at_assembly}). 

For the BHMFs, the ESD predictions initially exhibit a steeper decline at the low-mass end compared to the DGB results. This is primarily due to the absence of unresolved minor mergers in the ESD simulations. Our correction for this missing growth operates only after the first resolvable mergers occur in the ESD volumes (see Appendix~\ref{appendix_stochastic} for details). Once these mergers take place, the ESD BHMFs begin to converge toward the DGB predictions at higher BH masses. Consequently, at the massive end of the BHMFs, the discrepancy between the ESD and DGB predictions is significantly reduced. However, even at the high-mass end, the ESD BHMFs tend to remain slightly below their DGB counterparts. This likely reflects incomplete convergence of BH accretion at these resolutions; in particular, the peak gas densities that drive accretion tend to be underestimated at lower resolution~(this was also true for the \texttt{IllustrisTNG} simulations).

The differences between the ESD and DGB predictions (for a fixed seed model) are slightly larger than those reported in our previous works that employed BH repositioning \citep{2024MNRAS.529.3768B,2024MNRAS.531.4311B}. This is not unexpected, since repositioning leads to more efficient BH growth and therefore produces a larger population of massive BHs that can be used for more accurate calibration. Achieving comparable calibration accuracy under the subgrid-DF model would likely require a more extensive calibration effort using a larger suite of simulations, which we plan to pursue in future work. However, for the purposes of this paper, the differences between the DGB and ESD predictions remain much smaller than the differences between the seed models themselves. Therefore, these discrepancies do not affect the main conclusions of our work.

\section*{Acknowledgements}

AKB and PT acknowledge support from NSF-AST 2510738. AKB, AMG, and PT acknowledge support from NSF-AST 2346977 and the NSF-Simons AI Institute for Cosmic Origins which is supported by the National Science Foundation under Cooperative Agreement 2421782 and the Simons Foundation award MPS-AI-00010515. LB acknowledges support from NSF award AST-2307171 and NASA award 80NSSC22K0808. RW acknowledges funding of a Leibniz Junior Research Group (project number J131/2022). LH acknowledges support from the Simons Foundation under the ``Learning the Universe" initiative. The authors acknowledge Research Computing at The University of Virginia and The University of Florida for providing computational resources and technical support that have contributed to the results reported within this publication. URL: \url{https://rc.virginia.edu}
PN acknowledges support from the Gordon and Betty Moore Foundation and the John Templeton Foundation that fund the Black Hole Initiative (BHI) at Harvard University where she serves as an external PI. The authors used OpenAI GPT-5 to assist with simulation data-analysis scripts and editorial refinement of the manuscript text. The authors take full responsibility for the final content.

\section*{Data Availability}
The underlying data used in this work shall be made available
upon reasonable request to the corresponding author.

\bibliography{references}

@ARTICLE{2024ApJ...964...39G,
       author = {{Greene}, Jenny E. and {Labbe}, Ivo and {Goulding}, Andy D. and {Furtak}, Lukas J. and {Chemerynska}, Iryna and {Kokorev}, Vasily and {Dayal}, Pratika and {Volonteri}, Marta and {Williams}, Christina C. and {Wang}, Bingjie and {Setton}, David J. and {Burgasser}, Adam J. and {Bezanson}, Rachel and {Atek}, Hakim and {Brammer}, Gabriel and {Cutler}, Sam E. and {Feldmann}, Robert and {Fujimoto}, Seiji and {Glazebrook}, Karl and {de Graaff}, Anna and {Khullar}, Gourav and {Leja}, Joel and {Marchesini}, Danilo and {Maseda}, Michael V. and {Matthee}, Jorryt and {Miller}, Tim B. and {Naidu}, Rohan P. and {Nanayakkara}, Themiya and {Oesch}, Pascal A. and {Pan}, Richard and {Papovich}, Casey and {Price}, Sedona H. and {van Dokkum}, Pieter and {Weaver}, John R. and {Whitaker}, Katherine E. and {Zitrin}, Adi},
        title = "{UNCOVER Spectroscopy Confirms the Surprising Ubiquity of Active Galactic Nuclei in Red Sources at z > 5}",
      journal = {\apj},
         year = 2024,
        month = mar,
       volume = {964},
       number = {1},
          eid = {39},
        pages = {39},
          doi = {10.3847/1538-4357/ad1e5f},
archivePrefix = {arXiv},
       eprint = {2309.05714},
 primaryClass = {astro-ph.GA},
       adsurl = {https://ui.adsabs.harvard.edu/abs/2024ApJ...964...39G}
}

@ARTICLE{2025ApJ...993L..48W,
       author = {{Wu}, Ziyong and {Cen}, Renyue and {Teyssier}, Romain},
        title = "{How Fast Could Supermassive Black Holes Grow at the Epoch of Reionization?}",
      journal = {\apjl},
         year = 2025,
        month = nov,
       volume = {993},
       number = {2},
          eid = {L48},
        pages = {L48},
          doi = {10.3847/2041-8213/ae14d4},
archivePrefix = {arXiv},
       eprint = {2510.16532},
 primaryClass = {astro-ph.GA},
       adsurl = {https://ui.adsabs.harvard.edu/abs/2025ApJ...993L..48W}
}

@ARTICLE{2024Natur.636..594J,
       author = {{Juod{\v{z}}balis}, Ignas and {Maiolino}, Roberto and {Baker}, William M. and {Tacchella}, Sandro and {Scholtz}, Jan and {D'Eugenio}, Francesco and {Witstok}, Joris and {Schneider}, Raffaella and {Trinca}, Alessandro and {Valiante}, Rosa and {DeCoursey}, Christa and {Curti}, Mirko and {Carniani}, Stefano and {Chevallard}, Jacopo and {de Graaff}, Anna and {Arribas}, Santiago and {Bennett}, Jake S. and {Bourne}, Martin A. and {Bunker}, Andrew J. and {Charlot}, St{\'e}phane and {Jiang}, Brian and {Koudmani}, Sophie and {Perna}, Michele and {Robertson}, Brant and {Sijacki}, Debora and {{\"U}bler}, Hannah and {Williams}, Christina C. and {Willott}, Chris},
        title = "{A dormant overmassive black hole in the early Universe}",
      journal = {\nat},
         year = 2024,
        month = dec,
       volume = {636},
       number = {8043},
        pages = {594-597},
          doi = {10.1038/s41586-024-08210-5},
archivePrefix = {arXiv},
       eprint = {2403.03872},
 primaryClass = {astro-ph.GA},
       adsurl = {https://ui.adsabs.harvard.edu/abs/2024Natur.636..594J}
}

@ARTICLE{2026A&A...708A...7Z,
       author = {{Zana}, Tommaso and {Capelo}, Pedro R. and {Boresta}, Mairo and {Schneider}, Raffaella and {Lupi}, Alessandro and {Trinca}, Alessandro and {Mayer}, Lucio and {Valiante}, Rosa and {Graziani}, Luca},
        title = "{Super-Eddington accretion in protogalactic cores}",
      journal = {\aap},
         year = 2026,
        month = mar,
       volume = {708},
          eid = {A7},
        pages = {A7},
          doi = {10.1051/0004-6361/202557006},
archivePrefix = {arXiv},
       eprint = {2508.21114},
 primaryClass = {astro-ph.GA},
       adsurl = {https://ui.adsabs.harvard.edu/abs/2026A&A...708A...7Z}
}

@ARTICLE{2026NatAs.tmp...66L,
       author = {{Lambrides}, Erini and {Larson}, Rebecca L. and {Garofali}, Kristen and {Ptak}, Andrew and {Chiaberge}, Marco and {Long}, Arianna S. and {Hutchison}, Taylor A. and {Norman}, Colin and {McKinney}, Jed and {Akins}, Hollis B. and {Berg}, Danielle A. and {Chisholm}, John and {Civano}, Francesca and {Cloonan}, Aidan P. and {Endsley}, Ryan and {Faisst}, Andreas L. and {Gilli}, Roberto and {Gillman}, Steven and {Hirschmann}, Michaela and {Kartaltepe}, Jeyhan S. and {Kocevski}, Dale D. and {Kokorev}, Vasily and {Pacucci}, Fabio and {Richardson}, Chris T. and {Stiavelli}, Massimo and {Whalen}, Kelly E.},
        title = "{The case for super-Eddington accretion in JWST broad-line active galactic nuclei during the first billion years}",
      journal = {Nature Astronomy},
         year = 2026,
        month = apr,
          doi = {10.1038/s41550-026-02813-w},
archivePrefix = {arXiv},
       eprint = {2409.13047},
 primaryClass = {astro-ph.HE},
       adsurl = {https://ui.adsabs.harvard.edu/abs/2026NatAs.tmp...66L}
}

@ARTICLE{2025ApJ...978...92L,
       author = {{Labbe}, Ivo and {Greene}, Jenny E. and {Bezanson}, Rachel and {Fujimoto}, Seiji and {Furtak}, Lukas J. and {Goulding}, Andy D. and {Matthee}, Jorryt and {Naidu}, Rohan P. and {Oesch}, Pascal A. and {Atek}, Hakim and {Brammer}, Gabriel and {Chemerynska}, Iryna and {Coe}, Dan and {Cutler}, Sam E. and {Dayal}, Pratika and {Feldmann}, Robert and {Franx}, Marijn and {Glazebrook}, Karl and {Leja}, Joel and {Maseda}, Michael and {Marchesini}, Danilo and {Nanayakkara}, Themiya and {Nelson}, Erica J. and {Pan}, Richard and {Papovich}, Casey and {Price}, Sedona H. and {Suess}, Katherine A. and {Wang}, Bingjie and {Weaver}, John R. and {Whitaker}, Katherine E. and {Williams}, Christina C. and {Zitrin}, Adi},
        title = "{UNCOVER: Candidate Red Active Galactic Nuclei at 3 < z < 7 with JWST and ALMA}",
      journal = {\apj},
         year = 2025,
        month = jan,
       volume = {978},
       number = {1},
          eid = {92},
        pages = {92},
          doi = {10.3847/1538-4357/ad3551},
archivePrefix = {arXiv},
       eprint = {2306.07320},
 primaryClass = {astro-ph.GA},
       adsurl = {https://ui.adsabs.harvard.edu/abs/2025ApJ...978...92L}
}

@ARTICLE{2024ApJ...966L..30M,
       author = {{Mezcua}, Mar and {Pacucci}, Fabio and {Suh}, Hyewon and {Siudek}, Malgorzata and {Natarajan}, Priyamvada},
        title = "{Overmassive Black Holes at Cosmic Noon: Linking the Local and the High-redshift Universe}",
      journal = {\apjl},
         year = 2024,
        month = may,
       volume = {966},
       number = {2},
          eid = {L30},
        pages = {L30},
          doi = {10.3847/2041-8213/ad3c2a},
archivePrefix = {arXiv},
       eprint = {2404.05793},
 primaryClass = {astro-ph.GA},
       adsurl = {https://ui.adsabs.harvard.edu/abs/2024ApJ...966L..30M}
}

@ARTICLE{2013MNRAS.436..697B,
       author = {{Bernardi}, M. and {Meert}, A. and {Sheth}, R.~K. and {Vikram}, V. and {Huertas-Company}, M. and {Mei}, S. and {Shankar}, F.},
        title = "{The massive end of the luminosity and stellar mass functions: dependence on the fit to the light profile}",
      journal = {\mnras},
         year = 2013,
        month = nov,
       volume = {436},
       number = {1},
        pages = {697-704},
          doi = {10.1093/mnras/stt1607},
archivePrefix = {arXiv},
       eprint = {1304.7778},
 primaryClass = {astro-ph.CO},
       adsurl = {https://ui.adsabs.harvard.edu/abs/2013MNRAS.436..697B}
}

@ARTICLE{2025ApJ...978...77B,
       author = {{Burke}, Colin J. and {Natarajan}, Priyamvada and {Baldassare}, Vivienne F. and {Geha}, Marla},
        title = "{Multiwavelength Constraints on the Local Black Hole Occupation Fraction}",
      journal = {\apj},
         year = 2025,
        month = jan,
       volume = {978},
       number = {1},
          eid = {77},
        pages = {77},
          doi = {10.3847/1538-4357/ad94d9},
archivePrefix = {arXiv},
       eprint = {2410.11177},
 primaryClass = {astro-ph.GA},
       adsurl = {https://ui.adsabs.harvard.edu/abs/2025ApJ...978...77B}
}

@ARTICLE{2024ApJ...971L..29L,
       author = {{Liepold}, Emily R. and {Ma}, Chung-Pei},
        title = "{Big Galaxies and Big Black Holes: The Massive Ends of the Local Stellar and Black Hole Mass Functions and the Implications for Nanohertz Gravitational Waves}",
      journal = {\apjl},
         year = 2024,
        month = aug,
       volume = {971},
       number = {2},
          eid = {L29},
        pages = {L29},
          doi = {10.3847/2041-8213/ad66b8},
archivePrefix = {arXiv},
       eprint = {2407.14595},
 primaryClass = {astro-ph.GA},
       adsurl = {https://ui.adsabs.harvard.edu/abs/2024ApJ...971L..29L}
}

@ARTICLE{2023MNRAS.518.2123Z,
       author = {{Zhang}, Haowen and {Behroozi}, Peter and {Volonteri}, Marta and {Silk}, Joseph and {Fan}, Xiaohui and {Hopkins}, Philip F. and {Yang}, Jinyi and {Aird}, James},
        title = "{TRINITY I: self-consistently modelling the dark matter halo-galaxy-supermassive black hole connection from z = 0-10}",
      journal = {\mnras},
         year = 2023,
        month = jan,
       volume = {518},
       number = {2},
        pages = {2123-2163},
          doi = {10.1093/mnras/stac2633},
archivePrefix = {arXiv},
       eprint = {2105.10474},
 primaryClass = {astro-ph.GA},
       adsurl = {https://ui.adsabs.harvard.edu/abs/2023MNRAS.518.2123Z}
}

@ARTICLE{2024MNRAS.531.4503H,
       author = {{Hern{\'a}ndez-Y{\'e}venes}, J. and {Nagar}, N. and {Arratia}, V. and {Jarrett}, T.~H.},
        title = "{WISE2MBH: a scaling-based algorithm for probing supermassive black hole masses through WISE catalogues}",
      journal = {\mnras},
         year = 2024,
        month = jul,
       volume = {531},
       number = {4},
        pages = {4503-4523},
          doi = {10.1093/mnras/stae1372},
archivePrefix = {arXiv},
       eprint = {2405.18336},
 primaryClass = {astro-ph.GA},
       adsurl = {https://ui.adsabs.harvard.edu/abs/2024MNRAS.531.4503H}
}

@ARTICLE{2024ApJ...969...93C,
       author = {{Cho}, Hojin and {Woo}, Jong-Hak},
        title = "{Constraining the Low-mass End of the Black Hole Mass Function and the Active Fraction of the Intermediate-mass Black Holes}",
      journal = {\apj},
         year = 2024,
        month = jul,
       volume = {969},
       number = {2},
          eid = {93},
        pages = {93},
          doi = {10.3847/1538-4357/ad4966},
archivePrefix = {arXiv},
       eprint = {2405.09441},
 primaryClass = {astro-ph.GA},
       adsurl = {https://ui.adsabs.harvard.edu/abs/2024ApJ...969...93C}
}

@ARTICLE{2026ApJ...998...48P,
       author = {{Porras-Valverde}, Antonio J. and {Ricarte}, Angelo and {Natarajan}, Priyamvada and {Somerville}, Rachel S. and {Gabrielpillai}, Austen and {Yung}, L.~Y. Aaron},
        title = "{Tracking the Assembly of Supermassive Black Holes: A Comparison of Diverse Models across Cosmic Time}",
      journal = {\apj},
         year = 2026,
        month = feb,
       volume = {998},
       number = {1},
          eid = {48},
        pages = {48},
          doi = {10.3847/1538-4357/ae2fb1},
archivePrefix = {arXiv},
       eprint = {2504.11566},
 primaryClass = {astro-ph.GA},
       adsurl = {https://ui.adsabs.harvard.edu/abs/2026ApJ...998...48P}
}

@article{Xu_2023,
   title={Searching for the Nano-Hertz Stochastic Gravitational Wave Background with the Chinese Pulsar Timing Array Data Release I},
   volume={23},
   ISSN={1674-4527},
   url={http://dx.doi.org/10.1088/1674-4527/acdfa5},
   DOI={10.1088/1674-4527/acdfa5},
   number={7},
   journal={Research in Astronomy and Astrophysics},
   publisher={IOP Publishing},
   author={Xu, Heng and Chen, Siyuan and Guo, Yanjun and Jiang, Jinchen and Wang, Bojun and Xu, Jiangwei and Xue, Zihan and Nicolas Caballero, R. and Yuan, Jianping and Xu, Yonghua and Wang, Jingbo and Hao, Longfei and Luo, Jingtao and Lee, Kejia and Han, Jinlin and Jiang, Peng and Shen, Zhiqiang and Wang, Min and Wang, Na and Xu, Renxin and Wu, Xiangping and Manchester, Richard and Qian, Lei and Guan, Xin and Huang, Menglin and Sun, Chun and Zhu, Yan},
   year={2023},
   month=jun, pages={075024} }

@ARTICLE{2024A&A...685A..94E,
       author = {{EPTA Collaboration} and {InPTA Collaboration} and {Antoniadis}, J. and {Arumugam}, P. and {Arumugam}, S. and {Babak}, S. and {Bagchi}, M. and {Bak Nielsen}, A.-S. and {Bassa}, C.~G. and {Bathula}, A. and {Berthereau}, A. and {Bonetti}, M. and {Bortolas}, E. and {Brook}, P.~R. and {Burgay}, M. and {Caballero}, R.~N. and {Chalumeau}, A. and {Champion}, D.~J. and {Chanlaridis}, S. and {Chen}, S. and {Cognard}, I. and {Dandapat}, S. and {Deb}, D. and {Desai}, S. and {Desvignes}, G. and {Dhanda-Batra}, N. and {Dwivedi}, C. and {Falxa}, M. and {Ferdman}, R.~D. and {Franchini}, A. and {Gair}, J.~R. and {Goncharov}, B. and {Gopakumar}, A. and {Graikou}, E. and {Grie{\ss}meier}, J.-M. and {Gualandris}, A. and {Guillemot}, L. and {Guo}, Y.~J. and {Gupta}, Y. and {Hisano}, S. and {Hu}, H. and {Iraci}, F. and {Izquierdo-Villalba}, D. and {Jang}, J. and {Jawor}, J. and {Janssen}, G.~H. and {Jessner}, A. and {Joshi}, B.~C. and {Kareem}, F. and {Karuppusamy}, R. and {Keane}, E.~F. and {Keith}, M.~J. and {Kharbanda}, D. and {Kikunaga}, T. and {Kolhe}, N. and {Kramer}, M. and {Krishnakumar}, M.~A. and {Lackeos}, K. and {Lee}, K.~J. and {Liu}, K. and {Liu}, Y. and {Lyne}, A.~G. and {McKee}, J.~W. and {Maan}, Y. and {Main}, R.~A. and {Mickaliger}, M.~B. and {Ni{\c{t}}u}, I.~C. and {Nobleson}, K. and {Paladi}, A.~K. and {Parthasarathy}, A. and {Perera}, B.~B.~P. and {Perrodin}, D. and {Petiteau}, A. and {Porayko}, N.~K. and {Possenti}, A. and {Prabu}, T. and {Quelquejay Leclere}, H. and {Rana}, P. and {Samajdar}, A. and {Sanidas}, S.~A. and {Sesana}, A. and {Shaifullah}, G. and {Singha}, J. and {Speri}, L. and {Spiewak}, R. and {Srivastava}, A. and {Stappers}, B.~W. and {Surnis}, M. and {Susarla}, S.~C. and {Susobhanan}, A. and {Takahashi}, K. and {Tarafdar}, P. and {Theureau}, G. and {Tiburzi}, C. and {van der Wateren}, E. and {Vecchio}, A. and {Venkatraman Krishnan}, V. and {Verbiest}, J.~P.~W. and {Wang}, J. and {Wang}, L. and {Wu}, Z. and {Auclair}, P. and {Barausse}, E. and {Caprini}, C. and {Crisostomi}, M. and {Fastidio}, F. and {Khizriev}, T. and {Middleton}, H. and {Neronov}, A. and {Postnov}, K. and {Roper Pol}, A. and {Semikoz}, D. and {Smarra}, C. and {Steer}, D.~A. and {Truant}, R.~J. and {Valtolina}, S.},
        title = "{The second data release from the European Pulsar Timing Array. IV. Implications for massive black holes, dark matter, and the early Universe}",
      journal = {\aap},
         year = 2024,
        month = may,
       volume = {685},
          eid = {A94},
        pages = {A94},
          doi = {10.1051/0004-6361/202347433},
archivePrefix = {arXiv},
       eprint = {2306.16227},
 primaryClass = {astro-ph.CO},
       adsurl = {https://ui.adsabs.harvard.edu/abs/2024A&A...685A..94E}
}

@ARTICLE{2026ApJ...997..188M,
       author = {{Matt}, Cayenne and {G{\"u}ltekin}, Kayhan and {Kelley}, Luke Zoltan and {Blecha}, Laura and {Simon}, Joseph and {Agazie}, Gabriella and {Anumarlapudi}, Akash and {Archibald}, Anne M. and {Arzoumanian}, Zaven and {Baier}, Jeremy G. and {Baker}, Paul T. and {B{\'e}csy}, Bence and {Brazier}, Adam and {Brook}, Paul R. and {Burke-Spolaor}, Sarah and {Burnette}, Rand and {Case}, Robin and {Casey-Clyde}, J. Andrew and {Charisi}, Maria and {Chatterjee}, Shami and {Cohen}, Tyler and {Cordes}, James M. and {Cornish}, Neil J. and {Crawford}, Fronefield and {Cromartie}, H. Thankful and {Crowter}, Kathryn and {DeCesar}, Megan E. and {Demorest}, Paul B. and {Deng}, Heling and {Dey}, Lankeswar and {Dolch}, Timothy and {Ferrara}, Elizabeth C. and {Fiore}, William and {Fonseca}, Emmanuel and {Freedman}, Gabriel E. and {Gardiner}, Emiko C. and {Garver-Daniels}, Nate and {Gentile}, Peter A. and {Gersbach}, Kyle A. and {Glaser}, Joseph and {Good}, Deborah C. and {Harris}, C.~J. and {Hazboun}, Jeffrey S. and {Jennings}, Ross J. and {Johnson}, Aaron D. and {Jones}, Megan L. and {Kaplan}, David L. and {Kerr}, Matthew and {Key}, Joey S. and {Laal}, Nima and {Lam}, Michael T. and {Lamb}, William G. and {Larsen}, Bjorn and {Lazio}, T. Joseph W. and {Lewandowska}, Natalia and {Liu}, Tingting and {Lorimer}, Duncan R. and {Luo}, Jing and {Lynch}, Ryan S. and {Ma}, Chung-Pei and {Madison}, Dustin R. and {McEwen}, Alexander and {McKee}, James W. and {McLaughlin}, Maura A. and {McMann}, Natasha and {Meyers}, Bradley W. and {Meyers}, Patrick M. and {Mingarelli}, Chiara M.~F. and {Mitridate}, Andrea and {Ng}, Cherry and {Nice}, David J. and {Ocker}, Stella Koch and {Olum}, Ken D. and {Pennucci}, Timothy T. and {Perera}, Benetge B.~P. and {Petrov}, Polina and {Pol}, Nihan S. and {Radovan}, Henri A. and {Ransom}, Scott M. and {Ray}, Paul S. and {Romano}, Joseph D. and {Runnoe}, Jessie C. and {Saffer}, Alexander and {Sardesai}, Shashwat C. and {Schmiedekamp}, Ann and {Schmiedekamp}, Carl and {Schmitz}, Kai and {Shapiro-Albert}, Brent J. and {Siemens}, Xavier and {Fiscella}, Sophia V. Sosa and {Stairs}, Ingrid H. and {Stinebring}, Daniel R. and {Stovall}, Kevin and {Susobhanan}, Abhimanyu and {Swiggum}, Joseph K. and {Taylor}, Jacob and {Taylor}, Stephen R. and {Thompson}, Mercedes S. and {Turner}, Jacob E. and {Vallisneri}, Michele and {van Haasteren}, Rutger and {Vigeland}, Sarah J. and {Wahl}, Haley M. and {Wilson}, Kevin P. and {Witt}, Caitlin A. and {Wright}, David and {Young}, Olivia},
        title = "{Inferring M$_{BH}$─M$_{bulge}$ Evolution from the Gravitational-wave Background}",
      journal = {\apj},
         year = 2026,
        month = feb,
       volume = {997},
       number = {2},
          eid = {188},
        pages = {188},
          doi = {10.3847/1538-4357/ae2480},
archivePrefix = {arXiv},
       eprint = {2508.18126},
 primaryClass = {astro-ph.HE},
       adsurl = {https://ui.adsabs.harvard.edu/abs/2026ApJ...997..188M}
}

@ARTICLE{2024ApJ...963..129M,
       author = {{Matthee}, Jorryt and {Naidu}, Rohan P. and {Brammer}, Gabriel and {Chisholm}, John and {Eilers}, Anna-Christina and {Goulding}, Andy and {Greene}, Jenny and {Kashino}, Daichi and {Labbe}, Ivo and {Lilly}, Simon J. and {Mackenzie}, Ruari and {Oesch}, Pascal A. and {Weibel}, Andrea and {Wuyts}, Stijn and {Xiao}, Mengyuan and {Bordoloi}, Rongmon and {Bouwens}, Rychard and {van Dokkum}, Pieter and {Illingworth}, Garth and {Kramarenko}, Ivan and {Maseda}, Michael V. and {Mason}, Charlotte and {Meyer}, Romain A. and {Nelson}, Erica J. and {Reddy}, Naveen A. and {Shivaei}, Irene and {Simcoe}, Robert A. and {Yue}, Minghao},
        title = "{Little Red Dots: An Abundant Population of Faint Active Galactic Nuclei at z {\ensuremath{\sim}} 5 Revealed by the EIGER and FRESCO JWST Surveys}",
      journal = {\apj},
         year = 2024,
        month = mar,
       volume = {963},
       number = {2},
          eid = {129},
        pages = {129},
          doi = {10.3847/1538-4357/ad2345},
archivePrefix = {arXiv},
       eprint = {2306.05448},
 primaryClass = {astro-ph.GA},
       adsurl = {https://ui.adsabs.harvard.edu/abs/2024ApJ...963..129M}
}

@ARTICLE{2025ApJ...979..138H,
       author = {{Hainline}, Kevin N. and {Maiolino}, Roberto and {Juod{\v{z}}balis}, Ignas and {Scholtz}, Jan and {{\"U}bler}, Hannah and {D'Eugenio}, Francesco and {Helton}, Jakob M. and {Sun}, Yang and {Sun}, Fengwu and {Robertson}, Brant and {Tacchella}, Sandro and {Bunker}, Andrew J. and {Carniani}, Stefano and {Charlot}, Stephane and {Curtis-Lake}, Emma and {Egami}, Eiichi and {Johnson}, Benjamin D. and {Lin}, Xiaojing and {Lyu}, Jianwei and {P{\'e}rez-Gonz{\'a}lez}, Pablo G. and {Rinaldi}, Pierluigi and {Silcock}, Maddie S. and {Venturi}, Giacomo and {Williams}, Christina C. and {Willmer}, Christopher N.~A. and {Willott}, Chris and {Zhang}, Junyu and {Zhu}, Yongda},
        title = "{An Investigation into the Selection and Colors of Little Red Dots and Active Galactic Nuclei}",
      journal = {\apj},
         year = 2025,
        month = feb,
       volume = {979},
       number = {2},
          eid = {138},
        pages = {138},
          doi = {10.3847/1538-4357/ad9920},
archivePrefix = {arXiv},
       eprint = {2410.00100},
 primaryClass = {astro-ph.GA},
       adsurl = {https://ui.adsabs.harvard.edu/abs/2025ApJ...979..138H}
}

@ARTICLE{2025NatAs...9.1732S,
       author = {{Schindler}, Jan-Torge and {Hennawi}, Joseph F. and {Davies}, Frederick B. and {Bosman}, Sarah E.~I. and {Endsley}, Ryan and {Wang}, Feige and {Yang}, Jinyi and {Barth}, Aaron J. and {Eilers}, Anna-Christina and {Fan}, Xiaohui and {Kakiichi}, Koki and {Maseda}, Michael and {Pizzati}, Elia and {Nanni}, Riccardo},
        title = "{A little red dot at z = 7.3 within a large galaxy overdensity}",
      journal = {Nature Astronomy},
         year = 2025,
        month = nov,
       volume = {9},
       number = {11},
        pages = {1732-1744},
          doi = {10.1038/s41550-025-02660-1},
archivePrefix = {arXiv},
       eprint = {2411.11534},
 primaryClass = {astro-ph.GA},
       adsurl = {https://ui.adsabs.harvard.edu/abs/2025NatAs...9.1732S}
}

@ARTICLE{2026ApJ..1002..129B,
       author = {{Brooks}, Madisyn and {Trump}, Jonathan R. and {Simons}, Raymond C. and {Cole}, Justin and {Taylor}, Anthony J. and {Bagley}, Micaela B. and {Finkelstein}, Steven L. and {Davis}, Kelcey and {Amor{\'\i}n}, Ricardo O. and {Backhaus}, Bren E. and {Cleri}, Nikko J. and {Giavalisco}, Mauro and {Grogin}, Norman A. and {Hirschmann}, Michaela and {Holwerda}, Benne W. and {Huertas-Company}, Marc and {Kartaltepe}, Jeyhan S. and {Kocevski}, Dale D. and {Koekemoer}, Anton M. and {Lucas}, Ray A. and {Pacucci}, Fabio and {Wang}, Xin},
        title = "{Beyond the Monsters: A More Complete Census of Black Hole Activity at Cosmic Dawn}",
      journal = {\apj},
         year = 2026,
        month = may,
       volume = {1002},
       number = {2},
          eid = {129},
        pages = {129},
          doi = {10.3847/1538-4357/ae5652},
archivePrefix = {arXiv},
       eprint = {2511.19609},
 primaryClass = {astro-ph.GA},
       adsurl = {https://ui.adsabs.harvard.edu/abs/2026ApJ..1002..129B}
}

@ARTICLE{2025ApJ...985..169D,
       author = {{Durodola}, Emmanuel and {Pacucci}, Fabio and {Hickox}, Ryan C.},
        title = "{Exploring the Active Galactic Nucleus Fraction of a Sample of JWST's Little Red Dots at 4 < z < 8: Overmassive Black Holes Are Strongly Favored}",
      journal = {\apj},
         year = 2025,
        month = jun,
       volume = {985},
       number = {2},
          eid = {169},
        pages = {169},
          doi = {10.3847/1538-4357/adced2},
archivePrefix = {arXiv},
       eprint = {2406.10329},
 primaryClass = {astro-ph.GA},
       adsurl = {https://ui.adsabs.harvard.edu/abs/2025ApJ...985..169D}
}

@ARTICLE{2025ApJ...991...37A,
       author = {{Akins}, Hollis B. and {Casey}, Caitlin M. and {Lambrides}, Erini and {Allen}, Natalie and {Andika}, Irham T. and {Brinch}, Malte and {Champagne}, Jaclyn B. and {Cooper}, Olivia and {Ding}, Xuheng and {Drakos}, Nicole E. and {Faisst}, Andreas and {Finkelstein}, Steven L. and {Franco}, Maximilien and {Fujimoto}, Seiji and {Gentile}, Fabrizio and {Gillman}, Steven and {Gozaliasl}, Ghassem and {Harish}, Santosh and {Hayward}, Christopher C. and {Hirschmann}, Michaela and {Ilbert}, Olivier and {Kartaltepe}, Jeyhan S. and {Kocevski}, Dale D. and {Koekemoer}, Anton M. and {Kokorev}, Vasily and {Liu}, Daizhong and {Long}, Arianna S. and {McCracken}, Henry Joy and {McKinney}, Jed and {Onoue}, Masafusa and {Paquereau}, Louise and {Renzini}, Alvio and {Rhodes}, Jason and {Robertson}, Brant E. and {Shuntov}, Marko and {Silverman}, John D. and {Tanaka}, Takumi S. and {Toft}, Sune and {Trakhtenbrot}, Benny and {Valentino}, Francesco and {Zavala}, Jorge},
        title = "{COSMOS-Web: The Overabundance and Physical Nature of ``Little Red Dots''{\textemdash}Implications for Early Galaxy and SMBH Assembly}",
      journal = {\apj},
         year = 2025,
        month = sep,
       volume = {991},
       number = {1},
          eid = {37},
        pages = {37},
          doi = {10.3847/1538-4357/ade984},
archivePrefix = {arXiv},
       eprint = {2406.10341},
 primaryClass = {astro-ph.GA},
       adsurl = {https://ui.adsabs.harvard.edu/abs/2025ApJ...991...37A}
}

@ARTICLE{2025ApJ...994...40P,
       author = {{Pacucci}, Fabio and {Hernquist}, Lars and {Fujii}, Michiko},
        title = "{Little Red Dots are Nurseries of Massive Black Holes}",
      journal = {\apj},
         year = 2025,
        month = nov,
       volume = {994},
       number = {1},
          eid = {40},
        pages = {40},
          doi = {10.3847/1538-4357/ae1619},
archivePrefix = {arXiv},
       eprint = {2509.02664},
 primaryClass = {astro-ph.GA},
       adsurl = {https://ui.adsabs.harvard.edu/abs/2025ApJ...994...40P}
}

@ARTICLE{2024A&A...689A.128L,
       author = {{Lupi}, Alessandro and {Trinca}, Alessandro and {Volonteri}, Marta and {Dotti}, Massimo and {Mazzucchelli}, Chiara},
        title = "{Size matters: are we witnessing super-Eddington accretion in high-redshift black holes from JWST?}",
      journal = {\aap},
         year = 2024,
        month = sep,
       volume = {689},
          eid = {A128},
        pages = {A128},
          doi = {10.1051/0004-6361/202451249},
archivePrefix = {arXiv},
       eprint = {2406.17847},
 primaryClass = {astro-ph.HE},
       adsurl = {https://ui.adsabs.harvard.edu/abs/2024A&A...689A.128L}
}

@ARTICLE{TalPN2014,
       author = {{Alexander}, Tal and {Natarajan}, Priyamvada},
        title = "{Rapid growth of seed black holes in the early universe by supra-exponential accretion}",
      journal = {Science},
         year = 2014,
        month = sep,
       volume = {345},
       number = {6202},
        pages = {1330-1333},
          doi = {10.1126/science.1251053},
archivePrefix = {arXiv},
       eprint = {1408.1718},
 primaryClass = {astro-ph.GA},
       adsurl = {https://ui.adsabs.harvard.edu/abs/2014Sci...345.1330A}
}

@ARTICLE{2025arXiv250616145B,
       author = {{Barlow-Hall}, C.~L. and {Aird}, J.},
        title = "{Measurements of the z=4-10 X-ray Luminosity Function: the high space density of moderate-luminosity, obscured AGN}",
      journal = {arXiv e-prints},
         year = 2025,
        month = jun,
          eid = {arXiv:2506.16145},
        pages = {arXiv:2506.16145},
          doi = {10.48550/arXiv.2506.16145},
archivePrefix = {arXiv},
       eprint = {2506.16145},
 primaryClass = {astro-ph.HE},
       adsurl = {https://ui.adsabs.harvard.edu/abs/2025arXiv250616145B}
}

@ARTICLE{2015ApJ...813...82R,
       author = {{Reines}, Amy E. and {Volonteri}, Marta},
        title = "{Relations between Central Black Hole Mass and Total Galaxy Stellar Mass in the Local Universe}",
      journal = {\apj},
         year = 2015,
        month = nov,
       volume = {813},
       number = {2},
          eid = {82},
        pages = {82},
          doi = {10.1088/0004-637X/813/2/82},
archivePrefix = {arXiv},
       eprint = {1508.06274},
 primaryClass = {astro-ph.GA},
       adsurl = {https://ui.adsabs.harvard.edu/abs/2015ApJ...813...82R}
}

@ARTICLE{2025ApJ...991...81B,
       author = {{Bhowmick}, Aklant K. and {Blecha}, Laura and {Kelley}, Luke Z. and {Sivasankaran}, Aneesh and {Torrey}, Paul and {Weinberger}, Rainer and {Chen}, Nianyi and {Vogelsberger}, Mark and {Hernquist}, Lars and {Natarajan}, Priyamvada and {Di Matteo}, Tiziana},
        title = "{Dynamics of Low-mass Black Hole Seeds in the BRAHMA Simulations Using Subgrid Dynamical Friction: Impact on Merger-driven Black Hole Growth in the High-redshift Universe}",
      journal = {\apj},
         year = 2025,
        month = sep,
       volume = {991},
       number = {1},
          eid = {81},
        pages = {81},
          doi = {10.3847/1538-4357/adf96b},
archivePrefix = {arXiv},
       eprint = {2506.09184},
 primaryClass = {astro-ph.GA},
       adsurl = {https://ui.adsabs.harvard.edu/abs/2025ApJ...991...81B}
}

@ARTICLE{2024NatAs...8..126B,
       author = {{Bogd{\'a}n}, {\'A}kos and {Goulding}, Andy D. and {Natarajan}, Priyamvada and {Kov{\'a}cs}, Orsolya E. and {Tremblay}, Grant R. and {Chadayammuri}, Urmila and {Volonteri}, Marta and {Kraft}, Ralph P. and {Forman}, William R. and {Jones}, Christine and {Churazov}, Eugene and {Zhuravleva}, Irina},
        title = "{Evidence for heavy-seed origin of early supermassive black holes from a z {\ensuremath{\approx}} 10 X-ray quasar}",
      journal = {Nature Astronomy},
         year = 2024,
        month = jan,
       volume = {8},
       number = {1},
        pages = {126-133},
          doi = {10.1038/s41550-023-02111-9},
archivePrefix = {arXiv},
       eprint = {2305.15458},
 primaryClass = {astro-ph.GA},
       adsurl = {https://ui.adsabs.harvard.edu/abs/2024NatAs...8..126B}
}

@ARTICLE{2024arXiv240906772T,
       author = {{Taylor}, Anthony J. and {Finkelstein}, Steven L. and {Kocevski}, Dale D. and {Jeon}, Junehyoung and {Bromm}, Volker and {Amorin}, Ricardo O. and {Arrabal Haro}, Pablo and {Backhaus}, Bren E. and {Bagley}, Micaela B. and {Ba{\~n}ados}, Eduardo and {Bhatawdekar}, Rachana and {Brooks}, Madisyn and {Calabro}, Antonello and {Chavez Ortiz}, Oscar A. and {Cheng}, Yingjie and {Cleri}, Nikko J. and {Cole}, Justin W. and {Davis}, Kelcey and {Dickinson}, Mark and {Donnan}, Callum and {Dunlop}, James S. and {Ellis}, Richard S. and {Fernandez}, Vital and {Fontana}, Adriano and {Fujimoto}, Seiji and {Giavalisco}, Mauro and {Grazian}, Andrea and {Guo}, Jingsong and {Hathi}, Nimish P. and {Holwerda}, Benne W. and {Hirschmann}, Michaela and {Inayoshi}, Kohei and {Kartaltepe}, Jeyhan S. and {Khusanova}, Yana and {Koekemoer}, Anton M. and {Kokorev}, Vasily and {Larson}, Rebecca L. and {Leung}, Gene C.~K. and {Lucas}, Ray A. and {McLeod}, Derek J. and {Napolitano}, Lorenzo and {Onoue}, Masafusa and {Pacucci}, Fabio and {Papovich}, Casey and {P{\'e}rez-Gonz{\'a}lez}, Pablo G. and {Pirzkal}, Nor and {Somerville}, Rachel S. and {Trump}, Jonathan R. and {Wilkins}, Stephen M. and {Yung}, L.~Y. Aaron and {Zhang}, Haowen},
        title = "{Broad-Line AGN at $3.5<z<6$: The Black Hole Mass Function and a Connection with Little Red Dots}",
      journal = {arXiv e-prints},
         year = 2024,
        month = sep,
          eid = {arXiv:2409.06772},
        pages = {arXiv:2409.06772},
          doi = {10.48550/arXiv.2409.06772},
archivePrefix = {arXiv},
       eprint = {2409.06772},
 primaryClass = {astro-ph.GA},
       adsurl = {https://ui.adsabs.harvard.edu/abs/2024arXiv240906772T}
}

@ARTICLE{2026arXiv260104955C,
       author = {{Chon}, Sunmyon and {Hirano}, Shingo and {Ishiyama}, Tomoaki and {Chang}, Seok-Jun and {Springel}, Volker},
        title = "{Rapid emergence of overmassive black holes in the early Universe}",
      journal = {arXiv e-prints},
         year = 2026,
        month = jan,
          eid = {arXiv:2601.04955},
        pages = {arXiv:2601.04955},
          doi = {10.48550/arXiv.2601.04955},
archivePrefix = {arXiv},
       eprint = {2601.04955},
 primaryClass = {astro-ph.GA},
       adsurl = {https://ui.adsabs.harvard.edu/abs/2026arXiv260104955C}
}

@ARTICLE{2026ApJ..1000L..19Z,
       author = {{Zhang}, Saiyang and {Liu}, Boyuan and {Bromm}, Volker and {K{\"u}hnel}, Florian},
        title = "{Primordial Black Holes as Seeds for Extremely Overmassive Active Galactic Nuclei Observed by JWST}",
      journal = {\apjl},
         year = 2026,
        month = mar,
       volume = {1000},
       number = {1},
          eid = {L19},
        pages = {L19},
          doi = {10.3847/2041-8213/ae4bd0},
archivePrefix = {arXiv},
       eprint = {2512.14066},
 primaryClass = {astro-ph.GA},
       adsurl = {https://ui.adsabs.harvard.edu/abs/2026ApJ..1000L..19Z}
}

@ARTICLE{2025arXiv251109640P,
       author = {{Prole}, Lewis R. and {Regan}, John A. and {Mehta}, Daxal and {Pakmor}, Rudiger and {Koudmani}, Sophie and {Bourne}, Martin A. and {Glover}, Simon C.~O. and {Wise}, John H. and {Klessen}, Ralf S. and {Tremmel}, Michael and {Sijacki}, Debora and {Beckmann}, Ricarda S. and {Haehnelt}, Martin G. and {Brennan}, John and {van de Bor}, Pelle and {Clark}, Paul C.},
        title = "{The SEEDZ Simulations: Methodology and First Results on Massive Black Hole Seeding and Early Galaxy Growth}",
      journal = {arXiv e-prints},
         year = 2025,
        month = nov,
          eid = {arXiv:2511.09640},
        pages = {arXiv:2511.09640},
          doi = {10.48550/arXiv.2511.09640},
archivePrefix = {arXiv},
       eprint = {2511.09640},
 primaryClass = {astro-ph.GA},
       adsurl = {https://ui.adsabs.harvard.edu/abs/2025arXiv251109640P}
}

@ARTICLE{2025ApJ...979..127J,
       author = {{Jeon}, Junehyoung and {Bromm}, Volker and {Liu}, Boyuan and {Finkelstein}, Steven L.},
        title = "{Physical Pathways for JWST-observed Supermassive Black Holes in the Early Universe}",
      journal = {\apj},
         year = 2025,
        month = feb,
       volume = {979},
       number = {2},
          eid = {127},
        pages = {127},
          doi = {10.3847/1538-4357/ad9f3a},
archivePrefix = {arXiv},
       eprint = {2402.18773},
 primaryClass = {astro-ph.GA},
       adsurl = {https://ui.adsabs.harvard.edu/abs/2025ApJ...979..127J}
}

@ARTICLE{2007MNRAS.377L..64L,
       author = {{Lodato}, Giuseppe and {Natarajan}, Priyamvada},
        title = "{The mass function of high-redshift seed black holes}",
      journal = {\mnras},
         year = 2007,
        month = may,
       volume = {377},
       number = {1},
        pages = {L64-L68},
          doi = {10.1111/j.1745-3933.2007.00304.x},
archivePrefix = {arXiv},
       eprint = {astro-ph/0702340},
 primaryClass = {astro-ph},
       adsurl = {https://ui.adsabs.harvard.edu/abs/2007MNRAS.377L..64L}
}

@ARTICLE{2006MNRAS.371.1813L,
       author = {{Lodato}, Giuseppe and {Natarajan}, Priyamvada},
        title = "{Supermassive black hole formation during the assembly of pre-galactic discs}",
      journal = {\mnras},
         year = 2006,
        month = oct,
       volume = {371},
       number = {4},
        pages = {1813-1823},
          doi = {10.1111/j.1365-2966.2006.10801.x},
archivePrefix = {arXiv},
       eprint = {astro-ph/0606159},
 primaryClass = {astro-ph},
       adsurl = {https://ui.adsabs.harvard.edu/abs/2006MNRAS.371.1813L}
}

@ARTICLE{2019ApJ...887..245S,
       author = {{Schutte}, Zachary and {Reines}, Amy E. and {Greene}, Jenny E.},
        title = "{The Black Hole-Bulge Mass Relation Including Dwarf Galaxies Hosting Active Galactic Nuclei}",
      journal = {\apj},
         year = 2019,
        month = dec,
       volume = {887},
       number = {2},
          eid = {245},
        pages = {245},
          doi = {10.3847/1538-4357/ab35dd},
archivePrefix = {arXiv},
       eprint = {1908.00020},
 primaryClass = {astro-ph.GA},
       adsurl = {https://ui.adsabs.harvard.edu/abs/2019ApJ...887..245S}
}

@ARTICLE{2020ARA&A..58..257G,
       author = {{Greene}, Jenny E. and {Strader}, Jay and {Ho}, Luis C.},
        title = "{Intermediate-Mass Black Holes}",
      journal = {\araa},
         year = 2020,
        month = aug,
       volume = {58},
        pages = {257-312},
          doi = {10.1146/annurev-astro-032620-021835},
archivePrefix = {arXiv},
       eprint = {1911.09678},
 primaryClass = {astro-ph.GA},
       adsurl = {https://ui.adsabs.harvard.edu/abs/2020ARA&A..58..257G}
}

@ARTICLE{2014MNRAS.445..544S,
       author = {{Sugimura}, Kazuyuki and {Omukai}, Kazuyuki and {Inoue}, Akio K.},
        title = "{The critical radiation intensity for direct collapse black hole formation: dependence on the radiation spectral shape}",
      journal = {\mnras},
         year = 2014,
        month = nov,
       volume = {445},
       number = {1},
        pages = {544-553},
          doi = {10.1093/mnras/stu1778},
archivePrefix = {arXiv},
       eprint = {1407.4039},
 primaryClass = {astro-ph.GA},
       adsurl = {https://ui.adsabs.harvard.edu/abs/2014MNRAS.445..544S}
}

@ARTICLE{2010MNRAS.402.1249S,
       author = {{Shang}, Cien and {Bryan}, Greg L. and {Haiman}, Z.},
        title = "{Supermassive black hole formation by direct collapse: keeping protogalactic gas H$_{2}$ free in dark matter haloes with virial temperatures T$_{vir}$ > rsim {}10$^{4}$ K}",
      journal = {\mnras},
         year = 2010,
        month = feb,
       volume = {402},
       number = {2},
        pages = {1249-1262},
          doi = {10.1111/j.1365-2966.2009.15960.x},
archivePrefix = {arXiv},
       eprint = {0906.4773},
 primaryClass = {astro-ph.CO},
       adsurl = {https://ui.adsabs.harvard.edu/abs/2010MNRAS.402.1249S}
}

@ARTICLE{2025arXiv250316596N,
       author = {{Naidu}, Rohan P. and {Matthee}, Jorryt and {Katz}, Harley and {de Graaff}, Anna and {Oesch}, Pascal and {Smith}, Aaron and {Greene}, Jenny E. and {Brammer}, Gabriel and {Weibel}, Andrea and {Hviding}, Raphael and {Chisholm}, John and {Labb\textbackslash'e}, Ivo and {Simcoe}, Robert A. and {Witten}, Callum and {Atek}, Hakim and {Baggen}, Josephine F.~W. and {Belli}, Sirio and {Bezanson}, Rachel and {Boogaard}, Leindert A. and {Bose}, Sownak and {Covelo-Paz}, Alba and {Dayal}, Pratika and {Fudamoto}, Yoshinobu and {Furtak}, Lukas J. and {Giovinazzo}, Emma and {Goulding}, Andy and {Gronke}, Max and {Heintz}, Kasper E. and {Hirschmann}, Michaela and {Illingworth}, Garth and {Inoue}, Akio K. and {Johnson}, Benjamin D. and {Leja}, Joel and {Leonova}, Ecaterina and {McConachie}, Ian and {Maseda}, Michael V. and {Natarajan}, Priyamvada and {Nelson}, Erica and {Setton}, David J. and {Shivaei}, Irene and {Sobral}, David and {Stefanon}, Mauro and {Tacchella}, Sandro and {Toft}, Sune and {Torralba}, Alberto and {van Dokkum}, Pieter and {van der Wel}, Arjen and {Volonteri}, Marta and {Walter}, Fabian and {Wang}, Bingjie and {Watson}, Darach},
        title = "{A ``Black Hole Star'' Reveals the Remarkable Gas-Enshrouded Hearts of the Little Red Dots}",
      journal = {arXiv e-prints},
         year = 2025,
        month = mar,
          eid = {arXiv:2503.16596},
        pages = {arXiv:2503.16596},
          doi = {10.48550/arXiv.2503.16596},
archivePrefix = {arXiv},
       eprint = {2503.16596},
 primaryClass = {astro-ph.GA},
       adsurl = {https://ui.adsabs.harvard.edu/abs/2025arXiv250316596N}
}

@ARTICLE{2020MNRAS.499.4325R,
       author = {{Ramos Padilla}, Andr{\'e}s F. and {Ashby}, M.~L.~N. and {Smith}, Howard A. and {Mart{\'\i}nez-Galarza}, Juan R. and {Beverage}, Aliza G. and {Dietrich}, Jamie and {Higuera-G.}, Mario-A. and {Weiner}, Aaron S.},
        title = "{The AGN contribution to the UV-FIR luminosities of interacting galaxies and its role in identifying the main sequence}",
      journal = {\mnras},
         year = 2020,
        month = dec,
       volume = {499},
       number = {3},
        pages = {4325-4369},
          doi = {10.1093/mnras/staa2813},
archivePrefix = {arXiv},
       eprint = {2009.05614},
 primaryClass = {astro-ph.GA},
       adsurl = {https://ui.adsabs.harvard.edu/abs/2020MNRAS.499.4325R}
}

@ARTICLE{2026Natur.649..574R,
       author = {{Rusakov}, V. and {Watson}, D. and {Nikopoulos}, G.~P. and {Brammer}, G. and {Gottumukkala}, R. and {Harvey}, T. and {Heintz}, K.~E. and {Damgaard}, R. and {Sim}, S.~A. and {Sneppen}, A. and {Vijayan}, A.~P. and {Adams}, N. and {Austin}, D. and {Conselice}, C.~J. and {Goolsby}, C.~M. and {Toft}, S. and {Witstok}, J.},
        title = "{Little red dots as young supermassive black holes in dense ionized cocoons}",
      journal = {\nat},
         year = 2026,
        month = jan,
       volume = {649},
       number = {8097},
        pages = {574-579},
          doi = {10.1038/s41586-025-09900-4},
archivePrefix = {arXiv},
       eprint = {2503.16595},
 primaryClass = {astro-ph.GA},
       adsurl = {https://ui.adsabs.harvard.edu/abs/2026Natur.649..574R}
}

@ARTICLE{2025arXiv251007376J,
       author = {{Jones}, Brenda L. and {Kocevski}, Dale D. and {Pacucci}, Fabio and {Taylor}, Anthony J. and {Finkelstein}, Steven L. and {Buchner}, Johannes and {Trump}, Jonathan R. and {Somerville}, Rachel S. and {Hirschmann}, Michaela and {Yung}, L.~Y. Aaron and {Barro}, Guillermo and {Bell}, Eric F. and {Bisigello}, Laura and {Calabro}, Antonello and {Cleri}, Nikko J. and {Dekel}, Avishai and {Dickinson}, Mark and {Gandolfi}, Giovanni and {Giavalisco}, Mauro and {Grogin}, Norman A. and {Inayoshi}, Kohei and {Kartaltepe}, Jeyhan S. and {Koekemoer}, Anton M. and {Napolitano}, Lorenzo and {Onoue}, Masafusa and {Ravindranath}, Swara and {Rodighiero}, Giulia and {Wilkins}, Stephen M.},
        title = "{The $M_{\rm BH}-M_{*}$ Relationship at $3<z<7$: Big Black Holes in Little Red Dots}",
      journal = {arXiv e-prints},
         year = 2025,
        month = oct,
          eid = {arXiv:2510.07376},
        pages = {arXiv:2510.07376},
          doi = {10.48550/arXiv.2510.07376},
archivePrefix = {arXiv},
       eprint = {2510.07376},
 primaryClass = {astro-ph.GA},
       adsurl = {https://ui.adsabs.harvard.edu/abs/2025arXiv251007376J}
}

@ARTICLE{2007ApJ...670..249L,
       author = {{Lauer}, Tod R. and {Tremaine}, Scott and {Richstone}, Douglas and {Faber}, S.~M.},
        title = "{Selection Bias in Observing the Cosmological Evolution of the M$_{{\ensuremath{\bullet}}}$-{\ensuremath{\sigma}} and M$_{{\ensuremath{\bullet}}}$-L Relationships}",
      journal = {\apj},
         year = 2007,
        month = nov,
       volume = {670},
       number = {1},
        pages = {249-260},
          doi = {10.1086/522083},
archivePrefix = {arXiv},
       eprint = {0705.4103},
 primaryClass = {astro-ph},
       adsurl = {https://ui.adsabs.harvard.edu/abs/2007ApJ...670..249L}
}

@ARTICLE{2026arXiv260317967I,
       author = {{Iani}, E. and {Rinaldi}, P. and {Torralba}, A. and {Lyu}, J. and {Navarro-Carrera}, R. and {Rieke}, G.~H. and {Sun}, F. and {Willott}, C. and {Zhu}, Y. and {Alonso-Herrero}, A. and {Annunziatella}, M. and {Bergamini}, P. and {Caputi}, K. and {Catone}, M. and {Colina}, L. and {Cooper}, R. and {Costantin}, L. and {Crespo G{\'o}mez}, A. and {Desprez}, G. and {Di Cesare}, C. and {Hayes}, M.~J. and {Jermann}, I. and {Kotiwale}, G. and {Kramarenko}, I. and {Langeroodi}, D. and {Mascia}, S. and {Matthee}, J. and {Melinder}, J. and {Muzzin}, A. and {Navarrete}, B. and {Noirot}, G. and {{\"O}stlin}, G. and {Pacucci}, F. and {Rodighiero}, G. and {Sawicki}, M. and {Sun}, Y. and {Wu}, Z. and {Yang}, G.},
        title = "{JWST Reveals Two Overmassive Black Hole Candidates in Dwarf Galaxies at z $\approx$ 0.7: Pushing Black Hole Searches into the Dwarf-Galaxy Regime}",
      journal = {arXiv e-prints},
         year = 2026,
        month = mar,
          eid = {arXiv:2603.17967},
        pages = {arXiv:2603.17967},
archivePrefix = {arXiv},
       eprint = {2603.17967},
 primaryClass = {astro-ph.GA},
       adsurl = {https://ui.adsabs.harvard.edu/abs/2026arXiv260317967I}
}

@ARTICLE{2015MNRAS.451.1892A,
       author = {{Aird}, J. and {Coil}, A.~L. and {Georgakakis}, A. and {Nandra}, K. and {Barro}, G. and {P{\'e}rez-Gonz{\'a}lez}, P.~G.},
        title = "{The evolution of the X-ray luminosity functions of unabsorbed and absorbed AGNs out to z{\ensuremath{\sim}} 5}",
      journal = {\mnras},
         year = 2015,
        month = aug,
       volume = {451},
       number = {2},
        pages = {1892-1927},
          doi = {10.1093/mnras/stv1062},
archivePrefix = {arXiv},
       eprint = {1503.01120},
 primaryClass = {astro-ph.HE},
       adsurl = {https://ui.adsabs.harvard.edu/abs/2015MNRAS.451.1892A}
}

@ARTICLE{2023ApJ...957L...3P,
       author = {{Pacucci}, Fabio and {Nguyen}, Bao and {Carniani}, Stefano and {Maiolino}, Roberto and {Fan}, Xiaohui},
        title = "{JWST CEERS and JADES Active Galaxies at z = 4-7 Violate the Local M $_{{\textbullet}}$-M $_{{\ensuremath{\star}}}$ Relation at >3{\ensuremath{\sigma}}: Implications for Low-mass Black Holes and Seeding Models}",
      journal = {\apjl},
         year = 2023,
        month = nov,
       volume = {957},
       number = {1},
          eid = {L3},
        pages = {L3},
          doi = {10.3847/2041-8213/ad0158},
archivePrefix = {arXiv},
       eprint = {2308.12331},
 primaryClass = {astro-ph.GA},
       adsurl = {https://ui.adsabs.harvard.edu/abs/2023ApJ...957L...3P}
}

@ARTICLE{2015ApJ...799...98M,
       author = {{Miller}, Brendan P. and {Gallo}, Elena and {Greene}, Jenny E. and {Kelly}, Brandon C. and {Treu}, Tommaso and {Woo}, Jong-Hak and {Baldassare}, Vivienne},
        title = "{X-Ray Constraints on the Local Supermassive Black Hole Occupation Fraction}",
      journal = {\apj},
         year = 2015,
        month = jan,
       volume = {799},
       number = {1},
          eid = {98},
        pages = {98},
          doi = {10.1088/0004-637X/799/1/98},
archivePrefix = {arXiv},
       eprint = {1403.4246},
 primaryClass = {astro-ph.GA},
       adsurl = {https://ui.adsabs.harvard.edu/abs/2015ApJ...799...98M}
}

@ARTICLE{2017MNRAS.467.4739K,
       author = {{Kaviraj}, S. and {Laigle}, C. and {Kimm}, T. and {Devriendt}, J.~E.~G. and {Dubois}, Y. and {Pichon}, C. and {Slyz}, A. and {Chisari}, E. and {Peirani}, S.},
        title = "{The Horizon-AGN simulation: evolution of galaxy properties over cosmic time}",
      journal = {\mnras},
         year = 2017,
        month = jun,
       volume = {467},
       number = {4},
        pages = {4739-4752},
          doi = {10.1093/mnras/stx126},
archivePrefix = {arXiv},
       eprint = {1605.09379},
 primaryClass = {astro-ph.GA},
       adsurl = {https://ui.adsabs.harvard.edu/abs/2017MNRAS.467.4739K}
}

@ARTICLE{2020OJAp....3E..15R,
       author = {{Regan}, John A. and {Wise}, John H. and {Woods}, Tyrone E. and {Downes}, Turlough P. and {O'Shea}, Brian W. and {Norman}, Michael L.},
        title = "{The Formation of Very Massive Stars in Early Galaxies and Implications for Intermediate Mass Black Holes}",
      journal = {The Open Journal of Astrophysics},
         year = 2020,
        month = dec,
       volume = {3},
       number = {1},
          eid = {15},
        pages = {15},
          doi = {10.21105/astro.2008.08090},
archivePrefix = {arXiv},
       eprint = {2008.08090},
 primaryClass = {astro-ph.GA},
       adsurl = {https://ui.adsabs.harvard.edu/abs/2020OJAp....3E..15R}
}

@ARTICLE{2008MNRAS.388.1011M,
       author = {{Merloni}, Andrea and {Heinz}, Sebastian},
        title = "{A synthesis model for AGN evolution: supermassive black holes growth and feedback modes}",
      journal = {\mnras},
         year = 2008,
        month = aug,
       volume = {388},
       number = {3},
        pages = {1011-1030},
          doi = {10.1111/j.1365-2966.2008.13472.x},
archivePrefix = {arXiv},
       eprint = {0805.2499},
 primaryClass = {astro-ph},
       adsurl = {https://ui.adsabs.harvard.edu/abs/2008MNRAS.388.1011M}
}

@ARTICLE{2023ApJ...953L..29L,
       author = {{Larson}, Rebecca L. and {Finkelstein}, Steven L. and {Kocevski}, Dale D. and {Hutchison}, Taylor A. and {Trump}, Jonathan R. and {Arrabal Haro}, Pablo and {Bromm}, Volker and {Cleri}, Nikko J. and {Dickinson}, Mark and {Fujimoto}, Seiji and {Kartaltepe}, Jeyhan S. and {Koekemoer}, Anton M. and {Papovich}, Casey and {Pirzkal}, Nor and {Tacchella}, Sandro and {Zavala}, Jorge A. and {Bagley}, Micaela and {Behroozi}, Peter and {Champagne}, Jaclyn B. and {Cole}, Justin W. and {Jung}, Intae and {Morales}, Alexa M. and {Yang}, Guang and {Zhang}, Haowen and {Zitrin}, Adi and {Amor{\'\i}n}, Ricardo O. and {Burgarella}, Denis and {Casey}, Caitlin M. and {Ch{\'a}vez Ortiz}, {\'O}scar A. and {Cox}, Isabella G. and {Chworowsky}, Katherine and {Fontana}, Adriano and {Gawiser}, Eric and {Grazian}, Andrea and {Grogin}, Norman A. and {Harish}, Santosh and {Hathi}, Nimish P. and {Hirschmann}, Michaela and {Holwerda}, Benne W. and {Juneau}, St{\'e}phanie and {Leung}, Gene C.~K. and {Lucas}, Ray A. and {McGrath}, Elizabeth J. and {P{\'e}rez-Gonz{\'a}lez}, Pablo G. and {Rigby}, Jane R. and {Seill{\'e}}, Lise-Marie and {Simons}, Raymond C. and {de La Vega}, Alexander and {Weiner}, Benjamin J. and {Wilkins}, Stephen M. and {Yung}, L.~Y. Aaron and {Ceers Team}},
        title = "{A CEERS Discovery of an Accreting Supermassive Black Hole 570 Myr after the Big Bang: Identifying a Progenitor of Massive z > 6 Quasars}",
      journal = {\apjl},
         year = 2023,
        month = aug,
       volume = {953},
       number = {2},
          eid = {L29},
        pages = {L29},
          doi = {10.3847/2041-8213/ace619},
archivePrefix = {arXiv},
       eprint = {2303.08918},
 primaryClass = {astro-ph.GA},
       adsurl = {https://ui.adsabs.harvard.edu/abs/2023ApJ...953L..29L}
}

@ARTICLE{2024arXiv240403576K,
       author = {{Kocevski}, Dale D. and {Finkelstein}, Steven L. and {Barro}, Guillermo and {Taylor}, Anthony J. and {Calabr{\`o}}, Antonello and {Laloux}, Brivael and {Buchner}, Johannes and {Trump}, Jonathan R. and {Leung}, Gene C.~K. and {Yang}, Guang and {Dickinson}, Mark and {P{\'e}rez-Gonz{\'a}lez}, Pablo G. and {Pacucci}, Fabio and {Inayoshi}, Kohei and {Somerville}, Rachel S. and {McGrath}, Elizabeth J. and {Akins}, Hollis B. and {Bagley}, Micaela B. and {Bisigello}, Laura and {Bowler}, Rebecca A.~A. and {Carnall}, Adam and {Casey}, Caitlin M. and {Cheng}, Yingjie and {Cleri}, Nikko J. and {Costantin}, Luca and {Cullen}, Fergus and {Davis}, Kelcey and {Donnan}, Callum T. and {Dunlop}, James S. and {Ellis}, Richard S. and {Ferguson}, Henry C. and {Fujimoto}, Seiji and {Fontana}, Adriano and {Giavalisco}, Mauro and {Grazian}, Andrea and {Grogin}, Norman A. and {Hathi}, Nimish P. and {Hirschmann}, Michaela and {Huertas-Company}, Marc and {Holwerda}, Benne W. and {Illingworth}, Garth and {Juneau}, St{\'e}phanie and {Kartaltepe}, Jeyhan S. and {Koekemoer}, Anton M. and {Li}, Wenxiu and {Lucas}, Ray A. and {Magee}, Dan and {Mason}, Charlotte and {McLeod}, Derek J. and {McLure}, Ross J. and {Napolitano}, Lorenzo and {Papovich}, Casey and {Pirzkal}, Nor and {Rodighiero}, Giulia and {Santini}, Paola and {Wilkins}, Stephen M. and {Yung}, L.~Y. Aaron},
        title = "{The Rise of Faint, Red AGN at $z>4$: A Sample of Little Red Dots in the JWST Extragalactic Legacy Fields}",
      journal = {arXiv e-prints},
         year = 2024,
        month = apr,
          eid = {arXiv:2404.03576},
        pages = {arXiv:2404.03576},
          doi = {10.48550/arXiv.2404.03576},
archivePrefix = {arXiv},
       eprint = {2404.03576},
 primaryClass = {astro-ph.GA},
       adsurl = {https://ui.adsabs.harvard.edu/abs/2024arXiv240403576K}
}

@ARTICLE{2024arXiv240218773J,
       author = {{Jeon}, Junehyoung and {Bromm}, Volker and {Liu}, Boyuan and {Finkelstein}, Steven L.},
        title = "{Physical Pathways for JWST-Observed Supermassive Black Holes in the Early Universe}",
      journal = {arXiv e-prints},
         year = 2024,
        month = feb,
          eid = {arXiv:2402.18773},
        pages = {arXiv:2402.18773},
          doi = {10.48550/arXiv.2402.18773},
archivePrefix = {arXiv},
       eprint = {2402.18773},
 primaryClass = {astro-ph.GA},
       adsurl = {https://ui.adsabs.harvard.edu/abs/2024arXiv240218773J}
}

@ARTICLE{2013MNRAS.432.3438A,
       author = {{Agarwal}, Bhaskar and {Davis}, Andrew J. and {Khochfar}, Sadegh and {Natarajan}, Priyamvada and {Dunlop}, James S.},
        title = "{Unravelling obese black holes in the first galaxies}",
      journal = {\mnras},
         year = 2013,
        month = jul,
       volume = {432},
       number = {4},
        pages = {3438-3444},
          doi = {10.1093/mnras/stt696},
archivePrefix = {arXiv},
       eprint = {1302.6996},
 primaryClass = {astro-ph.CO},
       adsurl = {https://ui.adsabs.harvard.edu/abs/2013MNRAS.432.3438A}
}

@ARTICLE{2017ApJ...838..117N,
       author = {{Natarajan}, Priyamvada and {Pacucci}, Fabio and {Ferrara}, Andrea and {Agarwal}, Bhaskar and {Ricarte}, Angelo and {Zackrisson}, Erik and {Cappelluti}, Nico},
        title = "{Unveiling the First Black Holes With JWST:Multi-wavelength Spectral Predictions}",
      journal = {\apj},
         year = 2017,
        month = apr,
       volume = {838},
       number = {2},
          eid = {117},
        pages = {117},
          doi = {10.3847/1538-4357/aa6330},
archivePrefix = {arXiv},
       eprint = {1610.05312},
 primaryClass = {astro-ph.GA},
       adsurl = {https://ui.adsabs.harvard.edu/abs/2017ApJ...838..117N}
}

@ARTICLE{2024ApJ...960L...1N,
       author = {{Natarajan}, Priyamvada and {Pacucci}, Fabio and {Ricarte}, Angelo and {Bogd{\'a}n}, {\'A}kos and {Goulding}, Andy D. and {Cappelluti}, Nico},
        title = "{First Detection of an Overmassive Black Hole Galaxy UHZ1: Evidence for Heavy Black Hole Seed Formation from Direct Collapse}",
      journal = {\apjl},
         year = 2024,
        month = jan,
       volume = {960},
       number = {1},
          eid = {L1},
        pages = {L1},
          doi = {10.3847/2041-8213/ad0e76},
archivePrefix = {arXiv},
       eprint = {2308.02654},
 primaryClass = {astro-ph.HE},
       adsurl = {https://ui.adsabs.harvard.edu/abs/2024ApJ...960L...1N}
}

@misc{kho2026learninguniversehighredshifts,
      title={Learning the Universe at High Redshifts: Impact of Accretion Modeling on Early Black Hole Growth}, 
      author={Jonathan Kho and Aklant K. Bhowmick and Rainer Weinberger and Paul Torrey and Laura Blecha and Lars Hernquist and Greg L. Bryan and Alex M. Garcia and Niusha Ahvazi and Alejandro Saravia and Boon Kiat Oh},
      year={2026},
      eprint={2606.10036},
      archivePrefix={arXiv},
      primaryClass={astro-ph.GA},
      url={https://arxiv.org/abs/2606.10036}, 
}

@ARTICLE{2024ApJ...968...38K,
       author = {{Kokorev}, Vasily and {Caputi}, Karina I. and {Greene}, Jenny E. and {Dayal}, Pratika and {Trebitsch}, Maxime and {Cutler}, Sam E. and {Fujimoto}, Seiji and {Labb{\'e}}, Ivo and {Miller}, Tim B. and {Iani}, Edoardo and {Navarro-Carrera}, Rafael and {Rinaldi}, Pierluigi},
        title = "{A Census of Photometrically Selected Little Red Dots at 4 < z < 9 in JWST Blank Fields}",
      journal = {\apj},
         year = 2024,
        month = jun,
       volume = {968},
       number = {1},
          eid = {38},
        pages = {38},
          doi = {10.3847/1538-4357/ad4265},
archivePrefix = {arXiv},
       eprint = {2401.09981},
 primaryClass = {astro-ph.GA},
       adsurl = {https://ui.adsabs.harvard.edu/abs/2024ApJ...968...38K}
}

@ARTICLE{2023ApJ...957L...7K,
       author = {{Kokorev}, Vasily and {Fujimoto}, Seiji and {Labbe}, Ivo and {Greene}, Jenny E. and {Bezanson}, Rachel and {Dayal}, Pratika and {Nelson}, Erica J. and {Atek}, Hakim and {Brammer}, Gabriel and {Caputi}, Karina I. and {Chemerynska}, Iryna and {Cutler}, Sam E. and {Feldmann}, Robert and {Fudamoto}, Yoshinobu and {Furtak}, Lukas J. and {Goulding}, Andy D. and {de Graaff}, Anna and {Leja}, Joel and {Marchesini}, Danilo and {Miller}, Tim B. and {Nanayakkara}, Themiya and {Oesch}, Pascal A. and {Pan}, Richard and {Price}, Sedona H. and {Setton}, David J. and {Smit}, Renske and {Stefanon}, Mauro and {Wang}, Bingjie and {Weaver}, John R. and {Whitaker}, Katherine E. and {Williams}, Christina C. and {Zitrin}, Adi},
        title = "{UNCOVER: A NIRSpec Identification of a Broad-line AGN at z = 8.50}",
      journal = {\apjl},
         year = 2023,
        month = nov,
       volume = {957},
       number = {1},
          eid = {L7},
        pages = {L7},
          doi = {10.3847/2041-8213/ad037a},
archivePrefix = {arXiv},
       eprint = {2308.11610},
 primaryClass = {astro-ph.GA},
       adsurl = {https://ui.adsabs.harvard.edu/abs/2023ApJ...957L...7K}
}

@ARTICLE{2024A&A...685A..25A,
       author = {{Andika}, Irham T. and {Jahnke}, Knud and {Onoue}, Masafusa and {Silverman}, John D. and {Fitriana}, Itsna K. and {Bongiorno}, Angela and {Brinch}, Malte and {Casey}, Caitlin M. and {Faisst}, Andreas and {Gillman}, Steven and {Gozaliasl}, Ghassem and {Hayward}, Christopher C. and {Hirschmann}, Michaela and {Kocevski}, Dale and {Koekemoer}, Anton M. and {Kokorev}, Vasily and {Lambrides}, Erini and {Lee}, Minju M. and {Michael Rich}, Robert and {Trakhtenbrot}, Benny and {Megan Urry}, Claudia and {Wilkins}, Stephen M. and {Vijayan}, Aswin P.},
        title = "{Tracing the rise of supermassive black holes. A panchromatic search for faint, unobscured quasars at z {\ensuremath{\gtrsim}} 6 with COSMOS-Web and other surveys}",
      journal = {\aap},
         year = 2024,
        month = may,
       volume = {685},
          eid = {A25},
        pages = {A25},
          doi = {10.1051/0004-6361/202349025},
archivePrefix = {arXiv},
       eprint = {2401.11826},
 primaryClass = {astro-ph.GA},
       adsurl = {https://ui.adsabs.harvard.edu/abs/2024A&A...685A..25A}
}

@ARTICLE{2023arXiv230905714G,
       author = {{Greene}, Jenny E. and {Labbe}, Ivo and {Goulding}, Andy D. and {Furtak}, Lukas J. and {Chemerynska}, Iryna and {Kokorev}, Vasily and {Dayal}, Pratika and {Williams}, Christina C. and {Wang}, Bingjie and {Setton}, David J. and {Burgasser}, Adam J. and {Bezanson}, Rachel and {Atek}, Hakim and {Brammer}, Gabriel and {Cutler}, Sam E. and {Feldmann}, Robert and {Fujimoto}, Seiji and {Glazebrook}, Karl and {de Graaff}, Anna and {Leja}, Joel and {Marchesini}, Danilo and {Maseda}, Michael V. and {Matthee}, Jorryt and {Miller}, Tim B. and {Naidu}, Rohan P. and {Nanayakkara}, Themiya and {Oesch}, Pascal A. and {Pan}, Richard and {Papovich}, Casey and {Price}, Sedona H. and {van Dokkum}, Pieter and {Weaver}, John R. and {Whitaker}, Katherine E. and {Zitrin}, Adi},
        title = "{UNCOVER spectroscopy confirms a surprising ubiquity of AGN in red galaxies at $z>5$}",
      journal = {arXiv e-prints},
         year = 2023,
        month = sep,
          eid = {arXiv:2309.05714},
        pages = {arXiv:2309.05714},
          doi = {10.48550/arXiv.2309.05714},
archivePrefix = {arXiv},
       eprint = {2309.05714},
 primaryClass = {astro-ph.GA},
       adsurl = {https://ui.adsabs.harvard.edu/abs/2023arXiv230905714G}
}

@ARTICLE{2023arXiv230801230M,
       author = {{Maiolino}, Roberto and {Scholtz}, Jan and {Curtis-Lake}, Emma and {Carniani}, Stefano and {Baker}, William and {de Graaff}, Anna and {Tacchella}, Sandro and {{\"U}bler}, Hannah and {D'Eugenio}, Francesco and {Witstok}, Joris and {Curti}, Mirko and {Arribas}, Santiago and {Bunker}, Andrew J. and {Charlot}, St{\'e}phane and {Chevallard}, Jacopo and {Eisenstein}, Daniel J. and {Egami}, Eiichi and {Ji}, Zhiyuan and {Jones}, Gareth C. and {Lyu}, Jianwei and {Rawle}, Tim and {Robertson}, Brant and {Rujopakarn}, Wiphu and {Perna}, Michele and {Sun}, Fengwu and {Venturi}, Giacomo and {Williams}, Christina C. and {Willott}, Chris},
        title = "{JADES. The diverse population of infant Black Holes at 4<z<11: merging, tiny, poor, but mighty}",
      journal = {arXiv e-prints},
         year = 2023,
        month = aug,
          eid = {arXiv:2308.01230},
        pages = {arXiv:2308.01230},
          doi = {10.48550/arXiv.2308.01230},
archivePrefix = {arXiv},
       eprint = {2308.01230},
 primaryClass = {astro-ph.GA},
       adsurl = {https://ui.adsabs.harvard.edu/abs/2023arXiv230801230M}
}

@ARTICLE{2023ApJ...954L...4K,
       author = {{Kocevski}, Dale D. and {Onoue}, Masafusa and {Inayoshi}, Kohei and {Trump}, Jonathan R. and {Arrabal Haro}, Pablo and {Grazian}, Andrea and {Dickinson}, Mark and {Finkelstein}, Steven L. and {Kartaltepe}, Jeyhan S. and {Hirschmann}, Michaela and {Aird}, James and {Holwerda}, Benne W. and {Fujimoto}, Seiji and {Juneau}, St{\'e}phanie and {Amor{\'\i}n}, Ricardo O. and {Backhaus}, Bren E. and {Bagley}, Micaela B. and {Barro}, Guillermo and {Bell}, Eric F. and {Bisigello}, Laura and {Calabr{\`o}}, Antonello and {Cleri}, Nikko J. and {Cooper}, M.~C. and {Ding}, Xuheng and {Grogin}, Norman A. and {Ho}, Luis C. and {Hutchison}, Taylor A. and {Inoue}, Akio K. and {Jiang}, Linhua and {Jones}, Brenda and {Koekemoer}, Anton M. and {Li}, Wenxiu and {Li}, Zhengrong and {McGrath}, Elizabeth J. and {Molina}, Juan and {Papovich}, Casey and {P{\'e}rez-Gonz{\'a}lez}, Pablo G. and {Pirzkal}, Nor and {Wilkins}, Stephen M. and {Yang}, Guang and {Yung}, L.~Y. Aaron},
        title = "{Hidden Little Monsters: Spectroscopic Identification of Low-mass, Broad-line AGNs at z > 5 with CEERS}",
      journal = {\apjl},
         year = 2023,
        month = sep,
       volume = {954},
       number = {1},
          eid = {L4},
        pages = {L4},
          doi = {10.3847/2041-8213/ace5a0},
archivePrefix = {arXiv},
       eprint = {2302.00012},
 primaryClass = {astro-ph.GA},
       adsurl = {https://ui.adsabs.harvard.edu/abs/2023ApJ...954L...4K}
}

@ARTICLE{2023ApJ...959...39H,
       author = {{Harikane}, Yuichi and {Zhang}, Yechi and {Nakajima}, Kimihiko and {Ouchi}, Masami and {Isobe}, Yuki and {Ono}, Yoshiaki and {Hatano}, Shun and {Xu}, Yi and {Umeda}, Hiroya},
        title = "{A JWST/NIRSpec First Census of Broad-line AGNs at z = 4-7: Detection of 10 Faint AGNs with M $_{BH}$ {}10$^{6}$-{}10$^{8}$ M $_{{\ensuremath{\odot}}}$ and Their Host Galaxy Properties}",
      journal = {\apj},
         year = 2023,
        month = dec,
       volume = {959},
       number = {1},
          eid = {39},
        pages = {39},
          doi = {10.3847/1538-4357/ad029e},
archivePrefix = {arXiv},
       eprint = {2303.11946},
 primaryClass = {astro-ph.GA},
       adsurl = {https://ui.adsabs.harvard.edu/abs/2023ApJ...959...39H}
}

@ARTICLE{2023ApJ...942L..17O,
       author = {{Onoue}, Masafusa and {Inayoshi}, Kohei and {Ding}, Xuheng and {Li}, Wenxiu and {Li}, Zhengrong and {Molina}, Juan and {Inoue}, Akio K. and {Jiang}, Linhua and {Ho}, Luis C.},
        title = "{A Candidate for the Least-massive Black Hole in the First 1.1 Billion Years of the Universe}",
      journal = {\apjl},
         year = 2023,
        month = jan,
       volume = {942},
       number = {1},
          eid = {L17},
        pages = {L17},
          doi = {10.3847/2041-8213/aca9d3},
archivePrefix = {arXiv},
       eprint = {2209.07325},
 primaryClass = {astro-ph.GA},
       adsurl = {https://ui.adsabs.harvard.edu/abs/2023ApJ...942L..17O}
}

@ARTICLE{2024ApJ...961...76M,
       author = {{Mayer}, Lucio and {Capelo}, Pedro R. and {Zwick}, Lorenz and {Di Matteo}, Tiziana},
        title = "{Direct Formation of Massive Black Holes via Dynamical Collapse in Metal-enriched Merging Galaxies at z   10: Fully Cosmological Simulations}",
      journal = {\apj},
         year = 2024,
        month = jan,
       volume = {961},
       number = {1},
          eid = {76},
        pages = {76},
          doi = {10.3847/1538-4357/ad11cf},
archivePrefix = {arXiv},
       eprint = {2304.02066},
 primaryClass = {astro-ph.GA},
       adsurl = {https://ui.adsabs.harvard.edu/abs/2024ApJ...961...76M}
}

@ARTICLE{2023MNRAS.526L..94B,
       author = {{Begelman}, Mitchell C. and {Silk}, Joseph},
        title = "{Magnetic fields catalyse massive black hole formation and growth}",
      journal = {\mnras},
         year = 2023,
        month = nov,
       volume = {526},
       number = {1},
        pages = {L94-L99},
          doi = {10.1093/mnrasl/slad124},
archivePrefix = {arXiv},
       eprint = {2305.19081},
 primaryClass = {astro-ph.HE},
       adsurl = {https://ui.adsabs.harvard.edu/abs/2023MNRAS.526L..94B}
}

@ARTICLE{2019ApJ...872..104N,
       author = {{Nguyen}, Dieu D. and {Seth}, Anil C. and {Neumayer}, Nadine and {Iguchi}, Satoru and {Cappellari}, Michelle and {Strader}, Jay and {Chomiuk}, Laura and {Tremou}, Evangelia and {Pacucci}, Fabio and {Nakanishi}, Kouichiro and {Bahramian}, Arash and {Nguyen}, Phuong M. and {den Brok}, Mark and {Ahn}, Christopher C. and {Voggel}, Karina T. and {Kacharov}, Nikolay and {Tsukui}, Takafumi and {Ly}, Cuc K. and {Dumont}, Antoine and {Pechetti}, Renuka},
        title = "{Improved Dynamical Constraints on the Masses of the Central Black Holes in Nearby Low-mass Early-type Galactic Nuclei and the First Black Hole Determination for NGC 205}",
      journal = {\apj},
         year = 2019,
        month = feb,
       volume = {872},
       number = {1},
          eid = {104},
        pages = {104},
          doi = {10.3847/1538-4357/aafe7a},
archivePrefix = {arXiv},
       eprint = {1901.05496},
 primaryClass = {astro-ph.GA},
       adsurl = {https://ui.adsabs.harvard.edu/abs/2019ApJ...872..104N}
}

@misc{burke2024multiwavelengthconstraintslocalblack,
      title={Multi-wavelength constraints on the local black hole occupation fraction}, 
      author={Colin J. Burke and Priyamvada Natarajan and Vivienne F. Baldassare and Marla Geha},
      year={2024},
      eprint={2410.11177},
      archivePrefix={arXiv},
      primaryClass={astro-ph.GA},
      url={https://arxiv.org/abs/2410.11177}, 
}

@ARTICLE{2017MNRAS.470.1121T,
       author = {{Tremmel}, M. and {Karcher}, M. and {Governato}, F. and {Volonteri}, M. and {Quinn}, T.~R. and {Pontzen}, A. and {Anderson}, L. and {Bellovary}, J.},
        title = "{The Romulus cosmological simulations: a physical approach to the formation, dynamics and accretion models of SMBHs}",
      journal = {\mnras},
         year = 2017,
        month = sep,
       volume = {470},
       number = {1},
        pages = {1121-1139},
          doi = {10.1093/mnras/stx1160},
archivePrefix = {arXiv},
       eprint = {1607.02151},
 primaryClass = {astro-ph.GA},
       adsurl = {https://ui.adsabs.harvard.edu/abs/2017MNRAS.470.1121T}
}

@ARTICLE{2021MNRAS.503.1940H,
       author = {{Habouzit}, M{\'e}lanie and {Li}, Yuan and {Somerville}, Rachel S. and {Genel}, Shy and {Pillepich}, Annalisa and {Volonteri}, Marta and {Dav{\'e}}, Romeel and {Rosas-Guevara}, Yetli and {McAlpine}, Stuart and {Peirani}, S{\'e}bastien and {Hernquist}, Lars and {Angl{\'e}s-Alc{\'a}zar}, Daniel and {Reines}, Amy and {Bower}, Richard and {Dubois}, Yohan and {Nelson}, Dylan and {Pichon}, Christophe and {Vogelsberger}, Mark},
        title = "{Supermassive black holes in cosmological simulations I: M$_{BH}$ - M$_{{\ensuremath{\star}}}$ relation and black hole mass function}",
      journal = {\mnras},
         year = 2021,
        month = may,
       volume = {503},
       number = {2},
        pages = {1940-1975},
          doi = {10.1093/mnras/stab496},
archivePrefix = {arXiv},
       eprint = {2006.10094},
 primaryClass = {astro-ph.GA},
       adsurl = {https://ui.adsabs.harvard.edu/abs/2021MNRAS.503.1940H}
}

@ARTICLE{2024MNRAS.531.4584S,
       author = {{Scoggins}, Matthew T. and {Haiman}, Zolt{\'a}n},
        title = "{Diagnosing the massive-seed pathway to high-redshift black holes: statistics of the evolving black hole to host galaxy mass ratio}",
      journal = {\mnras},
         year = 2024,
        month = jul,
       volume = {531},
       number = {4},
        pages = {4584-4597},
          doi = {10.1093/mnras/stae1449},
archivePrefix = {arXiv},
       eprint = {2310.00202},
 primaryClass = {astro-ph.GA},
       adsurl = {https://ui.adsabs.harvard.edu/abs/2024MNRAS.531.4584S}
}

@ARTICLE{2023MNRAS.519.2155S,
       author = {{Scoggins}, Matthew T. and {Haiman}, Zolt{\'a}n and {Wise}, John H.},
        title = "{How long do high redshift massive black hole seeds remain outliers in black hole versus host galaxy relations?}",
      journal = {\mnras},
         year = 2023,
        month = feb,
       volume = {519},
       number = {2},
        pages = {2155-2168},
          doi = {10.1093/mnras/stac3715},
archivePrefix = {arXiv},
       eprint = {2205.09611},
 primaryClass = {astro-ph.GA},
       adsurl = {https://ui.adsabs.harvard.edu/abs/2023MNRAS.519.2155S}
}

@ARTICLE{2024arXiv240610329D,
       author = {{Durodola}, Emmanuel and {Pacucci}, Fabio and {Hickox}, Ryan C.},
        title = "{Exploring the AGN Fraction of a Sample of JWST's Little Red Dots at $5 < z < 8$: Overmassive Black Holes Are Strongly Favored}",
      journal = {arXiv e-prints},
         year = 2024,
        month = jun,
          eid = {arXiv:2406.10329},
        pages = {arXiv:2406.10329},
          doi = {10.48550/arXiv.2406.10329},
archivePrefix = {arXiv},
       eprint = {2406.10329},
 primaryClass = {astro-ph.GA},
       adsurl = {https://ui.adsabs.harvard.edu/abs/2024arXiv240610329D}
}

@ARTICLE{2024MNRAS.533.1907B,
       author = {{Bhowmick}, Aklant K. and {Blecha}, Laura and {Torrey}, Paul and {Somerville}, Rachel S. and {Kelley}, Luke Zoltan and {Vogelsberger}, Mark and {Weinberger}, Rainer and {Hernquist}, Lars and {Sivasankaran}, Aneesh},
        title = "{Growth of high-redshift supermassive black holes from heavy seeds in the BRAHMA cosmological simulations: implications of overmassive black holes}",
      journal = {\mnras},
         year = 2024,
        month = sep,
       volume = {533},
       number = {2},
        pages = {1907-1926},
          doi = {10.1093/mnras/stae1819},
archivePrefix = {arXiv},
       eprint = {2406.14658},
 primaryClass = {astro-ph.GA},
       adsurl = {https://ui.adsabs.harvard.edu/abs/2024MNRAS.533.1907B}
}

@ARTICLE{2026NatAs.tmp...21M,
       author = {{Mehta}, Daxal H. and {Regan}, John A. and {Prole}, Lewis},
        title = "{The growth of light seed black holes in the early Universe}",
      journal = {Nature Astronomy},
         year = 2026,
        month = jan,
          doi = {10.1038/s41550-025-02767-5},
archivePrefix = {arXiv},
       eprint = {2601.14395},
 primaryClass = {astro-ph.GA},
       adsurl = {https://ui.adsabs.harvard.edu/abs/2026NatAs.tmp...21M}
}

@ARTICLE{2021MNRAS.501.1413N,
       author = {{Natarajan}, Priyamvada},
        title = "{A new channel to form IMBHs throughout cosmic time}",
      journal = {\mnras},
         year = 2021,
        month = feb,
       volume = {501},
       number = {1},
        pages = {1413-1425},
          doi = {10.1093/mnras/staa3724},
archivePrefix = {arXiv},
       eprint = {2009.09156},
 primaryClass = {astro-ph.GA},
       adsurl = {https://ui.adsabs.harvard.edu/abs/2021MNRAS.501.1413N}
}

@ARTICLE{2023PhRvD.108h3012K,
       author = {{Kritos}, Konstantinos and {Berti}, Emanuele and {Silk}, Joseph},
        title = "{Massive black hole assembly in nuclear star clusters}",
      journal = {\prd},
         year = 2023,
        month = oct,
       volume = {108},
       number = {8},
          eid = {083012},
        pages = {083012},
          doi = {10.1103/PhysRevD.108.083012},
archivePrefix = {arXiv},
       eprint = {2212.06845},
 primaryClass = {astro-ph.HE},
       adsurl = {https://ui.adsabs.harvard.edu/abs/2023PhRvD.108h3012K}
}

@ARTICLE{2025A&A...695A..97D,
       author = {{Dekel}, Avishai and {Stone}, Nicholas C. and {Chowdhury}, Dhruba Dutta and {Gilbaum}, Shmuel and {Li}, Zhaozhou and {Mandelker}, Nir and {van den Bosch}, Frank C.},
        title = "{Growth of massive black holes in FFB galaxies at cosmic dawn}",
      journal = {\aap},
         year = 2025,
        month = mar,
       volume = {695},
          eid = {A97},
        pages = {A97},
          doi = {10.1051/0004-6361/202452393},
archivePrefix = {arXiv},
       eprint = {2409.18605},
 primaryClass = {astro-ph.GA},
       adsurl = {https://ui.adsabs.harvard.edu/abs/2025A&A...695A..97D}
}

@ARTICLE{2025ApJ...981...19L,
       author = {{Li}, Junyao and {Silverman}, John D. and {Shen}, Yue and {Volonteri}, Marta and {Jahnke}, Knud and {Zhuang}, Ming-Yang and {Scoggins}, Matthew T. and {Ding}, Xuheng and {Harikane}, Yuichi and {Onoue}, Masafusa and {Tanaka}, Takumi S.},
        title = "{Tip of the Iceberg: Overmassive Black Holes at 4 < z < 7 Found by JWST Are Not Inconsistent with the Local  Relation}",
      journal = {\apj},
         year = 2025,
        month = mar,
       volume = {981},
       number = {1},
          eid = {19},
        pages = {19},
          doi = {10.3847/1538-4357/ada603},
archivePrefix = {arXiv},
       eprint = {2403.00074},
 primaryClass = {astro-ph.GA},
       adsurl = {https://ui.adsabs.harvard.edu/abs/2025ApJ...981...19L}
}

@ARTICLE{2024ApJ...975..178K,
       author = {{Kokorev}, Vasily and {Chisholm}, John and {Endsley}, Ryan and {Finkelstein}, Steven L. and {Greene}, Jenny E. and {Akins}, Hollis B. and {Bromm}, Volker and {Casey}, Caitlin M. and {Fujimoto}, Seiji and {Labb{\'e}}, Ivo and {Larson}, Rebecca L.},
        title = "{Silencing the Giant: Evidence of Active Galactic Nucleus Feedback and Quenching in a Little Red Dot at z = 4.13}",
      journal = {\apj},
         year = 2024,
        month = nov,
       volume = {975},
       number = {2},
          eid = {178},
        pages = {178},
          doi = {10.3847/1538-4357/ad7d03},
archivePrefix = {arXiv},
       eprint = {2407.20320},
 primaryClass = {astro-ph.GA},
       adsurl = {https://ui.adsabs.harvard.edu/abs/2024ApJ...975..178K}
}

@ARTICLE{2024Natur.627...59M,
       author = {{Maiolino}, Roberto and {Scholtz}, Jan and {Witstok}, Joris and {Carniani}, Stefano and {D'Eugenio}, Francesco and {de Graaff}, Anna and {{\"U}bler}, Hannah and {Tacchella}, Sandro and {Curtis-Lake}, Emma and {Arribas}, Santiago and {Bunker}, Andrew and {Charlot}, St{\'e}phane and {Chevallard}, Jacopo and {Curti}, Mirko and {Looser}, Tobias J. and {Maseda}, Michael V. and {Rawle}, Timothy D. and {Rodr{\'\i}guez del Pino}, Bruno and {Willott}, Chris J. and {Egami}, Eiichi and {Eisenstein}, Daniel J. and {Hainline}, Kevin N. and {Robertson}, Brant and {Williams}, Christina C. and {Willmer}, Christopher N.~A. and {Baker}, William M. and {Boyett}, Kristan and {DeCoursey}, Christa and {Fabian}, Andrew C. and {Helton}, Jakob M. and {Ji}, Zhiyuan and {Jones}, Gareth C. and {Kumari}, Nimisha and {Laporte}, Nicolas and {Nelson}, Erica J. and {Perna}, Michele and {Sandles}, Lester and {Shivaei}, Irene and {Sun}, Fengwu},
        title = "{A small and vigorous black hole in the early Universe}",
      journal = {\nat},
         year = 2024,
        month = mar,
       volume = {627},
       number = {8002},
        pages = {59-63},
          doi = {10.1038/s41586-024-07052-5},
archivePrefix = {arXiv},
       eprint = {2305.12492},
 primaryClass = {astro-ph.GA},
       adsurl = {https://ui.adsabs.harvard.edu/abs/2024Natur.627...59M}
}

@ARTICLE{2023ApJ...955L..24G,
       author = {{Goulding}, Andy D. and {Greene}, Jenny E. and {Setton}, David J. and {Labbe}, Ivo and {Bezanson}, Rachel and {Miller}, Tim B. and {Atek}, Hakim and {Bogd{\'a}n}, {\'A}kos and {Brammer}, Gabriel and {Chemerynska}, Iryna and {Cutler}, Sam E. and {Dayal}, Pratika and {Fudamoto}, Yoshinobu and {Fujimoto}, Seiji and {Furtak}, Lukas J. and {Kokorev}, Vasily and {Khullar}, Gourav and {Leja}, Joel and {Marchesini}, Danilo and {Natarajan}, Priyamvada and {Nelson}, Erica and {Oesch}, Pascal A. and {Pan}, Richard and {Papovich}, Casey and {Price}, Sedona H. and {van Dokkum}, Pieter and {Wang}, Bingjie and {Weaver}, John R. and {Whitaker}, Katherine E. and {Zitrin}, Adi},
        title = "{UNCOVER: The Growth of the First Massive Black Holes from JWST/NIRSpec-Spectroscopic Redshift Confirmation of an X-Ray Luminous AGN at z = 10.1}",
      journal = {\apjl},
         year = 2023,
        month = sep,
       volume = {955},
       number = {1},
          eid = {L24},
        pages = {L24},
          doi = {10.3847/2041-8213/acf7c5},
archivePrefix = {arXiv},
       eprint = {2308.02750},
 primaryClass = {astro-ph.GA},
       adsurl = {https://ui.adsabs.harvard.edu/abs/2023ApJ...955L..24G}
}

@ARTICLE{2024ApJ...965L..21K,
       author = {{Kov{\'a}cs}, Orsolya E. and {Bogd{\'a}n}, {\'A}kos and {Natarajan}, Priyamvada and {Werner}, Norbert and {Azadi}, Mojegan and {Volonteri}, Marta and {Tremblay}, Grant R. and {Chadayammuri}, Urmila and {Forman}, William R. and {Jones}, Christine and {Kraft}, Ralph P.},
        title = "{A Candidate Supermassive Black Hole in a Gravitationally Lensed Galaxy at Z {\ensuremath{\approx}} 10}",
      journal = {\apjl},
         year = 2024,
        month = apr,
       volume = {965},
       number = {2},
          eid = {L21},
        pages = {L21},
          doi = {10.3847/2041-8213/ad391f},
archivePrefix = {arXiv},
       eprint = {2403.14745},
 primaryClass = {astro-ph.GA},
       adsurl = {https://ui.adsabs.harvard.edu/abs/2024ApJ...965L..21K}
}

@ARTICLE{2024PhRvL.133b1401A,
       author = {{Alonso-{\'A}lvarez}, Gonzalo and {Cline}, James M. and {Dewar}, Caitlyn},
        title = "{Self-Interacting Dark Matter Solves the Final Parsec Problem of Supermassive Black Hole Mergers}",
      journal = {\prl},
         year = 2024,
        month = jul,
       volume = {133},
       number = {2},
          eid = {021401},
        pages = {021401},
          doi = {10.1103/PhysRevLett.133.021401},
archivePrefix = {arXiv},
       eprint = {2401.14450},
 primaryClass = {astro-ph.CO},
       adsurl = {https://ui.adsabs.harvard.edu/abs/2024PhRvL.133b1401A}
}

@ARTICLE{2024A&A...690A.299F,
       author = {{Fischer}, Moritz S. and {Sagunski}, Laura},
        title = "{Dynamical friction from self-interacting dark matter}",
      journal = {\aap},
         year = 2024,
        month = oct,
       volume = {690},
          eid = {A299},
        pages = {A299},
          doi = {10.1051/0004-6361/202451304},
archivePrefix = {arXiv},
       eprint = {2405.19392},
 primaryClass = {astro-ph.CO},
       adsurl = {https://ui.adsabs.harvard.edu/abs/2024A&A...690A.299F}
}

@ARTICLE{2025JCAP...04..040Z,
       author = {{Ziparo}, F. and {Gallerani}, S. and {Ferrara}, A.},
        title = "{Primordial black holes as supermassive black hole seeds}",
      journal = {\jcap},
         year = 2025,
        month = apr,
       volume = {2025},
       number = {4},
          eid = {040},
        pages = {040},
          doi = {10.1088/1475-7516/2025/04/040},
archivePrefix = {arXiv},
       eprint = {2411.03448},
 primaryClass = {astro-ph.CO},
       adsurl = {https://ui.adsabs.harvard.edu/abs/2025JCAP...04..040Z}
}

@ARTICLE{2024arXiv241214248T,
       author = {{Trinca}, Alessandro and {Valiante}, Rosa and {Schneider}, Raffaella and {Juod{\v{z}}balis}, Ignas and {Maiolino}, Roberto and {Graziani}, Luca and {Lupi}, Alessandro and {Natarajan}, Priyamvada and {Volonteri}, Marta and {Zana}, Tommaso},
        title = "{Episodic super-Eddington accretion as a clue to Overmassive Black Holes in the early Universe}",
      journal = {arXiv e-prints},
         year = 2024,
        month = dec,
          eid = {arXiv:2412.14248},
        pages = {arXiv:2412.14248},
          doi = {10.48550/arXiv.2412.14248},
archivePrefix = {arXiv},
       eprint = {2412.14248},
 primaryClass = {astro-ph.GA},
       adsurl = {https://ui.adsabs.harvard.edu/abs/2024arXiv241214248T}
}

@ARTICLE{2023MNRAS.526.3250S,
       author = {{Schneider}, Raffaella and {Valiante}, Rosa and {Trinca}, Alessandro and {Graziani}, Luca and {Volonteri}, Marta and {Maiolino}, Roberto},
        title = "{Are we surprised to find SMBHs with JWST at z {\ensuremath{\geq}} 9?}",
      journal = {\mnras},
         year = 2023,
        month = dec,
       volume = {526},
       number = {3},
        pages = {3250-3261},
          doi = {10.1093/mnras/stad2503},
archivePrefix = {arXiv},
       eprint = {2305.12504},
 primaryClass = {astro-ph.GA},
       adsurl = {https://ui.adsabs.harvard.edu/abs/2023MNRAS.526.3250S}
}

@ARTICLE{2024A&A...691A.145M,
       author = {{Maiolino}, Roberto and {Scholtz}, Jan and {Curtis-Lake}, Emma and {Carniani}, Stefano and {Baker}, William and {de Graaff}, Anna and {Tacchella}, Sandro and {{\"U}bler}, Hannah and {D'Eugenio}, Francesco and {Witstok}, Joris and {Curti}, Mirko and {Arribas}, Santiago and {Bunker}, Andrew J. and {Charlot}, St{\'e}phane and {Chevallard}, Jacopo and {Eisenstein}, Daniel J. and {Egami}, Eiichi and {Ji}, Zhiyuan and {Jones}, Gareth C. and {Lyu}, Jianwei and {Rawle}, Tim and {Robertson}, Brant and {Rujopakarn}, Wiphu and {Perna}, Michele and {Sun}, Fengwu and {Venturi}, Giacomo and {Williams}, Christina C. and {Willott}, Chris},
        title = "{JADES: The diverse population of infant black holes at 4 < z < 11: Merging, tiny, poor, but mighty}",
      journal = {\aap},
         year = 2024,
        month = nov,
       volume = {691},
          eid = {A145},
        pages = {A145},
          doi = {10.1051/0004-6361/202347640},
archivePrefix = {arXiv},
       eprint = {2308.01230},
 primaryClass = {astro-ph.GA},
       adsurl = {https://ui.adsabs.harvard.edu/abs/2024A&A...691A.145M}
}

@ARTICLE{2025arXiv250512867L,
       author = {{Li}, Ruancun and {Ho}, Luis C. and {Chen}, Chang-Hao},
        title = "{The Dichotomy in the Nuclear and Host Galaxy Properties of High-redshift Quasars}",
      journal = {arXiv e-prints},
         year = 2025,
        month = may,
          eid = {arXiv:2505.12867},
        pages = {arXiv:2505.12867},
          doi = {10.48550/arXiv.2505.12867},
archivePrefix = {arXiv},
       eprint = {2505.12867},
 primaryClass = {astro-ph.GA},
       adsurl = {https://ui.adsabs.harvard.edu/abs/2025arXiv250512867L}
}

@ARTICLE{2025arXiv250205048L,
       author = {{Li}, Junyao and {Shen}, Yue and {Zhuang}, Ming-Yang},
        title = "{A prevalent population of normal-mass central black holes in high-redshift massive galaxies}",
      journal = {arXiv e-prints},
         year = 2025,
        month = feb,
          eid = {arXiv:2502.05048},
        pages = {arXiv:2502.05048},
          doi = {10.48550/arXiv.2502.05048},
archivePrefix = {arXiv},
       eprint = {2502.05048},
 primaryClass = {astro-ph.GA},
       adsurl = {https://ui.adsabs.harvard.edu/abs/2025arXiv250205048L}
}

@ARTICLE{2025arXiv250920452F,
       author = {{Fei}, Qinyue and {Fujimoto}, Seiji and {Naidu}, Rohan P. and {Chisholm}, John and {Atek}, Hakim and {Brammer}, Gabriel and {Asada}, Yoshihisa and {Bromm}, Volker and {Furtak}, Lukas J. and {Greene}, Jenny E. and {Hsiao}, Tiger Yu-Yang and {Jeon}, Junehyoung and {Kokorev}, Vasily and {Matthee}, Jorryt and {Natarajan}, Priyamvada and {Richard}, Johan and {Saldana-Lopez}, Alberto and {Schaerer}, Daniel and {Volonteri}, Marta and {Zitrin}, Adi},
        title = "{A GLIMPSE of Intermediate Mass Black holes in the epoch of reionization: Witnessing the Descendants of Direct Collapse?}",
      journal = {arXiv e-prints},
         year = 2025,
        month = sep,
          eid = {arXiv:2509.20452},
        pages = {arXiv:2509.20452},
          doi = {10.48550/arXiv.2509.20452},
archivePrefix = {arXiv},
       eprint = {2509.20452},
 primaryClass = {astro-ph.GA},
       adsurl = {https://ui.adsabs.harvard.edu/abs/2025arXiv250920452F}
}

@ARTICLE{2025arXiv250905434G,
       author = {{Greene}, Jenny E. and {Setton}, David J. and {Furtak}, Lukas J. and {Naidu}, Rohan P. and {Volonteri}, Marta and {Dayal}, Pratika and {Labbe}, Ivo and {van Dokkum}, Pieter and {Bezanson}, Rachel and {Brammer}, Gabriel and {Cutler}, Sam E. and {Glazebrook}, Karl and {de Graaff}, Anna and {Hirschmann}, Michaela and {Hviding}, Raphael E. and {Kokorev}, Vasily and {Leja}, Joel and {Liu}, Hanpu and {Ma}, Yilun and {Matthee}, Jorryt and {Nanayakkara}, Themiya and {Oesch}, Pascal A. and {Pan}, Richard and {Price}, Sedona H. and {Spilker}, Justin S. and {Wang}, Bingjie and {Weaver}, John R. and {Whitaker}, Katherine E. and {Williams}, Christina C. and {Zitrin}, Adi},
        title = "{What you see is what you get: empirically measured bolometric luminosities of Little Red Dots}",
      journal = {arXiv e-prints},
         year = 2025,
        month = sep,
          eid = {arXiv:2509.05434},
        pages = {arXiv:2509.05434},
          doi = {10.48550/arXiv.2509.05434},
archivePrefix = {arXiv},
       eprint = {2509.05434},
 primaryClass = {astro-ph.GA},
       adsurl = {https://ui.adsabs.harvard.edu/abs/2025arXiv250905434G}
}

@ARTICLE{2024OJAp....7E.107M,
       author = {{Mehta}, Daxal and {Regan}, John A. and {Prole}, Lewis},
        title = "{Growth of Light-Seed Black Holes in Gas-Rich Galaxies at High Redshift}",
      journal = {The Open Journal of Astrophysics},
         year = 2024,
        month = nov,
       volume = {7},
          eid = {107},
        pages = {107},
          doi = {10.33232/001c.126629},
archivePrefix = {arXiv},
       eprint = {2409.08326},
 primaryClass = {astro-ph.GA},
       adsurl = {https://ui.adsabs.harvard.edu/abs/2024OJAp....7E.107M}
}

@ARTICLE{2025MNRAS.538..518B,
       author = {{Bhowmick}, Aklant K. and {Blecha}, Laura and {Torrey}, Paul and {Somerville}, Rachel S. and {Kelley}, Luke Zoltan and {Weinberger}, Rainer and {Vogelsberger}, Mark and {Hernquist}, Lars and {Natarajan}, Priyamvada and {Kho}, Jonathan and {Di Matteo}, Tiziana},
        title = "{Signatures of black hole seeding in the local Universe: predictions from the BRAHMA cosmological simulations}",
      journal = {\mnras},
         year = 2025,
        month = mar,
       volume = {538},
       number = {1},
        pages = {518-536},
          doi = {10.1093/mnras/staf269},
archivePrefix = {arXiv},
       eprint = {2411.19332},
 primaryClass = {astro-ph.GA},
       adsurl = {https://ui.adsabs.harvard.edu/abs/2025MNRAS.538..518B}
}

@ARTICLE{2019AJ....157..236Y,
       author = {{Yang}, Jinyi and {Wang}, Feige and {Fan}, Xiaohui and {Yue}, Minghao and {Wu}, Xue-Bing and {Li}, Jiang-Tao and {Bian}, Fuyan and {Jiang}, Linhua and {Ba{\~n}ados}, Eduardo and {Beletsky}, Yuri},
        title = "{Exploring Reionization-era Quasars. IV. Discovery of Six New z {\ensuremath{\gtrsim}} 6.5 Quasars with DES, VHS, and unWISE Photometry}",
      journal = {\aj},
         year = 2019,
        month = jun,
       volume = {157},
       number = {6},
          eid = {236},
        pages = {236},
          doi = {10.3847/1538-3881/ab1be1},
archivePrefix = {arXiv},
       eprint = {1811.11915},
 primaryClass = {astro-ph.GA},
       adsurl = {https://ui.adsabs.harvard.edu/abs/2019AJ....157..236Y}
}

@ARTICLE{2020MNRAS.498.5652K,
       author = {{Kroupa}, Pavel and {Subr}, Ladislav and {Jerabkova}, Tereza and {Wang}, Long},
        title = "{Very high redshift quasars and the rapid emergence of supermassive black holes}",
      journal = {\mnras},
         year = 2020,
        month = nov,
       volume = {498},
       number = {4},
        pages = {5652-5683},
          doi = {10.1093/mnras/staa2276},
archivePrefix = {arXiv},
       eprint = {2007.14402},
 primaryClass = {astro-ph.GA},
       adsurl = {https://ui.adsabs.harvard.edu/abs/2020MNRAS.498.5652K}
}

@ARTICLE{2021MNRAS.tmp.1381D,
       author = {{Das}, Arpan and {Schleicher}, Dominik R.~G. and {Basu}, Shantanu and {Boekholt}, Tjarda C.~N.},
        title = "{Effect of mass loss due to stellar winds on the formation of supermassive black hole seeds in dense nuclear star clusters}",
      journal = {\mnras},
         year = 2021,
        month = may,
          doi = {10.1093/mnras/stab1428},
archivePrefix = {arXiv},
       eprint = {2105.03450},
 primaryClass = {astro-ph.GA},
       adsurl = {https://ui.adsabs.harvard.edu/abs/2021MNRAS.tmp.1381D}
}

@ARTICLE{2021MNRAS.503.1051D,
       author = {{Das}, Arpan and {Schleicher}, Dominik R.~G. and {Leigh}, Nathan W.~C. and {Boekholt}, Tjarda C.~N.},
        title = "{Formation of supermassive black hole seeds in nuclear star clusters via gas accretion and runaway collisions}",
      journal = {\mnras},
         year = 2021,
        month = may,
       volume = {503},
       number = {1},
        pages = {1051-1069},
          doi = {10.1093/mnras/stab402},
archivePrefix = {arXiv},
       eprint = {2012.01456},
 primaryClass = {astro-ph.GA},
       adsurl = {https://ui.adsabs.harvard.edu/abs/2021MNRAS.503.1051D}
}

@ARTICLE{2018ApJ...869L...9W,
       author = {{Wang}, Feige and {Yang}, Jinyi and {Fan}, Xiaohui and {Yue}, Minghao and {Wu}, Xue-Bing and {Schindler}, Jan-Torge and {Bian}, Fuyan and {Li}, Jiang-Tao and {Farina}, Emanuele P. and {Ba{\~n}ados}, Eduardo and {Davies}, Frederick B. and {Decarli}, Roberto and {Green}, Richard and {Jiang}, Linhua and {Hennawi}, Joseph F. and {Huang}, Yun-Hsin and {Mazzucchelli}, Chiara and {McGreer}, Ian D. and {Venemans}, Bram and {Walter}, Fabian and {Beletsky}, Yuri},
        title = "{The Discovery of a Luminous Broad Absorption Line Quasar at a Redshift of 7.02}",
      journal = {\apjl},
         year = 2018,
        month = dec,
       volume = {869},
       number = {1},
          eid = {L9},
        pages = {L9},
          doi = {10.3847/2041-8213/aaf1d2},
archivePrefix = {arXiv},
       eprint = {1810.11925},
 primaryClass = {astro-ph.GA},
       adsurl = {https://ui.adsabs.harvard.edu/abs/2018ApJ...869L...9W}
}

@ARTICLE{2018Natur.553..473B,
       author = {{Ba{\~n}ados}, Eduardo and {Venemans}, Bram P. and {Mazzucchelli}, Chiara and {Farina}, Emanuele P. and {Walter}, Fabian and {Wang}, Feige and {Decarli}, Roberto and {Stern}, Daniel and {Fan}, Xiaohui and {Davies}, Frederick B. and {Hennawi}, Joseph F. and {Simcoe}, Robert A. and {Turner}, Monica L. and {Rix}, Hans-Walter and {Yang}, Jinyi and {Kelson}, Daniel D. and {Rudie}, Gwen C. and {Winters}, Jan Martin},
        title = "{An 800-million-solar-mass black hole in a significantly neutral Universe at a redshift of 7.5}",
      journal = {\nat},
         year = 2018,
        month = jan,
       volume = {553},
       number = {7689},
        pages = {473-476},
          doi = {10.1038/nature25180},
archivePrefix = {arXiv},
       eprint = {1712.01860},
 primaryClass = {astro-ph.GA},
       adsurl = {https://ui.adsabs.harvard.edu/abs/2018Natur.553..473B}
}

@ARTICLE{2021ApJ...907L...1W,
       author = {{Wang}, Feige and {Yang}, Jinyi and {Fan}, Xiaohui and {Hennawi}, Joseph F. and {Barth}, Aaron J. and {Banados}, Eduardo and {Bian}, Fuyan and {Boutsia}, Konstantina and {Connor}, Thomas and {Davies}, Frederick B. and {Decarli}, Roberto and {Eilers}, Anna-Christina and {Farina}, Emanuele Paolo and {Green}, Richard and {Jiang}, Linhua and {Li}, Jiang-Tao and {Mazzucchelli}, Chiara and {Nanni}, Riccardo and {Schindler}, Jan-Torge and {Venemans}, Bram and {Walter}, Fabian and {Wu}, Xue-Bing and {Yue}, Minghao},
        title = "{A Luminous Quasar at Redshift 7.642}",
      journal = {\apjl},
         year = 2021,
        month = jan,
       volume = {907},
       number = {1},
          eid = {L1},
        pages = {L1},
          doi = {10.3847/2041-8213/abd8c6},
archivePrefix = {arXiv},
       eprint = {2101.03179},
 primaryClass = {astro-ph.GA},
       adsurl = {https://ui.adsabs.harvard.edu/abs/2021ApJ...907L...1W}
}

@ARTICLE{2019ApJ...872L...2M,
       author = {{Matsuoka}, Yoshiki and {Onoue}, Masafusa and {Kashikawa}, Nobunari and {Strauss}, Michael A. and {Iwasawa}, Kazushi and {Lee}, Chien-Hsiu and {Imanishi}, Masatoshi and {Nagao}, Tohru and {Akiyama}, Masayuki and {Asami}, Naoko and {Bosch}, James and {Furusawa}, Hisanori and {Goto}, Tomotsugu and {Gunn}, James E. and {Harikane}, Yuichi and {Ikeda}, Hiroyuki and {Izumi}, Takuma and {Kawaguchi}, Toshihiro and {Kato}, Nanako and {Kikuta}, Satoshi and {Kohno}, Kotaro and {Komiyama}, Yutaka and {Koyama}, Shuhei and {Lupton}, Robert H. and {Minezaki}, Takeo and {Miyazaki}, Satoshi and {Murayama}, Hitoshi and {Niida}, Mana and {Nishizawa}, Atsushi J. and {Noboriguchi}, Akatoki and {Oguri}, Masamune and {Ono}, Yoshiaki and {Ouchi}, Masami and {Price}, Paul A. and {Sameshima}, Hiroaki and {Schulze}, Andreas and {Shirakata}, Hikari and {Silverman}, John D. and {Sugiyama}, Naoshi and {Tait}, Philip J. and {Takada}, Masahiro and {Takata}, Tadafumi and {Tanaka}, Masayuki and {Tang}, Ji-Jia and {Toba}, Yoshiki and {Utsumi}, Yousuke and {Wang}, Shiang-Yu and {Yamashita}, Takuji},
        title = "{Discovery of the First Low-luminosity Quasar at z > 7}",
      journal = {\apjl},
         year = 2019,
        month = feb,
       volume = {872},
       number = {1},
          eid = {L2},
        pages = {L2},
          doi = {10.3847/2041-8213/ab0216},
archivePrefix = {arXiv},
       eprint = {1901.10487},
 primaryClass = {astro-ph.GA},
       adsurl = {https://ui.adsabs.harvard.edu/abs/2019ApJ...872L...2M}
}

@ARTICLE{2011Natur.474..616M,
       author = {{Mortlock}, Daniel J. and {Warren}, Stephen J. and {Venemans}, Bram P. and {Patel}, Mitesh and {Hewett}, Paul C. and {McMahon}, Richard G. and {Simpson}, Chris and {Theuns}, Tom and {Gonz{\'a}les-Solares}, Eduardo A. and {Adamson}, Andy and {Dye}, Simon and {Hambly}, Nigel C. and {Hirst}, Paul and {Irwin}, Mike J. and {Kuiper}, Ernst and {Lawrence}, Andy and {R{\"o}ttgering}, Huub J.~A.},
        title = "{A luminous quasar at a redshift of z = 7.085}",
      journal = {\nat},
         year = 2011,
        month = jun,
       volume = {474},
       number = {7353},
        pages = {616-619},
          doi = {10.1038/nature10159},
archivePrefix = {arXiv},
       eprint = {1106.6088},
 primaryClass = {astro-ph.CO},
       adsurl = {https://ui.adsabs.harvard.edu/abs/2011Natur.474..616M}
}

@ARTICLE{2018ApJS..237....5M,
       author = {{Matsuoka}, Yoshiki and {Iwasawa}, Kazushi and {Onoue}, Masafusa and {Kashikawa}, Nobunari and {Strauss}, Michael A. and {Lee}, Chien-Hsiu and {Imanishi}, Masatoshi and {Nagao}, Tohru and {Akiyama}, Masayuki and {Asami}, Naoko and {Bosch}, James and {Furusawa}, Hisanori and {Goto}, Tomotsugu and {Gunn}, James E. and {Harikane}, Yuichi and {Ikeda}, Hiroyuki and {Izumi}, Takuma and {Kawaguchi}, Toshihiro and {Kato}, Nanako and {Kikuta}, Satoshi and {Kohno}, Kotaro and {Komiyama}, Yutaka and {Lupton}, Robert H. and {Minezaki}, Takeo and {Miyazaki}, Satoshi and {Morokuma}, Tomoki and {Murayama}, Hitoshi and {Niida}, Mana and {Nishizawa}, Atsushi J. and {Oguri}, Masamune and {Ono}, Yoshiaki and {Ouchi}, Masami and {Price}, Paul A. and {Sameshima}, Hiroaki and {Schulze}, Andreas and {Shirakata}, Hikari and {Silverman}, John D. and {Sugiyama}, Naoshi and {Tait}, Philip J. and {Takada}, Masahiro and {Takata}, Tadafumi and {Tanaka}, Masayuki and {Tang}, Ji-Jia and {Toba}, Yoshiki and {Utsumi}, Yousuke and {Wang}, Shiang-Yu and {Yamashita}, Takuji},
        title = "{Subaru High-z Exploration of Low-luminosity Quasars (SHELLQs). IV. Discovery of 41 Quasars and Luminous Galaxies at 5.7 {\ensuremath{\leq}} z {\ensuremath{\leq}} 6.9}",
      journal = {\apjs},
         year = 2018,
        month = jul,
       volume = {237},
       number = {1},
          eid = {5},
        pages = {5},
          doi = {10.3847/1538-4365/aac724},
archivePrefix = {arXiv},
       eprint = {1803.01861},
 primaryClass = {astro-ph.GA},
       adsurl = {https://ui.adsabs.harvard.edu/abs/2018ApJS..237....5M}
}

@ARTICLE{2017MNRAS.468.4702R,
       author = {{Reed}, S.~L. and {McMahon}, R.~G. and {Martini}, P. and {Banerji}, M. and {Auger}, M. and {Hewett}, P.~C. and {Koposov}, S.~E. and {Gibbons}, S.~L.~J. and {Gonzalez-Solares}, E. and {Ostrovski}, F. and {Tie}, S.~S. and {Abdalla}, F.~B. and {Allam}, S. and {Benoit-L{\'e}vy}, A. and {Bertin}, E. and {Brooks}, D. and {Buckley-Geer}, E. and {Burke}, D.~L. and {Carnero Rosell}, A. and {Carrasco Kind}, M. and {Carretero}, J. and {da Costa}, L.~N. and {DePoy}, D.~L. and {Desai}, S. and {Diehl}, H.~T. and {Doel}, P. and {Evrard}, A.~E. and {Finley}, D.~A. and {Flaugher}, B. and {Fosalba}, P. and {Frieman}, J. and {Garc{\'\i}a-Bellido}, J. and {Gaztanaga}, E. and {Goldstein}, D.~A. and {Gruen}, D. and {Gruendl}, R.~A. and {Gutierrez}, G. and {James}, D.~J. and {Kuehn}, K. and {Kuropatkin}, N. and {Lahav}, O. and {Lima}, M. and {Maia}, M.~A.~G. and {Marshall}, J.~L. and {Melchior}, P. and {Miller}, C.~J. and {Miquel}, R. and {Nord}, B. and {Ogando}, R. and {Plazas}, A.~A. and {Romer}, A.~K. and {Sanchez}, E. and {Scarpine}, V. and {Schubnell}, M. and {Sevilla-Noarbe}, I. and {Smith}, R.~C. and {Sobreira}, F. and {Suchyta}, E. and {Swanson}, M.~E.~C. and {Tarle}, G. and {Tucker}, D.~L. and {Walker}, A.~R. and {Wester}, W.},
        title = "{Eight new luminous z {\ensuremath{\geq}} 6 quasars discovered via SED model fitting of VISTA, WISE and Dark Energy Survey Year 1 observations}",
      journal = {\mnras},
         year = 2017,
        month = jul,
       volume = {468},
       number = {4},
        pages = {4702-4718},
          doi = {10.1093/mnras/stx728},
archivePrefix = {arXiv},
       eprint = {1701.04852},
 primaryClass = {astro-ph.GA},
       adsurl = {https://ui.adsabs.harvard.edu/abs/2017MNRAS.468.4702R}
}

@article{2016Banados,
   title={THE PAN-STARRS1 DISTANT
                    z
                    > 5.6 QUASAR SURVEY: MORE THAN 100 QUASARS WITHIN THE FIRST GYR OF THE UNIVERSE},
   volume={227},
   ISSN={1538-4365},
   url={http://dx.doi.org/10.3847/0067-0049/227/1/11},
   DOI={10.3847/0067-0049/227/1/11},
   number={1},
   journal={The Astrophysical Journal Supplement Series},
   publisher={American Astronomical Society},
   author={Bañados, E. and Venemans, B. P. and Decarli, R. and Farina, E. P. and Mazzucchelli, C. and Walter, F. and Fan, X. and Stern, D. and Schlafly, E. and Chambers, K. C. and et al.},
   year={2016},
   month={Nov},
   pages={11}
}

@ARTICLE{2016ApJ...833..222J,
       author = {{Jiang}, Linhua and {McGreer}, Ian D. and {Fan}, Xiaohui and {Strauss}, Michael A. and {Ba{\~n}ados}, Eduardo and {Becker}, Robert H. and {Bian}, Fuyan and {Farnsworth}, Kara and {Shen}, Yue and {Wang}, Feige and {Wang}, Ran and {Wang}, Shu and {White}, Richard L. and {Wu}, Jin and {Wu}, Xue-Bing and {Yang}, Jinyi and {Yang}, Qian},
        title = "{The Final SDSS High-redshift Quasar Sample of 52 Quasars at z>5.7}",
      journal = {\apj},
         year = 2016,
        month = dec,
       volume = {833},
       number = {2},
          eid = {222},
        pages = {222},
          doi = {10.3847/1538-4357/833/2/222},
archivePrefix = {arXiv},
       eprint = {1610.05369},
 primaryClass = {astro-ph.GA},
       adsurl = {https://ui.adsabs.harvard.edu/abs/2016ApJ...833..222J}
}

@ARTICLE{2015MNRAS.453.2259V,
       author = {{Venemans}, B.~P. and {Verdoes Kleijn}, G.~A. and {Mwebaze}, J. and {Valentijn}, E.~A. and {Ba{\~n}ados}, E. and {Decarli}, R. and {de Jong}, J.~T.~A. and {Findlay}, J.~R. and {Kuijken}, K.~H. and {La Barbera}, F. and {McFarland}, J.~P. and {McMahon}, R.~G. and {Napolitano}, N. and {Sikkema}, G. and {Sutherland}, W.~J.},
        title = "{First discoveries of z {\ensuremath{\sim}} 6 quasars with the Kilo-Degree Survey and VISTA Kilo-Degree Infrared Galaxy survey}",
      journal = {\mnras},
         year = 2015,
        month = nov,
       volume = {453},
       number = {3},
        pages = {2259-2266},
          doi = {10.1093/mnras/stv1774},
archivePrefix = {arXiv},
       eprint = {1507.00726},
 primaryClass = {astro-ph.GA},
       adsurl = {https://ui.adsabs.harvard.edu/abs/2015MNRAS.453.2259V}
}

@ARTICLE{2010AJ....139..906W,
       author = {{Willott}, Chris J. and {Delorme}, Philippe and {Reyl{\'e}}, C{\'e}line and {Albert}, Loic and {Bergeron}, Jacqueline and {Crampton}, David and {Delfosse}, Xavier and {Forveille}, Thierry and {Hutchings}, John B. and {McLure}, Ross J. and {Omont}, Alain and {Schade}, David},
        title = "{The Canada-France High-z Quasar Survey: Nine New Quasars and the Luminosity Function at Redshift 6}",
      journal = {\aj},
         year = 2010,
        month = mar,
       volume = {139},
       number = {3},
        pages = {906-918},
          doi = {10.1088/0004-6256/139/3/906},
archivePrefix = {arXiv},
       eprint = {0912.0281},
 primaryClass = {astro-ph.CO},
       adsurl = {https://ui.adsabs.harvard.edu/abs/2010AJ....139..906W}
}

@ARTICLE{2001AJ....122.2833F,
       author = {{Fan}, Xiaohui and {Narayanan}, Vijay K. and {Lupton}, Robert H. and {Strauss}, Michael A. and {Knapp}, Gillian R. and {Becker}, Robert H. and {White}, Richard L. and {Pentericci}, Laura and {Leggett}, S.~K. and {Haiman}, Zolt{\'a}n and {Gunn}, James E. and {Ivezi{\'c}}, {\v{Z}}eljko and {Schneider}, Donald P. and {Anderson}, Scott F. and {Brinkmann}, J. and {Bahcall}, Neta A. and {Connolly}, Andrew J. and {Csabai}, Istv{\'a}n and {Doi}, Mamoru and {Fukugita}, Masataka and {Geballe}, Tom and {Grebel}, Eva K. and {Harbeck}, Daniel and {Hennessy}, Gregory and {Lamb}, Don Q. and {Miknaitis}, Gajus and {Munn}, Jeffrey A. and {Nichol}, Robert and {Okamura}, Sadanori and {Pier}, Jeffrey R. and {Prada}, Francisco and {Richards}, Gordon T. and {Szalay}, Alex and {York}, Donald G.},
        title = "{A Survey of z>5.8 Quasars in the Sloan Digital Sky Survey. I. Discovery of Three New Quasars and the Spatial Density of Luminous Quasars at z\raisebox{-0.5ex}\textasciitilde6}",
      journal = {\aj},
         year = 2001,
        month = dec,
       volume = {122},
       number = {6},
        pages = {2833-2849},
          doi = {10.1086/324111},
archivePrefix = {arXiv},
       eprint = {astro-ph/0108063},
 primaryClass = {astro-ph},
       adsurl = {https://ui.adsabs.harvard.edu/abs/2001AJ....122.2833F}
}

@ARTICLE{2022MNRAS.510..177B,
       author = {{Bhowmick}, Aklant K. and {Blecha}, Laura and {Torrey}, Paul and {Kelley}, Luke Zoltan and {Vogelsberger}, Mark and {Nelson}, Dylan and {Weinberger}, Rainer and {Hernquist}, Lars},
        title = "{Impact of gas spin and Lyman-Werner flux on black hole seed formation in cosmological simulations: implications for direct collapse}",
      journal = {\mnras},
         year = 2022,
        month = feb,
       volume = {510},
       number = {1},
        pages = {177-196},
          doi = {10.1093/mnras/stab3439},
archivePrefix = {arXiv},
       eprint = {2107.06899},
 primaryClass = {astro-ph.GA},
       adsurl = {https://ui.adsabs.harvard.edu/abs/2022MNRAS.510..177B}
}

@ARTICLE{2014MNRAS.442.2751T,
       author = {{Taylor}, P. and {Kobayashi}, C.},
        title = "{Seeding black holes in cosmological simulations}",
      journal = {\mnras},
         year = 2014,
        month = aug,
       volume = {442},
       number = {3},
        pages = {2751-2767},
          doi = {10.1093/mnras/stu983},
archivePrefix = {arXiv},
       eprint = {1405.4194},
 primaryClass = {astro-ph.GA},
       adsurl = {https://ui.adsabs.harvard.edu/abs/2014MNRAS.442.2751T}
}

@ARTICLE{2021MNRAS.507.2012B,
       author = {{Bhowmick}, Aklant K. and {Blecha}, Laura and {Torrey}, Paul and {Kelley}, Luke Zoltan and {Vogelsberger}, Mark and {Kosciw}, Kaitlyn and {Nelson}, Dylan and {Weinberger}, Rainer and {Hernquist}, Lars},
        title = "{Impact of gas-based seeding on supermassive black hole populations at z {\ensuremath{\geq}} 7}",
      journal = {\mnras},
         year = 2021,
        month = oct,
       volume = {507},
       number = {2},
        pages = {2012-2036},
          doi = {10.1093/mnras/stab2204},
archivePrefix = {arXiv},
       eprint = {2105.08055},
 primaryClass = {astro-ph.GA},
       adsurl = {https://ui.adsabs.harvard.edu/abs/2021MNRAS.507.2012B}
}

@ARTICLE{2013MNRAS.436.3031V,
       author = {{Vogelsberger}, Mark and {Genel}, Shy and {Sijacki}, Debora and {Torrey}, Paul and {Springel}, Volker and {Hernquist}, Lars},
        title = "{A model for cosmological simulations of galaxy formation physics}",
      journal = {\mnras},
         year = 2013,
        month = dec,
       volume = {436},
       number = {4},
        pages = {3031-3067},
          doi = {10.1093/mnras/stt1789},
archivePrefix = {arXiv},
       eprint = {1305.2913},
 primaryClass = {astro-ph.CO},
       adsurl = {https://ui.adsabs.harvard.edu/abs/2013MNRAS.436.3031V}
}

@ARTICLE{2008MNRAS.385.1443S,
       author = {{Smith}, Britton and {Sigurdsson}, Steinn and {Abel}, Tom},
        title = "{Metal cooling in simulations of cosmic structure formation}",
      journal = {\mnras},
         year = 2008,
        month = apr,
       volume = {385},
       number = {3},
        pages = {1443-1454},
          doi = {10.1111/j.1365-2966.2008.12922.x},
archivePrefix = {arXiv},
       eprint = {0706.0754},
 primaryClass = {astro-ph},
       adsurl = {https://ui.adsabs.harvard.edu/abs/2008MNRAS.385.1443S}
}

@ARTICLE{1996ApJS..105...19K,
       author = {{Katz}, Neal and {Weinberg}, David H. and {Hernquist}, Lars},
        title = "{Cosmological Simulations with TreeSPH}",
      journal = {\apjs},
         year = 1996,
        month = jul,
       volume = {105},
        pages = {19},
          doi = {10.1086/192305},
archivePrefix = {arXiv},
       eprint = {astro-ph/9509107},
 primaryClass = {astro-ph},
       adsurl = {https://ui.adsabs.harvard.edu/abs/1996ApJS..105...19K}
}

@ARTICLE{2018ApJ...865L...9V,
       author = {{Visbal}, Eli and {Haiman}, Zolt{\'a}n},
        title = "{Identifying Direct Collapse Black Hole Seeds through Their Small Host Galaxies}",
      journal = {\apjl},
         year = 2018,
        month = sep,
       volume = {865},
       number = {1},
          eid = {L9},
        pages = {L9},
          doi = {10.3847/2041-8213/aadf3a},
archivePrefix = {arXiv},
       eprint = {1809.01754},
 primaryClass = {astro-ph.GA},
       adsurl = {https://ui.adsabs.harvard.edu/abs/2018ApJ...865L...9V}
}

@ARTICLE{2018MNRAS.481.3278R,
       author = {{Ricarte}, Angelo and {Natarajan}, Priyamvada},
        title = "{The observational signatures of supermassive black hole seeds}",
      journal = {\mnras},
         year = 2018,
        month = dec,
       volume = {481},
       number = {3},
        pages = {3278-3292},
          doi = {10.1093/mnras/sty2448},
archivePrefix = {arXiv},
       eprint = {1809.01177},
 primaryClass = {astro-ph.GA},
       adsurl = {https://ui.adsabs.harvard.edu/abs/2018MNRAS.481.3278R}
}

@ARTICLE{2019Natur.566...85W,
       author = {{Wise}, John H. and {Regan}, John A. and {O'Shea}, Brian W. and {Norman}, Michael L. and {Downes}, Turlough P. and {Xu}, Hao},
        title = "{Formation of massive black holes in rapidly growing pre-galactic gas clouds}",
      journal = {\nat},
         year = 2019,
        month = jan,
       volume = {566},
       number = {7742},
        pages = {85-88},
          doi = {10.1038/s41586-019-0873-4},
archivePrefix = {arXiv},
       eprint = {1901.07563},
 primaryClass = {astro-ph.GA},
       adsurl = {https://ui.adsabs.harvard.edu/abs/2019Natur.566...85W}
}

@ARTICLE{2003ApJ...596...34B,
       author = {{Bromm}, Volker and {Loeb}, Abraham},
        title = "{Formation of the First Supermassive Black Holes}",
      journal = {\apj},
         year = 2003,
        month = oct,
       volume = {596},
       number = {1},
        pages = {34-46},
          doi = {10.1086/377529},
archivePrefix = {arXiv},
       eprint = {astro-ph/0212400},
 primaryClass = {astro-ph},
       adsurl = {https://ui.adsabs.harvard.edu/abs/2003ApJ...596...34B}
}

@ARTICLE{2016MNRAS.463..529H,
       author = {{Habouzit}, M{\'e}lanie and {Volonteri}, Marta and {Latif}, Muhammad and {Dubois}, Yohan and {Peirani}, S{\'e}bastien},
        title = "{On the number density of `direct collapse' black hole seeds}",
      journal = {\mnras},
         year = 2016,
        month = nov,
       volume = {463},
       number = {1},
        pages = {529-540},
          doi = {10.1093/mnras/stw1924},
archivePrefix = {arXiv},
       eprint = {1601.00557},
 primaryClass = {astro-ph.GA},
       adsurl = {https://ui.adsabs.harvard.edu/abs/2016MNRAS.463..529H}
}

@ARTICLE{2012MNRAS.422.2051N,
       author = {{Natarajan}, Priyamvada and {Volonteri}, Marta},
        title = "{The mass function of black holes 1<z<4.5: comparison of models with observations}",
      journal = {\mnras},
         year = 2012,
        month = may,
       volume = {422},
       number = {3},
        pages = {2051-2057},
          doi = {10.1111/j.1365-2966.2012.20708.x},
archivePrefix = {arXiv},
       eprint = {1107.4916},
 primaryClass = {astro-ph.CO},
       adsurl = {https://ui.adsabs.harvard.edu/abs/2012MNRAS.422.2051N}
}

@ARTICLE{2003PASP..115..763C,
       author = {{Chabrier}, Gilles},
        title = "{Galactic Stellar and Substellar Initial Mass Function}",
      journal = {\pasp},
         year = 2003,
        month = jul,
       volume = {115},
       number = {809},
        pages = {763-795},
          doi = {10.1086/376392},
archivePrefix = {arXiv},
       eprint = {astro-ph/0304382},
 primaryClass = {astro-ph},
       adsurl = {https://ui.adsabs.harvard.edu/abs/2003PASP..115..763C}
}

@ARTICLE{2024MNRAS.531.4311B,
       author = {{Bhowmick}, Aklant K. and {Blecha}, Laura and {Torrey}, Paul and {Kelley}, Luke Zoltan and {Weinberger}, Rainer and {Vogelsberger}, Mark and {Hernquist}, Lars and {Somerville}, Rachel S. and {Evans}, Analis Eolyn},
        title = "{Introducing the BRAHMA simulation suite: signatures of low-mass black hole seeding models in cosmological simulations}",
      journal = {\mnras},
         year = 2024,
        month = jul,
       volume = {531},
       number = {4},
        pages = {4311-4335},
          doi = {10.1093/mnras/stae1386},
archivePrefix = {arXiv},
       eprint = {2402.03626},
 primaryClass = {astro-ph.GA},
       adsurl = {https://ui.adsabs.harvard.edu/abs/2024MNRAS.531.4311B}
}

@ARTICLE{2024MNRAS.529.3768B,
       author = {{Bhowmick}, Aklant K. and {Blecha}, Laura and {Torrey}, Paul and {Weinberger}, Rainer and {Kelley}, Luke Zoltan and {Vogelsberger}, Mark and {Hernquist}, Lars and {Somerville}, Rachel S.},
        title = "{Representing low-mass black hole seeds in cosmological simulations: A new sub-grid stochastic seed model}",
      journal = {\mnras},
         year = 2024,
        month = apr,
       volume = {529},
       number = {4},
        pages = {3768-3792},
          doi = {10.1093/mnras/stae780},
archivePrefix = {arXiv},
       eprint = {2309.15341},
 primaryClass = {astro-ph.GA},
       adsurl = {https://ui.adsabs.harvard.edu/abs/2024MNRAS.529.3768B}
}

@ARTICLE{2018MNRAS.479.4056W,
       author = {{Weinberger}, Rainer and {Springel}, Volker and {Pakmor}, R{\"u}diger and {Nelson}, Dylan and {Genel}, Shy and {Pillepich}, Annalisa and {Vogelsberger}, Mark and {Marinacci}, Federico and {Naiman}, Jill and {Torrey}, Paul and {Hernquist}, Lars},
        title = "{Supermassive black holes and their feedback effects in the IllustrisTNG simulation}",
      journal = {\mnras},
         year = 2018,
        month = sep,
       volume = {479},
       number = {3},
        pages = {4056-4072},
          doi = {10.1093/mnras/sty1733},
archivePrefix = {arXiv},
       eprint = {1710.04659},
 primaryClass = {astro-ph.GA},
       adsurl = {https://ui.adsabs.harvard.edu/abs/2018MNRAS.479.4056W}
}

@ARTICLE{2014MNRAS.444.1518V,
       author = {{Vogelsberger}, Mark and {Genel}, Shy and {Springel}, Volker and {Torrey}, Paul and {Sijacki}, Debora and {Xu}, Dandan and {Snyder}, Greg and {Nelson}, Dylan and {Hernquist}, Lars},
        title = "{Introducing the Illustris Project: simulating the coevolution of dark and visible matter in the Universe}",
      journal = {\mnras},
         year = 2014,
        month = oct,
       volume = {444},
       number = {2},
        pages = {1518-1547},
          doi = {10.1093/mnras/stu1536},
archivePrefix = {arXiv},
       eprint = {1405.2921},
 primaryClass = {astro-ph.CO},
       adsurl = {https://ui.adsabs.harvard.edu/abs/2014MNRAS.444.1518V}
}

@ARTICLE{2012MNRAS.423.2533B,
       author = {{Barausse}, Enrico},
        title = "{The evolution of massive black holes and their spins in their galactic hosts}",
      journal = {\mnras},
         year = 2012,
        month = jul,
       volume = {423},
       number = {3},
        pages = {2533-2557},
          doi = {10.1111/j.1365-2966.2012.21057.x},
archivePrefix = {arXiv},
       eprint = {1201.5888},
 primaryClass = {astro-ph.CO},
       adsurl = {https://ui.adsabs.harvard.edu/abs/2012MNRAS.423.2533B}
}

@ARTICLE{2020MNRAS.491.4973D,
       author = {{DeGraf}, C. and {Sijacki}, D.},
        title = "{Cosmological simulations of massive black hole seeds: predictions for next-generation electromagnetic and gravitational wave observations}",
      journal = {\mnras},
         year = 2020,
        month = feb,
       volume = {491},
       number = {4},
        pages = {4973-4992},
          doi = {10.1093/mnras/stz3309},
archivePrefix = {arXiv},
       eprint = {1906.11271},
 primaryClass = {astro-ph.GA},
       adsurl = {https://ui.adsabs.harvard.edu/abs/2020MNRAS.491.4973D}
}

@ARTICLE{2006MNRAS.370..289B,
       author = {{Begelman}, Mitchell C. and {Volonteri}, Marta and {Rees}, Martin J.},
        title = "{Formation of supermassive black holes by direct collapse in pre-galactic haloes}",
      journal = {\mnras},
         year = 2006,
        month = jul,
       volume = {370},
       number = {1},
        pages = {289-298},
          doi = {10.1111/j.1365-2966.2006.10467.x},
archivePrefix = {arXiv},
       eprint = {astro-ph/0602363},
 primaryClass = {astro-ph},
       adsurl = {https://ui.adsabs.harvard.edu/abs/2006MNRAS.370..289B}
}

@ARTICLE{2018MNRAS.476.3523L,
       author = {{Luo}, Yang and {Ardaneh}, Kazem and {Shlosman}, Isaac and {Nagamine}, Kentaro and {Wise}, John H. and {Begelman}, Mitchell C.},
        title = "{Direct Collapse to Supermassive Black Hole Seeds with Radiative Transfer: Isolated Halos}",
      journal = {\mnras},
         year = 2018,
        month = may,
       volume = {476},
       number = {3},
        pages = {3523-3539},
          doi = {10.1093/mnras/sty362},
archivePrefix = {arXiv},
       eprint = {1801.08545},
 primaryClass = {astro-ph.GA},
       adsurl = {https://ui.adsabs.harvard.edu/abs/2018MNRAS.476.3523L}
}

@ARTICLE{2020MNRAS.492.4917L,
       author = {{Luo}, Yang and {Shlosman}, Isaac and {Nagamine}, Kentaro and {Fang}, Taotao},
        title = "{Direct collapse to supermassive black hole seeds: the critical conditions for suppression of H$_{2}$ cooling}",
      journal = {\mnras},
         year = 2020,
        month = mar,
       volume = {492},
       number = {4},
        pages = {4917-4926},
          doi = {10.1093/mnras/staa153},
archivePrefix = {arXiv},
       eprint = {1910.07556},
 primaryClass = {astro-ph.GA},
       adsurl = {https://ui.adsabs.harvard.edu/abs/2020MNRAS.492.4917L}
}

@ARTICLE{2014MNRAS.442.3616L,
       author = {{Lupi}, A. and {Colpi}, M. and {Devecchi}, B. and {Galanti}, G. and {Volonteri}, M.},
        title = "{Constraining the high-redshift formation of black hole seeds in nuclear star clusters with gas inflows}",
      journal = {\mnras},
         year = 2014,
        month = aug,
       volume = {442},
       number = {4},
        pages = {3616-3626},
          doi = {10.1093/mnras/stu1120},
archivePrefix = {arXiv},
       eprint = {1406.2325},
 primaryClass = {astro-ph.GA},
       adsurl = {https://ui.adsabs.harvard.edu/abs/2014MNRAS.442.3616L}
}

@ARTICLE{2011ApJ...740L..42D,
       author = {{Davies}, Melvyn B. and {Miller}, M. Coleman and {Bellovary}, Jillian M.},
        title = "{Supermassive Black Hole Formation Via Gas Accretion in Nuclear Stellar Clusters}",
      journal = {\apjl},
         year = 2011,
        month = oct,
       volume = {740},
       number = {2},
          eid = {L42},
        pages = {L42},
          doi = {10.1088/2041-8205/740/2/L42},
archivePrefix = {arXiv},
       eprint = {1106.5943},
 primaryClass = {astro-ph.CO},
       adsurl = {https://ui.adsabs.harvard.edu/abs/2011ApJ...740L..42D}
}

@ARTICLE{2001ApJ...550..372F,
       author = {{Fryer}, C.~L. and {Woosley}, S.~E. and {Heger}, A.},
        title = "{Pair-Instability Supernovae, Gravity Waves, and Gamma-Ray Transients}",
      journal = {\apj},
         year = 2001,
        month = mar,
       volume = {550},
       number = {1},
        pages = {372-382},
          doi = {10.1086/319719},
archivePrefix = {arXiv},
       eprint = {astro-ph/0007176},
 primaryClass = {astro-ph},
       adsurl = {https://ui.adsabs.harvard.edu/abs/2001ApJ...550..372F}
}

@ARTICLE{2001ApJ...551L..27M,
       author = {{Madau}, Piero and {Rees}, Martin J.},
        title = "{Massive Black Holes as Population III Remnants}",
      journal = {\apjl},
         year = 2001,
        month = apr,
       volume = {551},
       number = {1},
        pages = {L27-L30},
          doi = {10.1086/319848},
archivePrefix = {arXiv},
       eprint = {astro-ph/0101223},
 primaryClass = {astro-ph},
       adsurl = {https://ui.adsabs.harvard.edu/abs/2001ApJ...551L..27M}
}

@ARTICLE{2001MNRAS.328..726S,
       author = {{Springel}, Volker and {White}, Simon D.~M. and {Tormen}, Giuseppe and {Kauffmann}, Guinevere},
        title = "{Populating a cluster of galaxies - I. Results at z=0}",
      journal = {\mnras},
         year = 2001,
        month = dec,
       volume = {328},
       number = {3},
        pages = {726-750},
          doi = {10.1046/j.1365-8711.2001.04912.x},
archivePrefix = {arXiv},
       eprint = {astro-ph/0012055},
 primaryClass = {astro-ph},
       adsurl = {https://ui.adsabs.harvard.edu/abs/2001MNRAS.328..726S}
}

@ARTICLE{2023ApJ...951L...8A,
       author = {{Agazie}, Gabriella and {Anumarlapudi}, Akash and {Archibald}, Anne M. and {Arzoumanian}, Zaven and {Baker}, Paul T. and {B{\'e}csy}, Bence and {Blecha}, Laura and {Brazier}, Adam and {Brook}, Paul R. and {Burke-Spolaor}, Sarah and {Burnette}, Rand and {Case}, Robin and {Charisi}, Maria and {Chatterjee}, Shami and {Chatziioannou}, Katerina and {Cheeseboro}, Belinda D. and {Chen}, Siyuan and {Cohen}, Tyler and {Cordes}, James M. and {Cornish}, Neil J. and {Crawford}, Fronefield and {Cromartie}, H. Thankful and {Crowter}, Kathryn and {Cutler}, Curt J. and {Decesar}, Megan E. and {Degan}, Dallas and {Demorest}, Paul B. and {Deng}, Heling and {Dolch}, Timothy and {Drachler}, Brendan and {Ellis}, Justin A. and {Ferrara}, Elizabeth C. and {Fiore}, William and {Fonseca}, Emmanuel and {Freedman}, Gabriel E. and {Garver-Daniels}, Nate and {Gentile}, Peter A. and {Gersbach}, Kyle A. and {Glaser}, Joseph and {Good}, Deborah C. and {G{\"u}ltekin}, Kayhan and {Hazboun}, Jeffrey S. and {Hourihane}, Sophie and {Islo}, Kristina and {Jennings}, Ross J. and {Johnson}, Aaron D. and {Jones}, Megan L. and {Kaiser}, Andrew R. and {Kaplan}, David L. and {Kelley}, Luke Zoltan and {Kerr}, Matthew and {Key}, Joey S. and {Klein}, Tonia C. and {Laal}, Nima and {Lam}, Michael T. and {Lamb}, William G. and {Lazio}, T. Joseph W. and {Lewandowska}, Natalia and {Littenberg}, Tyson B. and {Liu}, Tingting and {Lommen}, Andrea and {Lorimer}, Duncan R. and {Luo}, Jing and {Lynch}, Ryan S. and {Ma}, Chung-Pei and {Madison}, Dustin R. and {Mattson}, Margaret A. and {McEwen}, Alexander and {McKee}, James W. and {McLaughlin}, Maura A. and {McMann}, Natasha and {Meyers}, Bradley W. and {Meyers}, Patrick M. and {Mingarelli}, Chiara M.~F. and {Mitridate}, Andrea and {Natarajan}, Priyamvada and {Ng}, Cherry and {Nice}, David J. and {Ocker}, Stella Koch and {Olum}, Ken D. and {Pennucci}, Timothy T. and {Perera}, Benetge B.~P. and {Petrov}, Polina and {Pol}, Nihan S. and {Radovan}, Henri A. and {Ransom}, Scott M. and {Ray}, Paul S. and {Romano}, Joseph D. and {Sardesai}, Shashwat C. and {Schmiedekamp}, Ann and {Schmiedekamp}, Carl and {Schmitz}, Kai and {Schult}, Levi and {Shapiro-Albert}, Brent J. and {Siemens}, Xavier and {Simon}, Joseph and {Siwek}, Magdalena S. and {Stairs}, Ingrid H. and {Stinebring}, Daniel R. and {Stovall}, Kevin and {Sun}, Jerry P. and {Susobhanan}, Abhimanyu and {Swiggum}, Joseph K. and {Taylor}, Jacob and {Taylor}, Stephen R. and {Turner}, Jacob E. and {Unal}, Caner and {Vallisneri}, Michele and {van Haasteren}, Rutger and {Vigeland}, Sarah J. and {Wahl}, Haley M. and {Wang}, Qiaohong and {Witt}, Caitlin A. and {Young}, Olivia and {Nanograv Collaboration}},
        title = "{The NANOGrav 15 yr Data Set: Evidence for a Gravitational-wave Background}",
      journal = {\apjl},
         year = 2023,
        month = jul,
       volume = {951},
       number = {1},
          eid = {L8},
        pages = {L8},
          doi = {10.3847/2041-8213/acdac6},
archivePrefix = {arXiv},
       eprint = {2306.16213},
 primaryClass = {astro-ph.HE},
       adsurl = {https://ui.adsabs.harvard.edu/abs/2023ApJ...951L...8A}
}

@ARTICLE{2026ApJ...997..187B,
       author = {{Bhowmick}, Aklant K. and {Blecha}, Laura and {Torrey}, Paul and {Kelley}, Luke Zoltan and {Natarajan}, Priyamvada and {Somerville}, Rachel S. and {Weinberger}, Rainer and {Garcia}, Alex M. and {Hernquist}, Lars and {Di Matteo}, Tiziana and {Kho}, Jonathan and {Vogelsberger}, Mark},
        title = "{Heavy Seeds and the First Black Holes: Insights from the BRAHMA Simulations}",
      journal = {\apj},
         year = 2026,
        month = feb,
       volume = {997},
       number = {2},
          eid = {187},
        pages = {187},
          doi = {10.3847/1538-4357/ae2607},
archivePrefix = {arXiv},
       eprint = {2510.01322},
 primaryClass = {astro-ph.GA},
       adsurl = {https://ui.adsabs.harvard.edu/abs/2026ApJ...997..187B}
}

@ARTICLE{2025MNRAS.542.2597C,
       author = {{Cenci}, Elia and {Habouzit}, Melanie},
        title = "{Little Red Dots as direct-collapse black hole nurseries}",
      journal = {\mnras},
         year = 2025,
        month = sep,
       volume = {542},
       number = {3},
        pages = {2597-2609},
          doi = {10.1093/mnras/staf1362},
archivePrefix = {arXiv},
       eprint = {2508.14897},
 primaryClass = {astro-ph.GA},
       adsurl = {https://ui.adsabs.harvard.edu/abs/2025MNRAS.542.2597C}
}

@ARTICLE{2024A&A...691A..52K,
       author = {{Killi}, Meghana and {Watson}, Darach and {Brammer}, Gabriel and {McPartland}, Conor and {Antwi-Danso}, Jacqueline and {Newshore}, Rosa and {Coe}, Dan and {Allen}, Natalie and {Fynbo}, Johan P.~U. and {Gould}, Katriona and {Heintz}, Kasper E. and {Rusakov}, Vadim and {Vejlgaard}, Simone},
        title = "{Deciphering the JWST spectrum of a 'little red dot' at z {\ensuremath{\sim}} 4.53: An obscured AGN and its star-forming host}",
      journal = {\aap},
         year = 2024,
        month = nov,
       volume = {691},
          eid = {A52},
        pages = {A52},
          doi = {10.1051/0004-6361/202348857},
archivePrefix = {arXiv},
       eprint = {2312.03065},
 primaryClass = {astro-ph.GA},
       adsurl = {https://ui.adsabs.harvard.edu/abs/2024A&A...691A..52K}
}

@ARTICLE{2017arXiv170200786A,
       author = {{Amaro-Seoane}, Pau and {Audley}, Heather and {Babak}, Stanislav and {Baker}, John and {Barausse}, Enrico and {Bender}, Peter and {Berti}, Emanuele and {Binetruy}, Pierre and {Born}, Michael and {Bortoluzzi}, Daniele and {Camp}, Jordan and {Caprini}, Chiara and {Cardoso}, Vitor and {Colpi}, Monica and {Conklin}, John and {Cornish}, Neil and {Cutler}, Curt and {Danzmann}, Karsten and {Dolesi}, Rita and {Ferraioli}, Luigi and {Ferroni}, Valerio and {Fitzsimons}, Ewan and {Gair}, Jonathan and {Gesa Bote}, Lluis and {Giardini}, Domenico and {Gibert}, Ferran and {Grimani}, Catia and {Halloin}, Hubert and {Heinzel}, Gerhard and {Hertog}, Thomas and {Hewitson}, Martin and {Holley-Bockelmann}, Kelly and {Hollington}, Daniel and {Hueller}, Mauro and {Inchauspe}, Henri and {Jetzer}, Philippe and {Karnesis}, Nikos and {Killow}, Christian and {Klein}, Antoine and {Klipstein}, Bill and {Korsakova}, Natalia and {Larson}, Shane L and {Livas}, Jeffrey and {Lloro}, Ivan and {Man}, Nary and {Mance}, Davor and {Martino}, Joseph and {Mateos}, Ignacio and {McKenzie}, Kirk and {McWilliams}, Sean T and {Miller}, Cole and {Mueller}, Guido and {Nardini}, Germano and {Nelemans}, Gijs and {Nofrarias}, Miquel and {Petiteau}, Antoine and {Pivato}, Paolo and {Plagnol}, Eric and {Porter}, Ed and {Reiche}, Jens and {Robertson}, David and {Robertson}, Norna and {Rossi}, Elena and {Russano}, Giuliana and {Schutz}, Bernard and {Sesana}, Alberto and {Shoemaker}, David and {Slutsky}, Jacob and {Sopuerta}, Carlos F. and {Sumner}, Tim and {Tamanini}, Nicola and {Thorpe}, Ira and {Troebs}, Michael and {Vallisneri}, Michele and {Vecchio}, Alberto and {Vetrugno}, Daniele and {Vitale}, Stefano and {Volonteri}, Marta and {Wanner}, Gudrun and {Ward}, Harry and {Wass}, Peter and {Weber}, William and {Ziemer}, John and {Zweifel}, Peter},
        title = "{Laser Interferometer Space Antenna}",
      journal = {arXiv e-prints},
         year = 2017,
        month = feb,
          eid = {arXiv:1702.00786},
        pages = {arXiv:1702.00786},
          doi = {10.48550/arXiv.1702.00786},
archivePrefix = {arXiv},
       eprint = {1702.00786},
 primaryClass = {astro-ph.IM},
       adsurl = {https://ui.adsabs.harvard.edu/abs/2017arXiv170200786A}
}

@ARTICLE{1985ApJ...292..371D,
       author = {{Davis}, M. and {Efstathiou}, G. and {Frenk}, C.~S. and {White}, S.~D.~M.},
        title = "{The evolution of large-scale structure in a universe dominated by cold dark matter}",
      journal = {\apj},
         year = 1985,
        month = may,
       volume = {292},
        pages = {371-394},
          doi = {10.1086/163168},
       adsurl = {https://ui.adsabs.harvard.edu/abs/1985ApJ...292..371D}
}

@ARTICLE{2013ARA&A..51..511K,
       author = {{Kormendy}, John and {Ho}, Luis C.},
        title = "{Coevolution (Or Not) of Supermassive Black Holes and Host Galaxies}",
      journal = {\araa},
         year = 2013,
        month = aug,
       volume = {51},
       number = {1},
        pages = {511-653},
          doi = {10.1146/annurev-astro-082708-101811},
archivePrefix = {arXiv},
       eprint = {1304.7762},
 primaryClass = {astro-ph.CO},
       adsurl = {https://ui.adsabs.harvard.edu/abs/2013ARA&A..51..511K}
}

@ARTICLE{1986Natur.324..446B,
       author = {{Barnes}, Josh and {Hut}, Piet},
        title = "{A hierarchical O(N log N) force-calculation algorithm}",
      journal = {\nat},
         year = 1986,
        month = dec,
       volume = {324},
       number = {6096},
        pages = {446-449},
          doi = {10.1038/324446a0},
       adsurl = {https://ui.adsabs.harvard.edu/abs/1986Natur.324..446B}
}

@ARTICLE{2018MNRAS.475..648P,
       author = {{Pillepich}, Annalisa and {Nelson}, Dylan and {Hernquist}, Lars and {Springel}, Volker and {Pakmor}, R{\"u}diger and {Torrey}, Paul and {Weinberger}, Rainer and {Genel}, Shy and {Naiman}, Jill P. and {Marinacci}, Federico and {Vogelsberger}, Mark},
        title = "{First results from the IllustrisTNG simulations: the stellar mass content of groups and clusters of galaxies}",
      journal = {\mnras},
         year = 2018,
        month = mar,
       volume = {475},
       number = {1},
        pages = {648-675},
          doi = {10.1093/mnras/stx3112},
archivePrefix = {arXiv},
       eprint = {1707.03406},
 primaryClass = {astro-ph.GA},
       adsurl = {https://ui.adsabs.harvard.edu/abs/2018MNRAS.475..648P}
}

@ARTICLE{2010MNRAS.401..791S,
       author = {{Springel}, Volker},
        title = "{E pur si muove: Galilean-invariant cosmological hydrodynamical simulations on a moving mesh}",
      journal = {\mnras},
         year = 2010,
        month = jan,
       volume = {401},
       number = {2},
        pages = {791-851},
          doi = {10.1111/j.1365-2966.2009.15715.x},
archivePrefix = {arXiv},
       eprint = {0901.4107},
 primaryClass = {astro-ph.CO},
       adsurl = {https://ui.adsabs.harvard.edu/abs/2010MNRAS.401..791S}
}

@ARTICLE{2020ApJS..248...32W,
       author = {{Weinberger}, Rainer and {Springel}, Volker and {Pakmor}, R{\"u}diger},
        title = "{The AREPO Public Code Release}",
      journal = {\apjs},
         year = 2020,
        month = jun,
       volume = {248},
       number = {2},
          eid = {32},
        pages = {32},
          doi = {10.3847/1538-4365/ab908c},
archivePrefix = {arXiv},
       eprint = {1909.04667},
 primaryClass = {astro-ph.IM},
       adsurl = {https://ui.adsabs.harvard.edu/abs/2020ApJS..248...32W}
}

@ARTICLE{2013ApJ...773...83X,
       author = {{Xu}, Hao and {Wise}, John H. and {Norman}, Michael L.},
        title = "{Population III Stars and Remnants in High-redshift Galaxies}",
      journal = {\apj},
         year = 2013,
        month = aug,
       volume = {773},
       number = {2},
          eid = {83},
        pages = {83},
          doi = {10.1088/0004-637X/773/2/83},
archivePrefix = {arXiv},
       eprint = {1305.1325},
 primaryClass = {astro-ph.CO},
       adsurl = {https://ui.adsabs.harvard.edu/abs/2013ApJ...773...83X}
}

@ARTICLE{2018MNRAS.480.3762S,
       author = {{Smith}, Britton D. and {Regan}, John A. and {Downes}, Turlough P. and {Norman}, Michael L. and {O'Shea}, Brian W. and {Wise}, John H.},
        title = "{The growth of black holes from Population III remnants in the Renaissance simulations}",
      journal = {\mnras},
         year = 2018,
        month = nov,
       volume = {480},
       number = {3},
        pages = {3762-3773},
          doi = {10.1093/mnras/sty2103},
archivePrefix = {arXiv},
       eprint = {1804.06477},
 primaryClass = {astro-ph.GA},
       adsurl = {https://ui.adsabs.harvard.edu/abs/2018MNRAS.480.3762S}
}

@ARTICLE{2007ApJ...654..731H,
       author = {{Hopkins}, Philip F. and {Richards}, Gordon T. and {Hernquist}, Lars},
        title = "{An Observational Determination of the Bolometric Quasar Luminosity Function}",
      journal = {\apj},
         year = 2007,
        month = jan,
       volume = {654},
       number = {2},
        pages = {731-753},
          doi = {10.1086/509629},
archivePrefix = {arXiv},
       eprint = {astro-ph/0605678},
 primaryClass = {astro-ph},
       adsurl = {https://ui.adsabs.harvard.edu/abs/2007ApJ...654..731H}
}

@ARTICLE{2020MNRAS.495.3252S,
       author = {{Shen}, Xuejian and {Hopkins}, Philip F. and {Faucher-Gigu{\`e}re}, Claude-Andr{\'e} and {Alexander}, D.~M. and {Richards}, Gordon T. and {Ross}, Nicholas P. and {Hickox}, R.~C.},
        title = "{The bolometric quasar luminosity function at z = 0-7}",
      journal = {\mnras},
         year = 2020,
        month = may,
       volume = {495},
       number = {3},
        pages = {3252-3275},
          doi = {10.1093/mnras/staa1381},
archivePrefix = {arXiv},
       eprint = {2001.02696},
 primaryClass = {astro-ph.GA},
       adsurl = {https://ui.adsabs.harvard.edu/abs/2020MNRAS.495.3252S}
}

@ARTICLE{2011MNRAS.415.2101H,
       author = {{Hahn}, Oliver and {Abel}, Tom},
        title = "{Multi-scale initial conditions for cosmological simulations}",
      journal = {\mnras},
         year = 2011,
        month = aug,
       volume = {415},
       number = {3},
        pages = {2101-2121},
          doi = {10.1111/j.1365-2966.2011.18820.x},
archivePrefix = {arXiv},
       eprint = {1103.6031},
 primaryClass = {astro-ph.CO},
       adsurl = {https://ui.adsabs.harvard.edu/abs/2011MNRAS.415.2101H}
}

@ARTICLE{2017MNRAS.465.3291W,
       author = {{Weinberger}, Rainer and {Springel}, Volker and {Hernquist}, Lars and {Pillepich}, Annalisa and {Marinacci}, Federico and {Pakmor}, R{\"u}diger and {Nelson}, Dylan and {Genel}, Shy and {Vogelsberger}, Mark and {Naiman}, Jill and {Torrey}, Paul},
        title = "{Simulating galaxy formation with black hole driven thermal and kinetic feedback}",
      journal = {\mnras},
         year = 2017,
        month = mar,
       volume = {465},
       number = {3},
        pages = {3291-3308},
          doi = {10.1093/mnras/stw2944},
archivePrefix = {arXiv},
       eprint = {1607.03486},
 primaryClass = {astro-ph.GA},
       adsurl = {https://ui.adsabs.harvard.edu/abs/2017MNRAS.465.3291W}
}

@ARTICLE{2018MNRAS.473.4077P,
       author = {{Pillepich}, Annalisa and {Springel}, Volker and {Nelson}, Dylan and {Genel}, Shy and {Naiman}, Jill and {Pakmor}, R{\"u}diger and {Hernquist}, Lars and {Torrey}, Paul and {Vogelsberger}, Mark and {Weinberger}, Rainer and {Marinacci}, Federico},
        title = "{Simulating galaxy formation with the IllustrisTNG model}",
      journal = {\mnras},
         year = 2018,
        month = jan,
       volume = {473},
       number = {3},
        pages = {4077-4106},
          doi = {10.1093/mnras/stx2656},
archivePrefix = {arXiv},
       eprint = {1703.02970},
 primaryClass = {astro-ph.GA},
       adsurl = {https://ui.adsabs.harvard.edu/abs/2018MNRAS.473.4077P}
}

@ARTICLE{2003MNRAS.339..289S,
       author = {{Springel}, Volker and {Hernquist}, Lars},
        title = "{Cosmological smoothed particle hydrodynamics simulations: a hybrid multiphase model for star formation}",
      journal = {\mnras},
         year = 2003,
        month = feb,
       volume = {339},
       number = {2},
        pages = {289-311},
          doi = {10.1046/j.1365-8711.2003.06206.x},
archivePrefix = {arXiv},
       eprint = {astro-ph/0206393},
 primaryClass = {astro-ph},
       adsurl = {https://ui.adsabs.harvard.edu/abs/2003MNRAS.339..289S}
}

@ARTICLE{2016MNRAS.462.2603P,
       author = {{Pakmor}, R{\"u}diger and {Pfrommer}, Christoph and {Simpson}, Christine M. and {Kannan}, Rahul and {Springel}, Volker},
        title = "{Semi-implicit anisotropic cosmic ray transport on an unstructured moving mesh}",
      journal = {\mnras},
         year = 2016,
        month = nov,
       volume = {462},
       number = {3},
        pages = {2603-2616},
          doi = {10.1093/mnras/stw1761},
archivePrefix = {arXiv},
       eprint = {1604.08587},
 primaryClass = {astro-ph.GA},
       adsurl = {https://ui.adsabs.harvard.edu/abs/2016MNRAS.462.2603P}
}

@ARTICLE{2011MNRAS.418.1392P,
       author = {{Pakmor}, Ruediger and {Bauer}, Andreas and {Springel}, Volker},
        title = "{Magnetohydrodynamics on an unstructured moving grid}",
      journal = {\mnras},
         year = 2011,
        month = dec,
       volume = {418},
       number = {2},
        pages = {1392-1401},
          doi = {10.1111/j.1365-2966.2011.19591.x},
archivePrefix = {arXiv},
       eprint = {1108.1792},
 primaryClass = {astro-ph.IM},
       adsurl = {https://ui.adsabs.harvard.edu/abs/2011MNRAS.418.1392P}
}

@ARTICLE{2023MNRAS.519.5543M,
       author = {{Ma}, Linhao and {Hopkins}, Philip F. and {Kelley}, Luke Zoltan and {Faucher-Gigu{\`e}re}, Claude-Andr{\'e}},
        title = "{A new discrete dynamical friction estimator based on N-body simulations}",
      journal = {\mnras},
         year = 2023,
        month = mar,
       volume = {519},
       number = {4},
        pages = {5543-5553},
          doi = {10.1093/mnras/stad036},
archivePrefix = {arXiv},
       eprint = {2208.12275},
 primaryClass = {astro-ph.GA},
       adsurl = {https://ui.adsabs.harvard.edu/abs/2023MNRAS.519.5543M}
}

@ARTICLE{2016A&A...594A..13P,
       author = {{Planck Collaboration} and {Ade}, P.~A.~R. and {Aghanim}, N. and {Arnaud}, M. and {Ashdown}, M. and {Aumont}, J. and {Baccigalupi}, C. and {Banday}, A.~J. and {Barreiro}, R.~B. and {Bartlett}, J.~G. and {Bartolo}, N. and {Battaner}, E. and {Battye}, R. and {Benabed}, K. and {Beno{\^\i}t}, A. and {Benoit-L{\'e}vy}, A. and {Bernard}, J. -P. and {Bersanelli}, M. and {Bielewicz}, P. and {Bock}, J.~J. and {Bonaldi}, A. and {Bonavera}, L. and {Bond}, J.~R. and {Borrill}, J. and {Bouchet}, F.~R. and {Boulanger}, F. and {Bucher}, M. and {Burigana}, C. and {Butler}, R.~C. and {Calabrese}, E. and {Cardoso}, J. -F. and {Catalano}, A. and {Challinor}, A. and {Chamballu}, A. and {Chary}, R. -R. and {Chiang}, H.~C. and {Chluba}, J. and {Christensen}, P.~R. and {Church}, S. and {Clements}, D.~L. and {Colombi}, S. and {Colombo}, L.~P.~L. and {Combet}, C. and {Coulais}, A. and {Crill}, B.~P. and {Curto}, A. and {Cuttaia}, F. and {Danese}, L. and {Davies}, R.~D. and {Davis}, R.~J. and {de Bernardis}, P. and {de Rosa}, A. and {de Zotti}, G. and {Delabrouille}, J. and {D{\'e}sert}, F. -X. and {Di Valentino}, E. and {Dickinson}, C. and {Diego}, J.~M. and {Dolag}, K. and {Dole}, H. and {Donzelli}, S. and {Dor{\'e}}, O. and {Douspis}, M. and {Ducout}, A. and {Dunkley}, J. and {Dupac}, X. and {Efstathiou}, G. and {Elsner}, F. and {En{\ss}lin}, T.~A. and {Eriksen}, H.~K. and {Farhang}, M. and {Fergusson}, J. and {Finelli}, F. and {Forni}, O. and {Frailis}, M. and {Fraisse}, A.~A. and {Franceschi}, E. and {Frejsel}, A. and {Galeotta}, S. and {Galli}, S. and {Ganga}, K. and {Gauthier}, C. and {Gerbino}, M. and {Ghosh}, T. and {Giard}, M. and {Giraud-H{\'e}raud}, Y. and {Giusarma}, E. and {Gjerl{\o}w}, E. and {Gonz{\'a}lez-Nuevo}, J. and {G{\'o}rski}, K.~M. and {Gratton}, S. and {Gregorio}, A. and {Gruppuso}, A. and {Gudmundsson}, J.~E. and {Hamann}, J. and {Hansen}, F.~K. and {Hanson}, D. and {Harrison}, D.~L. and {Helou}, G. and {Henrot-Versill{\'e}}, S. and {Hern{\'a}ndez-Monteagudo}, C. and {Herranz}, D. and {Hildebrandt}, S.~R. and {Hivon}, E. and {Hobson}, M. and {Holmes}, W.~A. and {Hornstrup}, A. and {Hovest}, W. and {Huang}, Z. and {Huffenberger}, K.~M. and {Hurier}, G. and {Jaffe}, A.~H. and {Jaffe}, T.~R. and {Jones}, W.~C. and {Juvela}, M. and {Keih{\"a}nen}, E. and {Keskitalo}, R. and {Kisner}, T.~S. and {Kneissl}, R. and {Knoche}, J. and {Knox}, L. and {Kunz}, M. and {Kurki-Suonio}, H. and {Lagache}, G. and {L{\"a}hteenm{\"a}ki}, A. and {Lamarre}, J. -M. and {Lasenby}, A. and {Lattanzi}, M. and {Lawrence}, C.~R. and {Leahy}, J.~P. and {Leonardi}, R. and {Lesgourgues}, J. and {Levrier}, F. and {Lewis}, A. and {Liguori}, M. and {Lilje}, P.~B. and {Linden-V{\o}rnle}, M. and {L{\'o}pez-Caniego}, M. and {Lubin}, P.~M. and {Mac{\'\i}as-P{\'e}rez}, J.~F. and {Maggio}, G. and {Maino}, D. and {Mandolesi}, N. and {Mangilli}, A. and {Marchini}, A. and {Maris}, M. and {Martin}, P.~G. and {Martinelli}, M. and {Mart{\'\i}nez-Gonz{\'a}lez}, E. and {Masi}, S. and {Matarrese}, S. and {McGehee}, P. and {Meinhold}, P.~R. and {Melchiorri}, A. and {Melin}, J. -B. and {Mendes}, L. and {Mennella}, A. and {Migliaccio}, M. and {Millea}, M. and {Mitra}, S. and {Miville-Desch{\^e}nes}, M. -A. and {Moneti}, A. and {Montier}, L. and {Morgante}, G. and {Mortlock}, D. and {Moss}, A. and {Munshi}, D. and {Murphy}, J.~A. and {Naselsky}, P. and {Nati}, F. and {Natoli}, P. and {Netterfield}, C.~B. and {N{\o}rgaard-Nielsen}, H.~U. and {Noviello}, F. and {Novikov}, D. and {Novikov}, I. and {Oxborrow}, C.~A. and {Paci}, F. and {Pagano}, L. and {Pajot}, F. and {Paladini}, R. and {Paoletti}, D. and {Partridge}, B. and {Pasian}, F. and {Patanchon}, G. and {Pearson}, T.~J. and {Perdereau}, O. and {Perotto}, L. and {Perrotta}, F. and {Pettorino}, V. and {Piacentini}, F. and {Piat}, M. and {Pierpaoli}, E. and {Pietrobon}, D. and {Plaszczynski}, S. and {Pointecouteau}, E. and {Polenta}, G. and {Popa}, L. and {Pratt}, G.~W. and {Pr{\'e}zeau}, G. and {Prunet}, S. and {Puget}, J. -L. and {Rachen}, J.~P. and {Reach}, W.~T. and {Rebolo}, R. and {Reinecke}, M. and {Remazeilles}, M. and {Renault}, C. and {Renzi}, A. and {Ristorcelli}, I. and {Rocha}, G. and {Rosset}, C. and {Rossetti}, M. and {Roudier}, G. and {Rouill{\'e} d'Orfeuil}, B. and {Rowan-Robinson}, M. and {Rubi{\~n}o-Mart{\'\i}n}, J.~A. and {Rusholme}, B. and {Said}, N. and {Salvatelli}, V. and {Salvati}, L. and {Sandri}, M. and {Santos}, D. and {Savelainen}, M. and {Savini}, G. and {Scott}, D. and {Seiffert}, M.~D. and {Serra}, P. and {Shellard}, E.~P.~S. and {Spencer}, L.~D. and {Spinelli}, M. and {Stolyarov}, V. and {Stompor}, R. and {Sudiwala}, R. and {Sunyaev}, R. and {Sutton}, D. and {Suur-Uski}, A. -S. and {Sygnet}, J. -F. and {Tauber}, J.~A. and {Terenzi}, L. and {Toffolatti}, L. and {Tomasi}, M. and {Tristram}, M. and {Trombetti}, T. and {Tucci}, M. and {Tuovinen}, J. and {T{\"u}rler}, M. and {Umana}, G. and {Valenziano}, L. and {Valiviita}, J. and {Van Tent}, F. and {Vielva}, P. and {Villa}, F. and {Wade}, L.~A. and {Wandelt}, B.~D. and {Wehus}, I.~K. and {White}, M. and {White}, S.~D.~M. and {Wilkinson}, A. and {Yvon}, D. and {Zacchei}, A. and {Zonca}, A.},
        title = "{Planck 2015 results. XIII. Cosmological parameters}",
      journal = {\aap},
         year = 2016,
        month = sep,
       volume = {594},
          eid = {A13},
        pages = {A13},
          doi = {10.1051/0004-6361/201525830},
archivePrefix = {arXiv},
       eprint = {1502.01589},
 primaryClass = {astro-ph.CO},
       adsurl = {https://ui.adsabs.harvard.edu/abs/2016A&A...594A..13P}
}

@ARTICLE{2014MNRAS.438.1985T,
       author = {{Torrey}, Paul and {Vogelsberger}, Mark and {Genel}, Shy and {Sijacki}, Debora and {Springel}, Volker and {Hernquist}, Lars},
        title = "{A model for cosmological simulations of galaxy formation physics: multi-epoch validation}",
      journal = {\mnras},
         year = 2014,
        month = mar,
       volume = {438},
       number = {3},
        pages = {1985-2004},
          doi = {10.1093/mnras/stt2295},
archivePrefix = {arXiv},
       eprint = {1305.4931},
 primaryClass = {astro-ph.CO},
       adsurl = {https://ui.adsabs.harvard.edu/abs/2014MNRAS.438.1985T}
}

\end{document}